\documentclass[a4paper,fleqn,manuscript,usenatbib]{mnras}

\usepackage{caption}
\usepackage{amsmath}
\usepackage{amssymb}
\usepackage{multicol}
\usepackage{rotating}
\usepackage{lscape}
\usepackage{graphicx}
\usepackage{color}
\usepackage{placeins}
\usepackage{xcolor}
\usepackage{float}
\usepackage{url}
\usepackage{verbatim}
\usepackage{lipsum,adjustbox}
\usepackage{subcaption}

\usepackage{tikz}
\usetikzlibrary{shapes,snakes,calc,positioning}
\usetikzlibrary{arrows,positioning} 

\tikzset{
 myarrow/.style={->, >=latex', shorten >=1pt, thick},
    >=stealth',
    punkt/.style={
           rectangle,
           rounded corners,
           draw=black, very thick,
           text width=6.5em,
           minimum height=2em,
           text centered},
    pil/.style={
           ->,
           thick,
           shorten <=2pt,
           shorten >=2pt,}
}

\newcommand{\beq}{\begin{equation}}
\newcommand{\eeq}{\end{equation}}
\newcommand{\beqa}{\begin{eqnarray}}
\newcommand{\eeqa}{\end{eqnarray}}

\definecolor{gray}{gray}{0.55}

\def\Mpc{\, h^{-1} \, {\rm Mpc}}

\newcommand{\kmsmpc}{\mbox{km\,s$^{-1}$\,Mpc$^{-1}$}}
\newcommand{\mbi}[1]{\mbox{\boldmath$#1$}}

\newcommand{\mat}[1]{\mbox{\rm\bf #1}}
\newcommand{\lsim}{\mbox{${\,\hbox{\hbox{$ < $}\kern -0.8em \lower 1.0ex\hbox{$\sim$}}\,}$}}
\newcommand{\gsim}{\mbox{${\,\hbox{\hbox{$ > $}\kern -0.8em \lower 1.0ex\hbox{$\sim$}}\,}$}}

\newcommand{\dd}{{\rm d}}

\def\beqn{\vspace{2mm}
\begin{eqnarray}} 
\def\eeqn{\vspace{2mm} 
\end{eqnarray}}

 \def\la{\langle}

\newcommand{\be}{\begin{equation}}
\newcommand{\ee}{\end{equation}}
\newcommand{\ba}{\begin{eqnarray}}
\newcommand{\ea}{\end{eqnarray}}
\newcommand{\brr}{\begin{array}}
 
\newcommand{\err}{\end{array}}
\newcommand{\bc}{\begin{center}}
\newcommand{\ec}{\end{center}}

\newcommand{\hperm}{\,h\,{\rm Mpc}^{-1}}

\title[Cosmic Flows and Cosmic Web from LRGs]{The Clustering of Galaxies in the  Completed SDSS-III Baryon Oscillation Spectroscopic Survey: Cosmic Flows and Cosmic Web  from  Luminous Red Galaxies}

\author[Ata, Kitaura et al.]{ \hspace{-0.35cm} Metin~Ata$^1$\thanks{E-mail:mata@aip.de}, Francisco-Shu Kitaura$^{1,2,3,4,5}$\thanks{E-mail:kitaura@aip.de}, Chia-Hsun Chuang$^{6,1}$, \and Sergio Rodr{\'i}guez-Torres$^{6,7,8}$,    Raul E.~Angulo$^9$,  Simone Ferraro$^{2,3}$,   Hector Gil-Mar{\'in}$^{10,11,12}$, \and  Patrick McDonald$^{2}$, Carlos Hern{\'a}ndez Monteagudo$^9$, Volker M{\"u}ller$^1$,  Gustavo Yepes$^{8}$, \and   Mathieu Autefage$^1$, Falk Baumgarten$^1$,  Florian Beutler$^{12}$,    Joel R.~Brownstein$^{13}$, \and  Angela Burden$^{14}$,  Daniel J.~Eisenstein$^{15}$,   Hong Guo$^{16}$, Shirley Ho$^{17}$,    Cameron McBride$^{15}$, \and Mark Neyrinck$^{18}$,   Matthew D.~Olmstead$^{19}$,  Nikhil Padmanabhan$^{14}$,      Will J.~Percival$^{12}$,  \and    Francisco Prada$^{6,7,20}$,   Graziano Rossi$^{21}$, Ariel G.~S{\'a}nchez$^{22}$,   David Schlegel$^{3}$,  \and  Donald P. Schneider$^{23,24}$, Hee-Jong Seo$^{25}$, Alina Streblyanska$^{4}$,     Jeremy Tinker$^{26}$, \and  Rita Tojeiro$^{27}$,  Mariana Vargas-Magana$^{28}$\\ 
{\it Affiliations are listed at the end of the paper}
\vspace{-.5cm}
}

\date{\today.\vspace{-.5cm}}

\begin{document}
\maketitle

\begin{abstract}

We present a Bayesian phase-space reconstruction of the cosmic large-scale matter density and velocity fields from the SDSS-III Baryon Oscillations Spectroscopic Survey Data Release 12 (BOSS DR12) CMASS  galaxy clustering catalogue.
We rely on a given $\Lambda$CDM cosmology, a mesh resolution in the range of 6-10 $h^{-1}$ Mpc, and a lognormal-Poisson model with a redshift dependent nonlinear bias. The bias parameters are derived from the data and a general renormalised perturbation theory approach.
We use combined Gibbs and Hamiltonian sampling, implemented in the \textsc{argo} code, to iteratively reconstruct the dark matter density field and the coherent peculiar velocities of individual galaxies, correcting hereby for coherent redshift space distortions (RSD). 
Our tests relying on accurate $N$-body based mock galaxy catalogues, show unbiased real space power spectra of the nonlinear density field up to $k\sim0.2\, h$ Mpc$^{-1}$, and vanishing quadrupoles down to $r\sim20\,h^{-1}$ Mpc. We also demonstrate that the nonlinear cosmic web can be obtained from the tidal field tensor based on the Gaussian component of the reconstructed density field. We find that the reconstructed velocities have a statistical correlation coefficient compared to the true velocities of each individual lightcone mock galaxy of $r\sim0.68$  including about 10\% of satellite galaxies with virial motions (about $r=0.75$ without satellites). The power spectra of the velocity divergence agree well with theoretical predictions up to $k\sim0.2\,h\,{\rm Mpc}^{-1}$.
This work will be especially useful to improve, e.g. BAO reconstructions, kinematic Sunyaev-Zeldovich (kSZ), integrated Sachs-Wolfe (ISW) measurements, or environmental studies.
\end{abstract}

\begin{keywords}
%\vspace{-1.cm}
  cosmology: theory -- large-scale structure of the Universe -- catalogues -- galaxies: statistics -- methods: numerical
\end{keywords}

\section{Introduction}

The large-scale structure of the Universe is a key observable probe to study cosmology.
Galaxy redshift surveys provide a three dimensional picture of the distribution of luminous tracers across the history of the Universe after cosmic dawn.
The recovery of this information relies on accurate modelling of effects including the survey geometry, radial selection functions, galaxy bias, and redshift space distortions caused by the peculiar motions of galaxies. 

%\pagebreak

Many studies require reliable reconstructions of the large-scale gravitational potential from which also the coherent peculiar velocities can be derived. This is the case of the integrated Sachs Wolfe effect \cite[see e.g.,][]{2008ApJ...683L..99G,2013A&A...556A..51I}, the kinematic Sunyaev-Zeldovich effect \citep[see e.g.,][]{2016A&A...586A.140P,Carlos15,Schaan2015}, the cosmic flows  \citep[e.g.][]{2009MNRAS.392..743W,2010ApJ...709..483L,2012MNRAS.424..472B,Courtois_flows_2012,kit2mrs,2016MNRAS.456.4247H}, or the baryon acoustic oscillations (BAO) reconstructions  \citep[see e.g.][]{ESS07,PXE12,AAB14,RSH15}. Also environmental studies of galaxies demonstrated to benefit from accurate density and velocity reconstructions \citep[see][]{Nuza14}.

In addition, a number of works have suggested nonlinear transformations, Gaussianising the density field to obtain improved cosmological constraints \citep[][]{2009ApJ...698L..90N,2011ApJ...731..116N,2011PhRvD..84b3523Y,2011MNRAS.418..145J,2014MNRAS.439L..11C,2016PhRvD..93b3525S}. Also, linearised density fields can yield improved displacement and peculiar velocity fields \citep[][]{kitlin,kitaura_vel,Falck12}.

Nevertheless, all these studies are affected by redshift space distortions and the sparsity of the signal, which must be handled carefully \citep[][]{2016MNRAS.457.3652M}.  Indeed, \citet[][]{2012JCAP...03..004S} has pointed out that if not properly modeled, nonlinear transformations on density fields including redshift space distortions can lead to biased results. Such a careful modeling is one motivation for the current work.

The  inferred galaxy line-of-sight position is a combination of the so-called  Hubble flow, i.e. their real distance, and their peculiar motion. The modifications produced by this effect are referred to as redshift space distortions (RSD). They can  be used to constrain the nature of gravity and cosmological parameters  \citep[see e.g.][for recent studies]{BerNarWei01,Zhang07,Jain08,Guz08,NesPer08,Song09a,Song09b,PerWhi08,McDSel09,WhiSonPer09,Song10,Zhao10,Song11}.
The measurement of RSD have in fact become a common technique \citep[][]{CFW95,Pea01,Per04,Ang08,Oku08,Guz08,WiggleZRSD,JenBauPas11,KwaLewLin12,SamPerRac12,Reid12,OkuSelDes12,Chuang2013,Chuang2013b,Chuang2013c,Chuang2016,Samushia13,Zheng13,Bla13,Tor13,Beutler14,Samushia14,San14,Bel14,Toj14,Oku14,Beutler14,WangY2014,Alam2015}.
These studies are usually based on the large-scale anisotropic clustering displayed by the galaxy distribution in redshift space, although $N$-body based models for fitting the data to smaller scales have been presented in \citet[][]{Reid14,Guo2015,Guo2015b,Guo2016}. 
A recent study suggested to measure the growth rate from density reconstructions  \citep{Granettvimos}. However, instead of correcting redshift space distortions, these were included in the power spectrum used to recover the density field in redshift space.

Different approaches have been proposed in the literature to recover the peculiar velocity field from galaxy distributions   \cite[][]{1991ApJ...372..380Y,1993ApJ...405L..47G,zaroubi,1995MNRAS.272..885F,1996ApJ...473...22D,1997MNRAS.285..793C,1999MNRAS.308..763M,2002MNRAS.335...53B,2008MNRAS.383.1292L,2012MNRAS.424..472B,2012MNRAS.420.1809W,kit2mrs,2016MNRAS.456.4247H}, based on various density-velocity relations  \citep[see][]{1991ApJ...379....6N,1992ApJ...390L..61B,1998MNRAS.300.1027C,1999MNRAS.309..543B,2000MNRAS.316..464K,2005ApJ...635L.113M,2008MNRAS.391.1796B,2015MNRAS.449.3407J,kitaura_vel,2016MNRAS.457.2773N}.

The main objective of this paper is to perform a self-consistent inference analysis of the density and peculiar velocity field on large scales   accounting for all the above mentioned  systematic effects (survey geometry, radial selection function, galaxy bias, RSD, non-Gaussian statistics, shot noise).
We will rely on the lognormal-Poisson model within the Bayesian framework  \citep[][]{kitaura_log} to infer the density field from the galaxy distribution. 
Lognormal-Poisson Bayesian inference performed independently on each density cell reduces to a sufficient statistic characterizing the density field at the two-point level \citep[][]{2014MNRAS.439L..11C}, but including the density covariance matrix as done here carries additional statistical power.
Furthermore we will iteratively solve for redshift space distortions relying on linear theory \citep[][]{Kitaura:2015bca}. 

 More complex priors describing the density field than the lognormal assumption can be used  \citep[][]{Coles1991}, based on perturbation theory \citep[][]{Kitaura2013,Jasche2013,Wang2013,HKG13}, or even on particle mesh approaches \citep[][]{Wang2014}. Also the likelihood describing the statistical distribution of galaxies can be improved modelling the deviation from Poissonity \citep[][]{Ata15}. Moreover, the relation between the density and the peculiar velocity field could be more accurately modelled including tidal field tensors \citep[][]{kitaura_vel}.
In this work, we want to focus however, on the simplest and most efficient models, which permit us to make the least assumptions with the smallest number of parameters. We leave a more complex nonlinear analysis for future work.
In fact we will see that with simple models we can recover the large-scale density and peculiar velocity in the presence of light-cone, survey mask, and selection function effects, with a given $\Lambda$CDM cosmology, having chosen the resolution  at which our models apply (6-10 $h^{-1}$ Mpc).
The majority of previous Bayesian density field reconstructions applied to galaxy redshift surveys did not correct for the anisotropic redshift space distortions  \citep[see e.g.][]{2004MNRAS.352..939E,kitaura_sdss,jasche_sdss,Jascheborg,Granettvimos}.   We aim at filling that gap in this work, and think that the approach presented in this work could become standard in the analysis of galaxy surveys due to  its efficiency, simplicity, and its critical accuracy isotropizing the galaxy distribution while dealing with survey masks, selection functions and bias.

Ongoing and future surveys, such as the BOSS\footnote{\url{http://www.sdss3.org/surveys/boss.php}} \citep[][]{boss2011,BSA12,AAA15}, eBOSS \citep{Dawsonboss}, DESI\footnote{\url{http://desi.lbl.gov/}}/BigBOSS \citep[][]{bigboss2011}, DES\footnote{\url{http://www.darkenergysurvey.org}} \citep[][]{des2013}, LSST \footnote{\url{http://www.lsst.org/lsst/}} \citep[][]{lsst2012}, J-PAS\footnote{\url{http://j-pas.org/}} \citep[][]{jpas2014}, 4MOST\footnote{\url{https://www.4most.eu/}} \citep[][]{4most} or Euclid\footnote{\url{http://www.euclid-ec.org}} \citep[][]{euclid2009}, will require special data analysis techniques, like the one presented here, to extract the maximum available cosmological information.

The structure of this paper is as follows: in Sec.~\ref{sec:method} we present the main aspects of our reconstruction method and the {\textsc{argo}}-code ({\bf A}lgorithm for {\bf R}econstructing the {\bf G}alaxy traced {\bf O}verdensities). We emphasize the challenges of dealing with a galaxy redshift survey including cosmic evolution and the novel improvements to this work. In Sec.~\ref{sec:data} we describe the BOSS CMASS DR12 data and the mock galaxy catalogues used in this study.  In Sec.~\ref{sec:results} we show and evaluate the results of our application. We finally present the conclusions in Sec.~\ref{sec:conc}.

\section{Method}
\label{sec:method}

Our basic approach relies on an iterative Gibbs-sampling method, as proposed in \citet[][]{kitaura,kitaura_lyman} and presented in more detail in \citet[][]{Kitaura:2015bca}. The first step samples linear density fields defined on a mesh $\mbi\delta_{\rm L}$ with $N_{\rm c}$ cells compatible with the number counts on that mesh $\mbi N_{\rm G}$  of the galaxy distribution in real space \{\mbi r\}.  The second step obtains the real space distribution for each galaxy given its observed redshift space $s^{\rm obs}$ position required for the first step,  from sampling the peculiar velocities $\{\mbi v\left(\mbi\delta_{\rm L},f_\Omega\right)\}$ (with the growth rate given by $f_{\Omega}\equiv d\log D(a)/d\log a$, and $D(a)$ being the growth factor for a scale factor $a=1/(1+z)$ or redshift $z$), assuming that the density field and the growth rate $f_\Omega$ are known.  The Gibbs-sampling conditional probablity distribution functions can be written as follows showing the quantities, linear densities ${\mbi\delta_{\rm L}}$ and a set of galaxies in real space  $\{\mbi r\}$, which are sampled from the corresponding conditional PDFs:
\ba
{\mbi\delta_{\rm L}}&\curvearrowleft& \mathcal P_{\delta}\left({\mbi\delta_{\rm L}}|{N_{\rm G} \left(\{\mbi r\}\right)},\mbi w,{\mat C}_{\rm L}\left(\{p_{\rm c}\}\right),\{b_{\rm p}\}\right)\,{,}\label{eq:sig} \\
{\{\mbi r\}}&\curvearrowleft& \mathcal P_{r}\left(\{\mbi r\}|\{\mbi s^{\rm obs}\},\{\mbi v\left(\mbi\delta_{\rm L},f_\Omega\right)\}\right)\,{,}\label{eq:samvel} 
\ea
which is equivalent to sample from the following joint probability distribution function:
\ba
\mathcal P_{\rm joint}\left(\mbi\delta_{\rm L}, {\{\mbi r\}}|\{\mbi s^{\rm obs}\},\mbi w,\mat C_{\rm L}(\{p_{\rm c}\}),\{b_{\rm p}\},f_{\Omega}\right)\,{.}
\ea 
To account for the angular completeness (survey mask) and radial selection function, we need to compute the 3D completeness $\mbi w$ defined on the same mesh, as the density field \citep[see e.g.][]{kitaura_sdss}.  Also we have to assume a given covariance matrix  ${\mat C}_{\rm L} \equiv \langle \mbi\delta^\dagger_{\rm L}\mbi\delta_{\rm L} \rangle$ (a $N_{\rm c}\times N_{\rm c}$ matrix), determined by a set of cosmological parameters $\{p_{\rm c}\}$ within a $\Lambda$CDM framework.
We aim at recovering the dark matter density field which governs the dynamics of galaxies. Since galaxies are biased tracers, we have to assume some parametrised model relating the density field to the galaxy density field with a set of bias parameters $\{b_{\rm p}\}$.
{\color{black} We note that assuming a wrong growth rate will yield an anisotropic reconstructed density field. A recent work investigated this by jointly sampling the anisotrpic power spectrum including the growth rate and the redshift space density field \citep[see][]{Granettvimos}. }

After these probabilities reach their so-called stationary distribution, the drawn samples are represetatives of the target distribution.
In the following we define Eqs.~\ref{eq:sig} and \ref{eq:samvel} in detail and describe our sampling strategy.

\subsection{Density sampling}

The posterior probability distribution of Eq.~\ref{eq:sig} is sampled using a Hamiltonian Monte Carlo (HMC) technique \citep[see][]{duane}. For a comprehensive review see \citet[][]{Neal2012}. This technique has been applied in cosmology in a number of works \citep[see e.g.][]{2008MNRAS.389.1284T,jasche_hamil,jasche_sdss,kitaura_lyman,kit2mrs,Kitaura2013,Wang2013,Wang2014,Ata15,Jasche2013}.  To apply this technique to our Bayesian reconstruction model, we need to define the posterior distribution function through the product of  a prior $\pi$ (see  Sec.~\ref{sec:prior}) and a likelihood $\mathcal L$ (see Sec.~\ref{sec:likel}) which up to a normalisation is given by 
\ba
\lefteqn{\mathcal P_{\delta}\left({\mbi\delta_{\rm L}}|{N_{\rm G} \left(\{\mbi r\}\right)},\mbi w,\mat C\left(\{p_{\rm c}\}\right),\{b_{\rm p}\}\right) \propto}\\&&\pi(\mbi\delta_{\rm L}|\mat C\left(\{p_{\rm c}\}\right))\times\mathcal L(\mbi N_{\rm G}|\mbi\rho^{\rm obs}_{{\rm G}},\{b_{\rm p}\})\,,
\label{eq:denssampling}
\ea
with $\mbi\rho^{\rm obs}_{{\rm G}}$ being the expected number counts per volume element. 
The overall sampling strategy then is enclosed in Sec.~\ref{sec:sampling}. 

\subsubsection{Lognormal Prior}

\label{sec:prior}

As a prior we rely on the lognormal structure formation model introduced in \citet{Coles1991}. This model gives an accurate  description of the matter statistics (of the cosmic evolved density contrast $\delta\equiv\rho/\bar{\rho}-1$) on scales larger than about 6-10 $h^{-1}$ Mpc \citep[see e.g.][]{kitaura_sdss}. 
In such a model one considers that the logarithmically transformed density field $\mbi\delta_{\rm L}$ is a good representation of the linear density field
\ba
\mbi \delta_{\rm L} \equiv \log\left(1+\mbi \delta \right) - \mu \, ,
\label{eq:logn}
\ea
with 
\ba
\mu\equiv\langle\log\left(1+\mbi \delta \right)\rangle\,,
\label{eq:mu}
\ea
and is Gaussian distributed with zero mean and a given covariance matrix $\mat C_{\rm L}$ 
\ba
-\ln {\pi}(\delta_{\rm L}|\mat C_{\rm L}\left(\{p_{\rm c}\}\right))=\frac{1}{2}\mbi\delta_{\rm L}^\dagger\,\mat C_{\rm L}^{-1}\,\mbi\delta_{\rm L}+c\,{,}
\label{eq:gauss}
\ea
with $c$ being some normalisation constant of the prior.
This model yields, however, a poor description of the three-point statistics \citep[see][]{WTM14,ChuangComp15}, and will have a different mean field $\mu$ depending on the higher order statistics of the dark matter field. The  mean field computed based on the density field, as obtained from $N$-body simulations using the definition in Eq.~\ref{eq:mu}, can strongly deviate  from the theoretical prediction for lognormal fields $\mu=-\sigma^2/2$ depending on the resolution (with $\sigma^2$ being the variance of the field $\mbi \delta_{\rm L}$).
In fact, if one expands the logarithm of the density field in a series with the first term being the linear density field followed by all the higher order terms $\mbi \delta^+$ \cite[see][]{kitlin} 
\ba 
\log(1+\mbi \delta) = \mbi \delta_{\rm L} + \mbi \delta^+\,,
\label{eq:norm}
\ea
one finds that the mean field depends on the order of the expansion
\ba
\mu\equiv\langle \log( 1+\mbi \delta) \rangle  = \langle \mbi \delta^+\rangle \,.
\ea
In practice, the data will determine the mean field $\mu$. In unobserved regions, the mean field should be given by the theoretical lognormal value ($\mu=-\sigma^2/2$). In observed regions, the number density and completeness will determine the value of the mean field. Since galaxy redshift surveys have in general a varying completeness as a function of distance, the assumption of a unique mean field can introduce an artificial radial selection function.
For this reason we suggest to follow \citet[][]{kitaura_lyman} and iteratively sample the mean field from the reconstructed linear density field assuming large enough volumes $\langle\delta\rangle=0=\langle{\rm e}^{\delta_{\rm L}+\mu}-1\rangle$, i.e., $\mu=-\ln(\langle {\rm e}^{\delta_{\rm L}}\rangle)$.
The assumption that volume averages of the linear and nonlinear density field vanish in the ensemble average, does not imply that this happens for the individual reconstructions, which will be drawn from our posterior analysis allowing for cosmic variance.
 We will  consider, as a crucial novel contribution, individual redshift  $z$ and completeness $w$ bins
\ba
\mu_{(z,w)}=-\ln(\langle\rm e^{\delta_{\rm L}}\rangle_{(z,w)})\,.  %EDIT
\label{eq:mu2}
\ea
This can be expressed as an additional Gibbs-sampling step
\ba
{\mu_{(z,w)}}&\curvearrowleft& \mathcal P_{\mu}\left(\mu_{(z,w)}|\mbi\delta_{\rm L}(\mbi r, z), \mbi w\right)\,{.}\label{eq:mu3}
\ea
In this way we account for redshift and  completeness dependent renormalised lognormal priors.
In practice, since the evolution of the three-point statistics can be considered to be negligible within the covered redshift range for CMASS galaxies \citep[][]{2016MNRAS.456.4156K}, we will perform the ensemble average only in completeness bins.

\subsubsection{Likelihood and data model}
\label{sec:likel}

The likelihood describes the data model. In our case the probability to draw a particular number of galaxy counts $N_{{\rm G}i}$ per cell $i$, given an expected number count per cell  $\rho^{\rm obs}_{{\rm G}i}$, is modelled by the Poisson distribution function
\ba
-\ln \mathcal L(\mbi N_{\rm G}|\mbi\rho^{\rm obs}_{{\rm G}},\{b_{\rm p}\})=\sum_i^{N{\rm c}}\left(-N_{{\rm G}i}\ln\rho^{\rm obs}_{{\rm G}i}+\rho^{\rm obs}_{{\rm G}i}\right)+c \, ,
\label{eq:like}
\ea
with $N_{\rm c}$ being the total number of cells of the mesh, and $c$ being some normalisation constant of the likelihood.
This expectation value  is connected to the underlying matter density $\delta_{i}$ by the particular chosen bias model $\mathcal{B}(\mbi \rho_{\rm G}| \mbi \delta)$.
In particular, we rely on a power-law bias (linear in the log-density field) connecting the galaxy density field to the underlying dark matter density $\rho_{\rm G}\propto(1+\delta)^{b}$ \citep[][]{delaTorre2012}. More complex biasing models can be found in the literature \citep[][]{1993ApJ...413..447F,Cen-Ostriker-93,Mcdonald09,2014MNRAS.439L..21K,2014MNRAS.441..646N,Ahn15}. In fact threshold bias can be very relevant to describe the three-point statistics of the galaxy field \citep[][]{Kitetal15,2016MNRAS.456.4156K}, and stochastic bias \citep[][]{2014MNRAS.439L..21K} is crucial to properly describe the clustering on small scales. All these bias components have been investigated within a Bayesian framework in \citet[][]{Ata15}. We will, however, focus in this work on the two-point statistics on large scales ($k\lsim0.2\,h$ Mpc$^{-1}$), and neglect such deviations.
The bias model needs to account for cosmic evolution.  In linear theory and within $\Lambda$CDM this is described by the growth factor: 
\be
D(z)= \frac{H(z)}{H_0} {\int\limits_z^\infty dz'\frac{(1+z')}{H^3(z')}}/{\int\limits_0^\infty dz'\frac{(1+z')}{H^3(z')}}\,, 
\ee
permitting one to relate the density field at a given redshift to a reference redshift  $z_{\rm ref}$: $\delta_i(z_{\rm ref}) = G(z_{\rm ref},z_i)\, \delta_i(z_i)$ with 
\ba
G(z_{\rm ref},z_i)\equiv{D(z_i)}/{D(z_{\rm ref})}\,.
\ea
%$z_{\rm ref}=0.0877$
The reference redshift must be chosen to be lower than the lowest redshift in the considered volume to ensure that the growth factor ratio $G(z_{\rm ref},z_i)\equiv{D(z_i)}/{D(z_{\rm ref})}$ remains below one. Otherwise, negative densities will arise in low density cells, causing singularities in the lognormal model.
Another important ingredient in our model is the angular mask and radial selection function describing the three dimensional completeness $\mbi w$, which can be seen as a response function between the signal and the data: $\rho^{\rm obs}_{{\rm G}i} \equiv w_i \rho_{{\rm G}i}\propto w_i \mathcal{B}(\mbi \rho_{\rm G}| \mbi \delta)|_i$ \citep[see e.g.][]{kitaura}.
One needs to consider now, that only when the bias is linear the proportionality factor is given by the mean number density $\bar{N}\equiv\langle\mbi\rho_{\rm G}\rangle$: $\rho_{{\rm G}i} = \bar{N}\,(1+b_{\rm L}\delta)$, with $b_{\rm L}$ being the linear bias. This model is inconvenient for bias larger than one, as it is the case of luminous red galaxies, since negative densities could arise.  In the general case, the proportionality constant will be given by the bias model \citep[][]{2014MNRAS.439L..21K}
\ba
\gamma(z)\equiv\bar{N}/\langle\mathcal{B}(\mbi \rho_{\rm G}|\mbi \delta)\rangle_{(z)}\, ,
\label{eq:gamma}
\ea
 which we suggest to iteratively sample from the reconstructed  density field in redshift bins. 
 If we  instead use a model  defined as 
\be
\rho_{{\rm G}i} \equiv \bar{N}\left(1+ \mathcal{B}\left(\mbi \rho_{\rm G}| \mbi \delta\right)|_i-\langle\mathcal{B}\left(\mbi \rho_{\rm G}| \mbi \delta\right)\rangle\right)\nonumber\,,
\ee
which also ensures the correct mean number density by construction, negative expected number counts are allowed, which we want to avoid. For this reason we will rely on the following bias model:
\ba
\rho^{\rm obs}_{{\rm G}i} \equiv w_i  \gamma(z_i) (1+G(z_i,z_{\rm ref})\delta_i)^{b_{\rm L}(z_i)f_{\rm b}} \, ,
\label{eq:lambda}
\ea
where we have included a bias correction factor $f_{\rm b}$, which accounts for the deviation between linear and power-law bias.
With this model, the sampling of the normalisation constant can be expressed as an additional Gibbs-sampling step
\ba
{\gamma_{(z)}}&\curvearrowleft& \mathcal P_{\gamma}\left(\gamma_{(z)}|\bar{N},\mbi\delta,G(z,z_{\rm ref}),b_{\rm L}(z),f_{\rm b}\right)\,{.}\label{eq:gamma}
\ea
Given a redshift $z$  one can define  the ratio between the galaxy correlation function in redshift space at $z$ ($\xi^s_{\rm G}(z)$) and the matter correlation function in real space at $z_{\rm ref}$ ($\xi_{\rm M}(z_{\rm ref})$) as
\ba
c^s_{\rm L}(z)\equiv\sqrt{\xi^s_{\rm G}(z)/\xi_{\rm M}(z_{\rm ref})}\,.
\ea 
The quantity $\xi^s_{\rm G}(z)$ can be obtained from the data without having to assume any bias, nor growth rate.
Furthermore, one can use the Kaiser factor  \citep[$K=1+2/3f_{\Omega}/b_{\rm L}+1/5(f_{\Omega}/b_{\rm L})^2$, with $f_{\Omega}$ being the growth rate,][]{Kaiser-87} to relate the galaxy correlation function in redshift space to the matter real space correlation function
\ba
\xi_{\rm G}^{s}(z)&=&K(z)\,\xi_{\rm G}(z)\nonumber\\
&=&K(z)\,b^2_{\rm L}(z)\,G^2(z,z_{\rm ref})\,\xi_{\rm M}(z_{\rm ref})\,.
\ea
From the last two equations we find a quadratic expression for $b_{\rm L}(z)$ for each redshift $z$
\ba
b_{\rm L}^2(z)+\frac{2}{3}f_\Omega(z)b_{\rm L}(z)+\frac{1}{5}f_\Omega^2(z)-\frac{(c^{s}_{\rm L}(z))^2}{G^2(z,z_{\rm ref})}=0\,,
\ea
 with only one positive solution, leaving the bias correction factor $f_{\rm b}$ as a  potential free parameter in our model (see the renormalised perturbation theory based derivation below)
\ba
\label{eq:linearbias}
b_{\rm L}(z)=-\frac{1}{3}f_{\Omega}(z)+\sqrt{-\frac{4}{45}f_\Omega(z)^2+(c^{s}_{\rm L}(z))^2\left(\frac{D(z_{\rm ref})}{D(z)}\right)^2}\,.
\ea

By coincidence, the  bias measured in redshift space  on large scales $c^s_{\rm L}(z)=1.84 \pm 0.1$ (with respect to the dark matter power spectrum at redshift $z=0.57$)  is constant for CMASS galaxies across the considered redshift range \citep[see, e.g.,][]{sergio15}. Nevertheless, the (real space) linear  bias $b_{\rm L}(z)$ is not, as it needs to precisely compensate for the growth of structures (growth factor) and the evolving growth rates, ranging between 2.00 and 2.30. 
The nonlinear bias correction factor $f_{\rm b}$ is expected to be less than ``one'', since we are using the linear bias in the power-law. 
One can predict $f_{\rm b}$ from renormalised perturbation theory, which in general, will be a function of redshift.
Let us Taylor expand our bias expression (Eq.~\ref{eq:lambda}) to third order
\ba
\label{eq:biaspow}
\lefteqn{\delta_{\rm g}(z_i)\equiv\frac{\rho_{\rm g}}{\bar{\rho}_{\rm g}}(z_i)-1\simeq b_{\rm L}(z_i)f_{\rm b}(z_i)\delta(z_i)}\\
&&\hspace{-0.75cm} +\frac{1}{2}b_{\rm L}(z_i)f_{\rm b}(z_i)(b_{\rm L}(z_i)f_{\rm b}(z_i)-1)\left(\left(\delta(z_i)\right)^2-\sigma^2(z_i)\right)+\nonumber\\\
&&\hspace{-0.75cm} \frac{1}{3!} b_{\rm L}(z_i)f_{\rm b}(z_i)(b_{\rm L}(z_i)f_{\rm b}(z_i)-1)(b_{\rm L}(z_i)f_{\rm b}(z_i)-2)\left(\delta(z_i)\right)^3\nonumber\,,
\ea
with $\delta(z_i)=G(z_i,z_{\rm ref})\delta(z_{\rm ref})$.
The usual expression for the perturbatively expanded overdensity field to third order ignoring nonlocal terms is given by
\be
\label{eq:biaspt}
\delta_{\rm g}(z_i)=c_\delta(z_i)\delta(z_i)+\frac{1}{2}c_{\delta^2}(z_i)(\delta^2(z_i)-\sigma^2(z_i))+\frac{1}{3!}c_{\delta^3}(z_i)\delta^3(z_i)\,.
\ee
Correspondingly, one can show that the observed, renormalised, linear bias is given by \citep[see][]{Mcdonald09}
\ba
\label{eq:renorm}
b_\delta(z_i)=c_\delta(z_i)+\frac{34}{21}c_{\delta^2}(z_i)\sigma^2(z_i)+\frac{1}{2}c_{\delta^3}(z_i)\sigma^2(z_i)\,.
\ea
By considering that in our case the observable linear bias is expected to be  given by $b_{\rm L}(z_i)$ and identifying the coefficients \{$c_\delta=f_{\rm b}b_{\rm L}$, $c_{\delta^2}=f_{\rm b}b_{\rm L}(f_{\rm b}b_{\rm L}-1)$, $c_{\delta^3}=f_{\rm b}b_{\rm L}(f_{\rm b}b_{\rm L}-1)(f_{\rm b}b_{\rm L}-2)$\} from Eqs.~\ref{eq:biaspow} and \ref{eq:biaspt} one can derive the following cubic equation
 for $f_{\rm b}$
\ba
\lefteqn{f_{\rm b}^3\,\left(\frac{1}{2}b^3_{\rm L}(z_i)\sigma^2(z_i)\right)}\\
&&\hspace{-0.75cm}+f_{\rm b}^2\,\left(-\frac{3}{2} b^2_{\rm L}(z_i)\sigma^2(z_i)+\frac{34}{21}b^2_{\rm L}(z_i)\sigma^2(z_i)\right)\nonumber\\
&&\hspace{-0.75cm}+f_{\rm b}\,b_{\rm L}(z_i)\left(1+\left(-\frac{34}{21}+1\right)\sigma^2(z_i)\right)-b_{\rm L}(z_i)=0 \nonumber\,.
\ea 
Let us consider the case of a cell resolution of $6.25\,h^{-1}\,{\rm Mpc}$.
The only real solutions for redshift $z=0.57$ ($G=0.78$) and $b_{\rm L}=2.1 \pm 0.1$, are $f_{\rm b}=0.62 \pm 0.01$ including the variance from the nonlinear transformed field ($\sigma^2(\delta)=1.75$),   and $f_{\rm b}=0.71 \pm 0.02$ including the variance from the linear field ($\sigma^2(\delta_{\rm L})=0.91$). This gives us a hint of the uncertainty in the nonlinear expansion.  Let us, hence, quote as the theoretical prediction for the bias correction factor the average between both mean values with the uncertainty given by the difference between them $f_{\rm b}=0.66 \pm 0.1$. These results show  little variation ($\pm0.01$) across the redshift range  (see \S \ref{sec:mocks}). 
Leaving $f_{\rm b}$ as a free parameter and sampling it to match the power spectrum on large scales yields $f_{\rm b}=0.7 \pm 0.05$ (see \S \ref{sec:results}). Although there is an additional uncertainty associated to this measure, since the result depends on the particular $k$ mode range used in the goodness of fit.
Therefore, one can conclude that the theoretical predictions  account for the nonlinear correction within the associated uncertainties on large scales in terms of the two point statistics.
{\color{black}
We include only delta bias terms in Eqs.~\ref{eq:biaspt}, \ref{eq:renorm},  because these equations describe the model we implemented, represented by Eqs. \ref{eq:lambda}, \ref{eq:biaspow}, where we did not include any tidal bias. 
As shown by \citet[][]{Mcdonald09}, the only effect of tidal bias terms in the low $k$ (large scale) limit is to renormalize the standard linear delta bias (and shot noise). We are therefore implicitly including these effects if present in the data when we fit for the bias (Eq.~\ref{eq:linearbias}), i.e., our model is complete in the low-$k$ limit. 
As we go to higher $k$, i.e., smaller scales, tidal bias can have a non-trivial effect in the model \citep[][]{Mcdonald09}, along with various other non-linear effects which enter at the same order in perturbation theory (i.e., non-linear gravitational evolution, higher order density bias different from that implied by Eq.~\ref{eq:lambda}, non-linearity/biases related to the redshift space transformation). These effects could be included in future models for higher accuracy. 
}

\subsubsection{Hamiltonian Monte Carlo of the linear density field}
\label{sec:sampling}

In this section we recap the Hamiltonian Monte Carlo sampling technique (HMC) to sample the matter density within the Bayesian framework.
This technique requires the gradients of the lognormal-Poisson model, as introduced in \citet[][]{kitaura_log}. The HMC technique was first applied to this model with a linear bias in \citet[][]{jasche_hamil} and later with more complex bias relations and likelihoods in \citet[][]{Ata15}.
In this approach one defines a potential energy $U(\mbi x)$, given by the negative logarithm of the posterior distribution function, and a kinetic energy $\mbi K(\mbi p)$
\ba 
U(\mbi x) &=& - \ln{\mathcal{P}(\mbi x)} \\
\mathcal{H(\mbi x, \mbi p)} &=& U(\mbi x) + K(\mbi p) \, ,
\ea
where the Hamiltonian $\mathcal{H(\mbi x, \mbi p)}$ is given by the sum of the potential and the kinetic energy.
In this formalism we use $\mbi x$ as a pseudo spatial variable (in our case the linear density field $\mbi \delta_{\rm L}$) and $\mbi p$ as the conjugate momentum.
HMC requires the computation of the negative logarithm of Eq.~\ref{eq:denssampling} and its derivatives with respect to the sampled quantity (the linear density field {\mbox{\boldmath$\delta_{\rm L}$}} in our case). 
The {\it kinetic energy} term is constructed on the nuisance parameters given by the {\it momenta} $\mbi p$ and {\it mass} variance $\mat M$: 
\be
  \label{eq:kin}
  K(\mbi p)\equiv\frac{1}{2}\sum_{ij}p_iM_{ij}^{-1}p_j\,.
\ee
The {\it canonical} distribution function defined by the Hamiltonian (or the joint distribution function of the signal and {\it momenta}) is then given by:
\ba
  \label{eq:joint}
  P(\mbi x,\mbi p)&=&\frac{1}{Z_H}\exp(-\mathcal{H}(\mbi s,\mbi p)) \nonumber\\
  &=&\left[\frac{1}{Z_K}\exp(-K(\mbi p))\right]\left[\frac{1}{Z_E}\exp(-U(\mbi x))\right] \nonumber\\
  &=&P(\mbi p)P(\mbi x) \,,
\ea
with $Z_H$, $Z_K$ and $Z_E$ being the {\it partition} functions so that the probability distribution functions are normalised to one. In particular, the normalisation of the Gaussian distribution for the {\it momenta} is represented by the {\it kinetic partition} function $Z_K$.  The Hamiltonian sampling technique does not require the terms which are independent of the configuration coordinates as we will show below.

From Eq.~(\ref{eq:joint}) it can be noticed that in case we have a method to sample
from the joint distribution function  $P(\mbi x,\mbi p)$, marginalizing over the
momenta we can in fact, sample  the posterior $P(\mbi x)$.

The Hamiltonian dynamics provides such a method. We can define a dynamics on {\it phase-space} (positions and momenta) with the introduction of a {\it time} parameter $t$. The Hamiltonian equations of motion are given by:
\ba
  \label{eq:EoM1}
  \frac{dx_i}{dt}& =& \frac{\partial \mathcal{H}}{\partial p_i}=\sum_j M^{-1}_{ij} p_j\, ,\\
  \label{eq:EoM2}
  \frac{dp_i}{dt}& =& - \frac{\partial \mathcal{H}}{\partial x_i} = - \frac{\partial U(x)}{\partial x_i}\, .
\ea
To sample the posterior one has to solve these equations for randomly drawn {\it momenta} according to the kinetic term defined by Eq.~(\ref{eq:kin}). This is done by drawing  Gaussian samples with a variance given by the {\it mass} $\mat M$ which can tune the efficiency of the sampler \citep[see][]{jasche_hamil}. We rely on the Fourier formulation to capture the correlation function through the power spectrum and include some preconditioning diagonal matrices to speed up the algorithm. The marginalization over the {\it momenta} occurs by drawing new {\it momenta} for each Hamiltonian step disregarding the ones of the previous step.

It is not possible to follow the dynamics exactly, as one has to use a discretized version of the equations of motion. It is convenient to use the {\it leapfrog} scheme which has the properties of being {\it time}-reversible and conserve {\it phase-space} volume being necessary conditions to ensure {\it ergodicity}:
\ba
  \label{eq:leap1}
  p_i\left(t+\frac{\epsilon}{2}\right) &=& p_i(t) -\frac{\epsilon}{2} \left .\frac{\partial  U(\mbi x)}{\partial x_l} \right |_{x_i(t)} \, , \\
\label{eq:leap2}
  x_i\left(t+\epsilon \right) &=& x_i(t) +\epsilon\sum_jM_{ij}^{-1}\,p_j\left(t+\frac{\epsilon}{2}\right)  \, , \\
  p_i\left(t+\epsilon\right) &=& p_i\left(t+\frac{\epsilon}{2}\right) -\frac{\epsilon}{2} \left .\frac{\partial  U(\mbi x)}{\partial x_l} \right |_{x_i\left(t+\epsilon \right)} \, . 
\ea
The dynamics of this system are followed for a period of {\it time} $\Delta \tau$, with a value of $\epsilon$ small enough to give acceptable errors and for $N_\tau=\Delta\tau/\epsilon$ iterations. In practice $\epsilon$ and $N_\tau$ are randomly drawn from a uniform distribution to avoid resonant trajectories \citep[see][]{neal1993}.

The solution of the equations of motion will move the system from an initial state  \((\mbi s,\mbi p)\)  to a final state \((\mbi s',\mbi p')\) after each sampling step.
Although the Hamiltonian equations of motion are energy conserving, our approximate solution is not. Moreover, the starting guess will not be drawn from the correct distribution and a {\it burn-in} phase will be needed. For these reasons a Metropolis-Hastings acceptance step has to be introduced in which the new {\it phase-space} state \((\mbi x',\mbi p')\) is accepted with probability:
\be
\label{eq:acc}
{P}_A = {\rm min}\left[1,{\rm exp}(-\delta \mathcal{H})\right]\, ,
\ee
with $\delta \mathcal{H}\equiv \mathcal{H}(\mbi x',\mbi p')-\mathcal{H}(\mbi x,\mbi p)$.

In particular, the required lognormal-Poisson gradients for the prior and likelihood including cosmic evolution are given by
\ba
  - \frac{\partial}{\partial {\mbi \delta_{\rm L}}} \ln \pi =   \mat C_{\rm L}^{-1} \mbi \delta_{\rm L} \, ,
\ea
and 
\ba
-\frac{\partial \ln {\cal L}}{\partial \mbi \delta_{\rm L}}|_i=\left(-\frac{N_{{\rm G}i}}{\rho^{\rm obs}_{{\rm G}i}}+1\right) \cdot \frac{b_{\rm L}(z)f_{\rm b}\, G(z,z_{\rm ref})(1+\delta_i)}{1+G(z,z_{\rm ref})\delta_i} \rho^{\rm obs}_{{\rm G}i}\, ,
\ea
respectively.
The linear density field is defined at the reference redshift $z_{\rm ref}$.

\begin{figure*}
 \begin{tabular}{cc}
\hspace{-.3cm}
   \includegraphics[width=11cm]{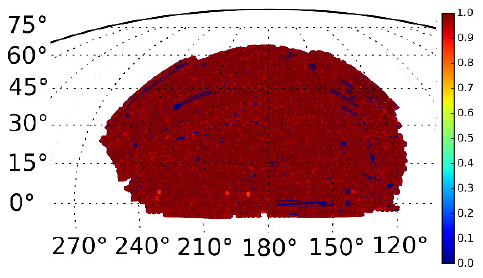}
   \put(-320,100){\rotatebox[]{90}{\text{DEC}}}
   \put(-165,0){\rotatebox[]{0}{\text{RA}}}
\hspace{-1.3cm}
   \includegraphics[width=7.2cm]{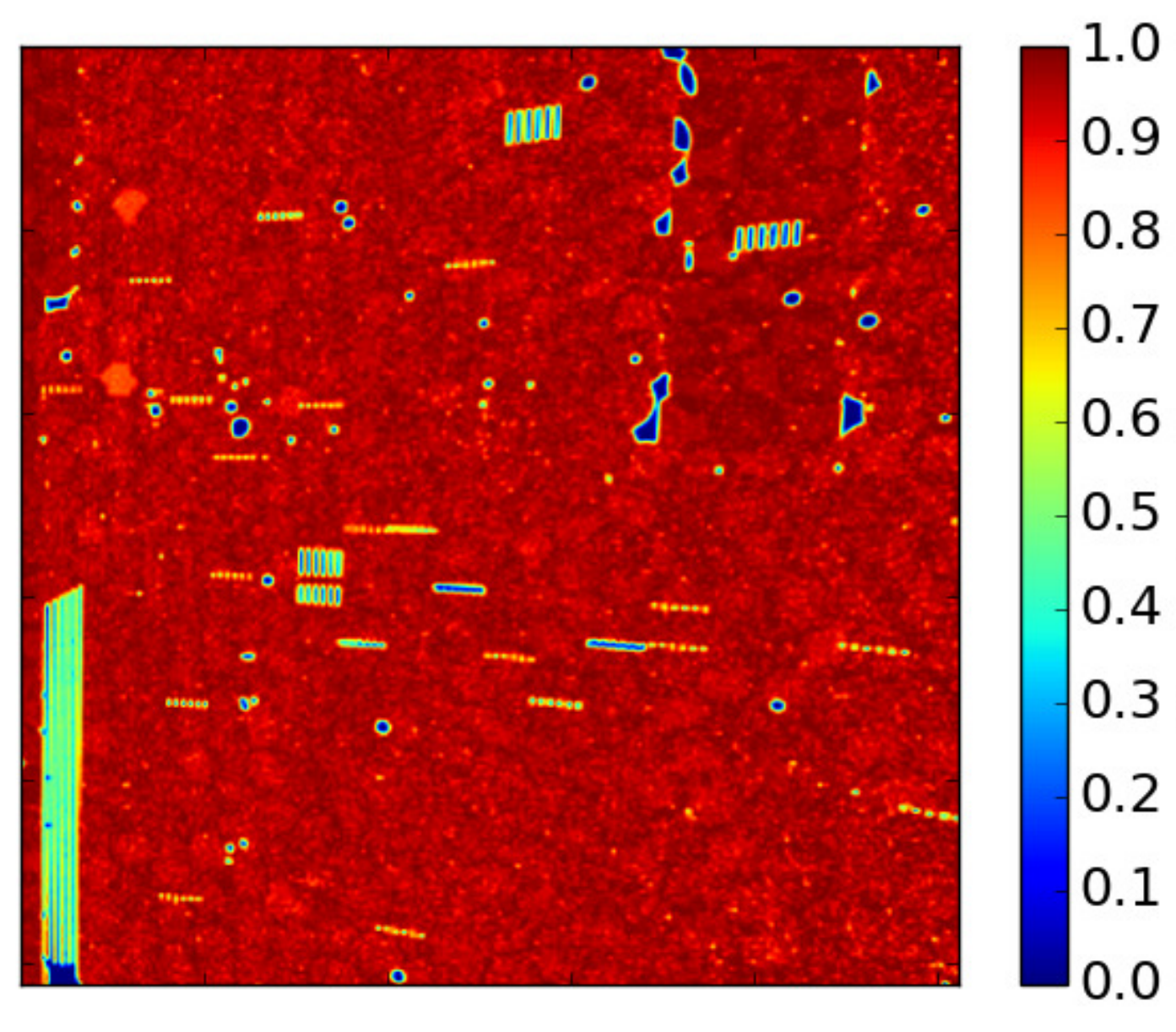}
 \end{tabular}
	\caption{{\bf Left panel:} Angular mask (right ascension RA vs declination DEC), ranging from zero to one, showing the completeness on the sky of the SDSS-III BOSS DR12 survey.  {\bf Right panel:}  Slice (in the $x-y$ plane)  of the 3D-projected angular mask on a volume of 1250 $h^{=1}$ Mpc side.}
 \label{fig:mask}
\end{figure*}

\subsection{Velocity sampling}
\label{sec:vel}

The peculiar motions of galaxies can be divided into two categories: coherent flows \citep[][]{Kaiser-87} and quasi-virialised or dispersed velocities. While the former are well constrained by the large-scale density field, the latter become relevant on smaller nonlinear scales \citep[see e.g.][]{Reid14}.    
Thus, one can write the total velocity field as the sum of the curl-free  coherent bulk flow, which can  directly be inferred from the large-scale density field within linear theory, and the dispersion term $\mbi v_{\rm disp}$
\ba
{\mbi v}(\mbi r,z) &=&  -f_{\Omega}(a)\,H(a)\,a\,  \nabla\nabla^{-2} \delta(\mbi r,z) +\mbi v_{\rm disp}\,,\label{eq:vel}
\ea
where $H$ is the Hubble constant. A simple way of including the dispersion term is to randomly draw it from a Gaussian  with a particular standard deviation. One may consider about 50 km $s^{-1}$ (see \S \ref{sec:results}), the typical 1-$\sigma$ uncertainty within linear theory \citep[][]{kitaura_vel}. More precise and sofisticated ways of dealing with quasi-virialised RSD are left for future work \citep[see e.g.][]{HKG13,2015MNRAS.449.3407J,Kitaura:2015bca}.
 Here we aim at focussing on the coherent flows on the limit of vanishing dispersions (see \S \ref{sec:results} for a comparison study with and w/o dispersion).
 In practice we are restricting our study to resolutions in the range between 6 and 10 $h^{-1}$ Mpc, which yield robust results on large scales \citep[see][]{Kitaura:2015bca}. 
Tidal field corrections could be included in the model \citep[see][]{kitaura_vel}. Also one could try to get improved velocity reconstructions from the linear component rather than from the nonlinear one as we do here \citep[see][]{Falck12,kitlin}. Nevertheless, there is a (nearly constant) bias from the lognormal transformation present in the linear density field, which we want to avoid to reduce the number of parameters \citep[see][]{2009ApJ...698L..90N}. 
The mapping between real space and redshift space positions for each individual galaxy is described by
\ba
{\mbi r^{j+1}} &=& {\mbi s^{\rm obs}} - \left(\frac{{\mbi v}\left(\mbi r^j,z\right)  \cdot {\mbi {\hat r}}}{H(a)\,a}\right){\mbi {\hat r}}, \label{eq:RSD} 
\ea
 where $j$ and $j+1$ are two subsequent Gibbs-sampling iterations, and ${\mbi {\hat r}}$ denotes the unit vector in line of sight direction.
The peculiar velocity needs to be evaluated in real space, which requires an iterative sampling scheme.
Each galaxy requires in principle a peculiar velocity field computed at that redshift, as the growth rate changes with redshift. In practice we construct a number of peculiar velocity fields defined on the same mesh but at different redshifts, i.e., from density fields multiplied with the corresponding growth factors and rates.
Each galaxy will get a peculiar velocity field assigned interpolated to its position within the cell taken from the peculiar velocity mesh at the corresponding redshift bin.

\begin{figure} 
  \begin{tabular}{c}
   \includegraphics[width=8cm]{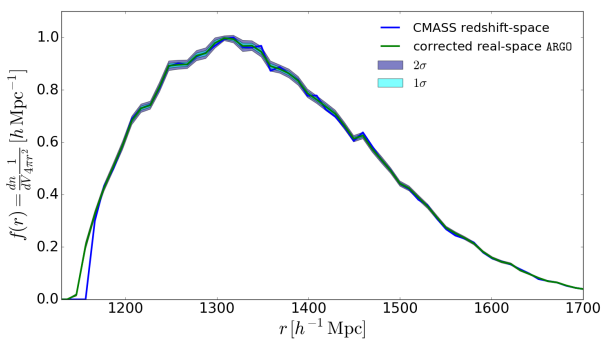}
   \end{tabular}
 \caption{Radial selection function $f(r)$ for a subvolume of the CMASS galaxy survey normalised to unity before and after RSD corrections with \textsc{argo}. The mean is calculated by calculating $f(r)$ for $2000$ reconstructions.}
 \label{fig:fr_argo}
\end{figure}

\section{Input data}
\label{sec:data}

  \begin{figure*}
  \begin{tabular}{ccc}
   \hspace{-1.cm}
\begin{subfigure}{.3\textwidth}
   \includegraphics[width=6.3cm]{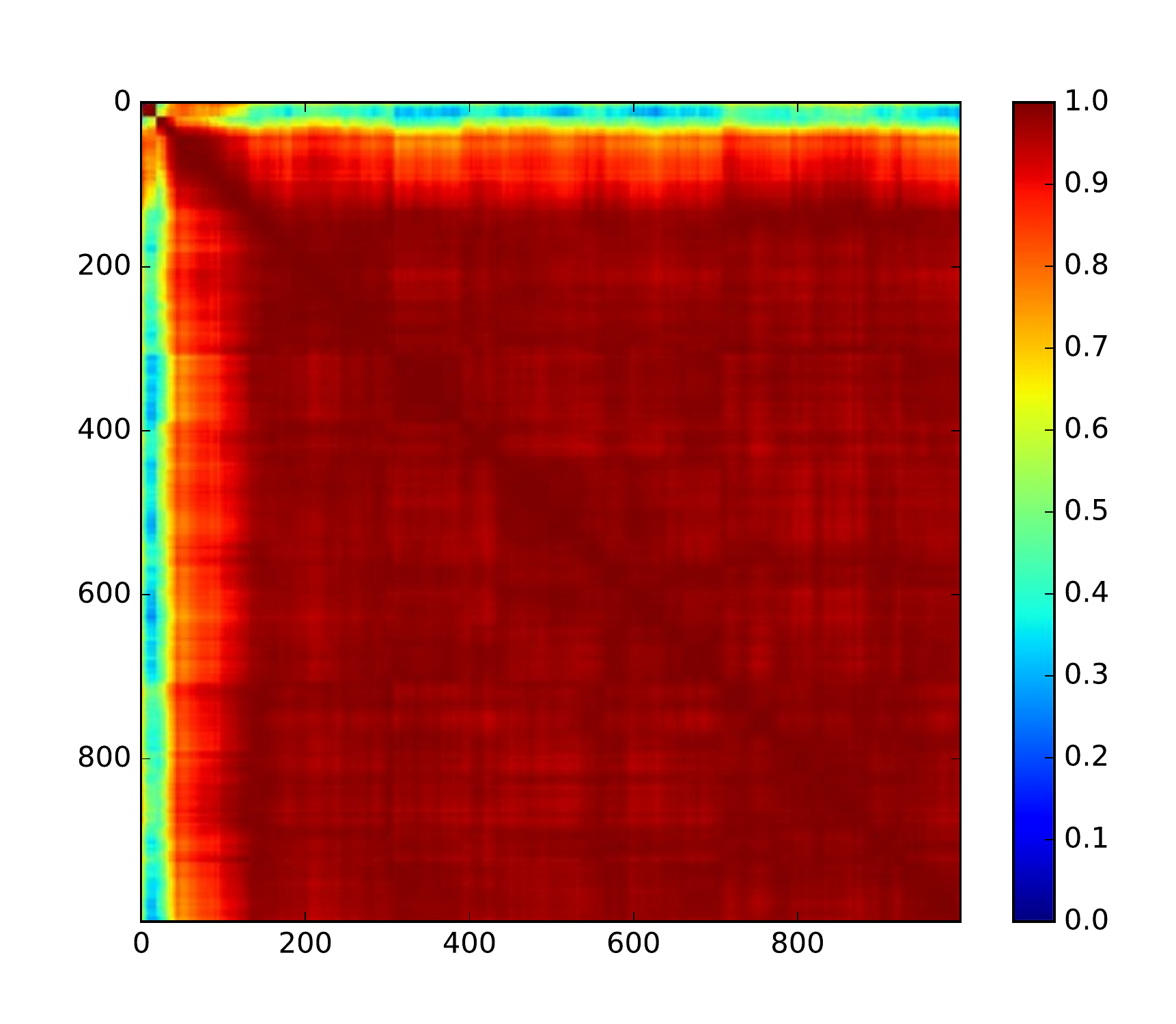}
   \put(-180,70){\rotatebox[]{90}{\text{$P_j$}}}
   \put(-100,0){\rotatebox[]{0}{\text{$P_i$}}}
   \hspace{-1.cm}
\end{subfigure}
\begin{subfigure}{.3\textwidth}
\vspace{-0.3cm}
   \includegraphics[width=6.3cm]{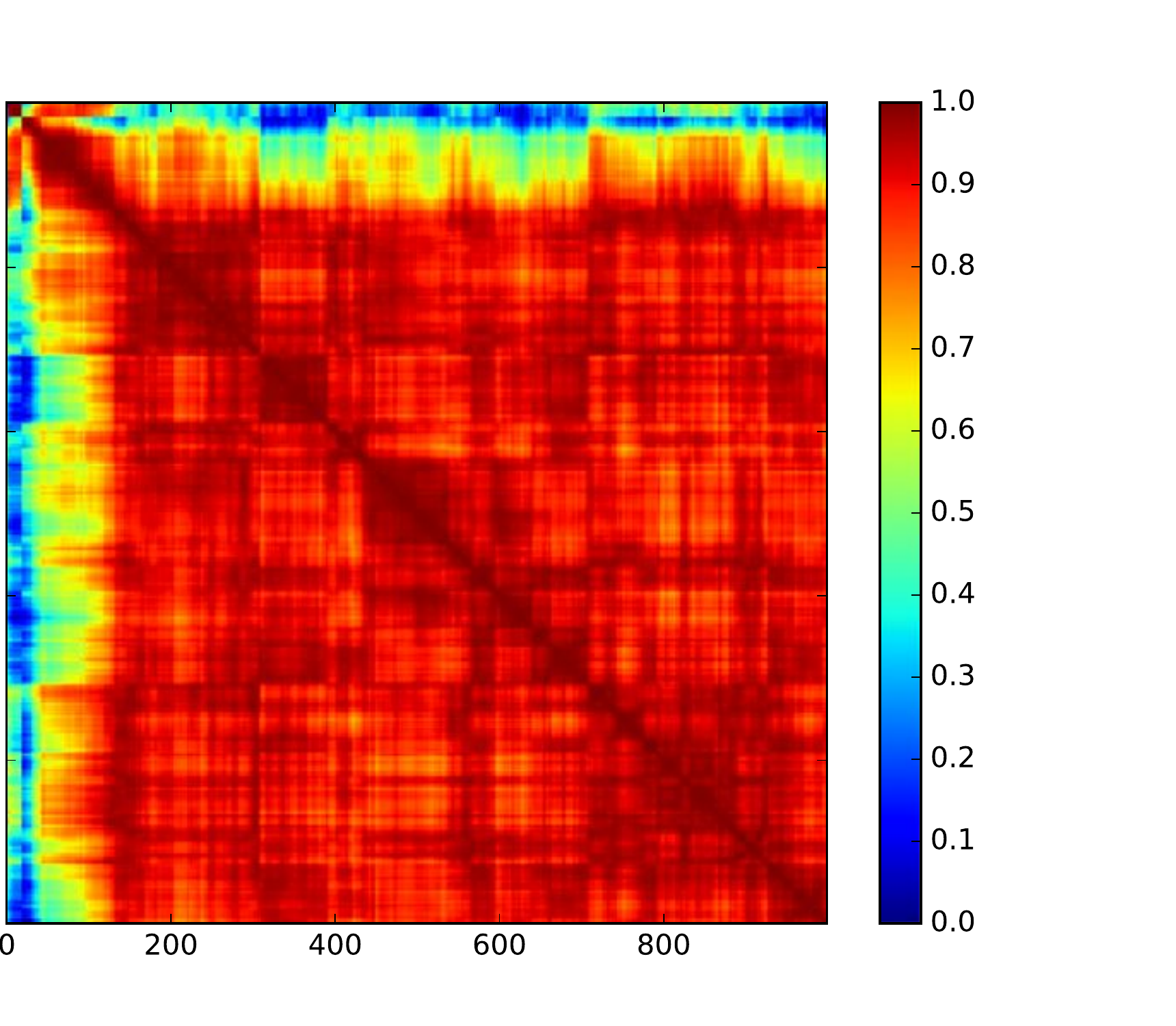}
   \put(-115,0){\rotatebox[]{0}{\text{$P_i$}}}
\end{subfigure}
\hspace{0.7cm}
\begin{subfigure}{.3\textwidth}
\vspace{0.cm}
   \includegraphics[width=5.8cm]{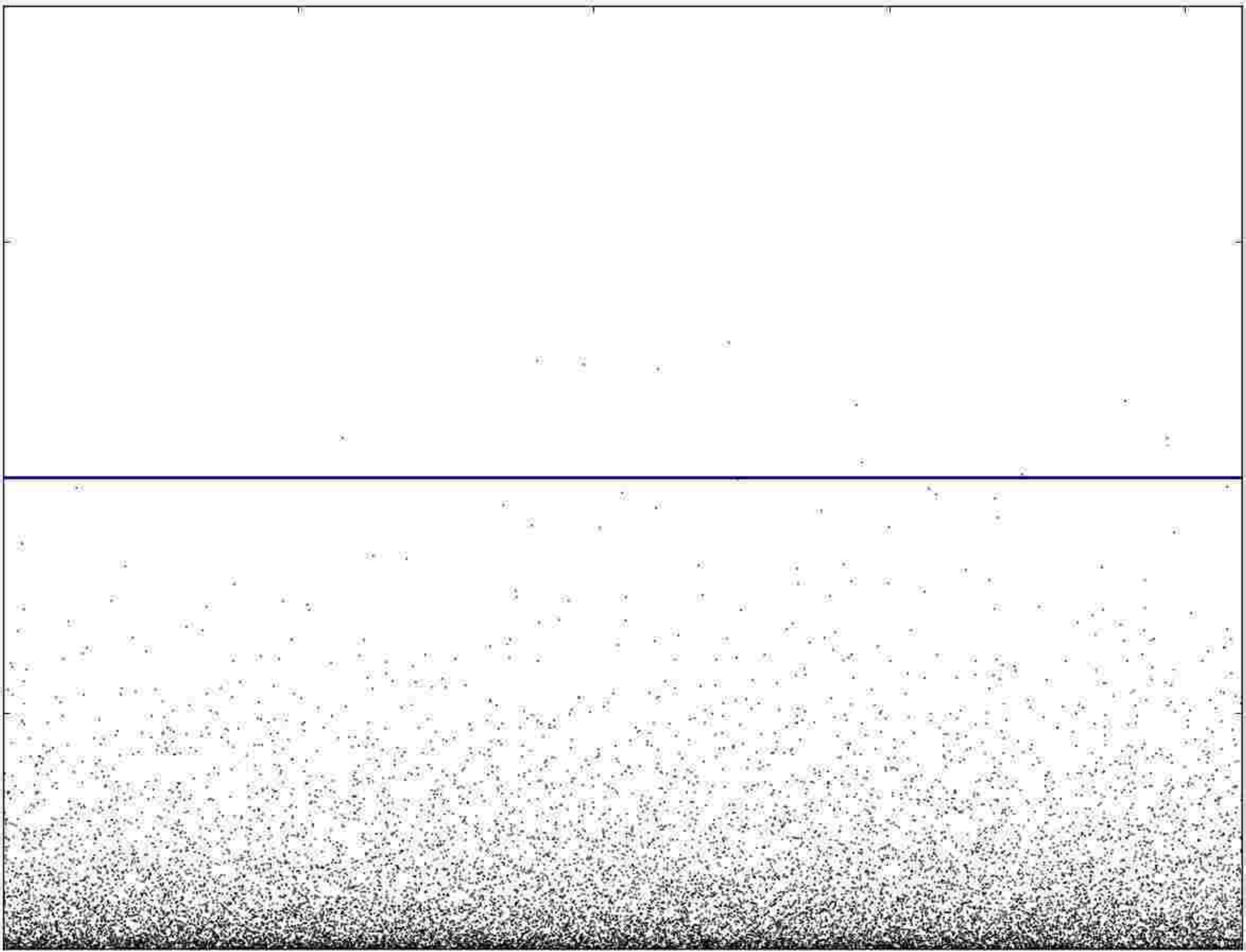}
   \put(-187,60){\rotatebox[]{90}{\text{$\hat P-1$}}}
   \put(-90,-12){\rotatebox[]{0}{\text{$i_{\rm c}\times10^{-6}$}}}
   \put(-178,124){\tiny 0.20}
   \put(-178,93){\tiny 0.15}
   \put(-178,62){\tiny 0.10}
   \put(-178,30){\tiny 0.05}
   \put(-178,0){\tiny 0.00}
   \put(-166,-4){\tiny 0.0}
   \put(-128,-4){\tiny 0.5}
   \put(-90,-4){\tiny 1.0}
   \put(-52,-4){\tiny 1.5}
   \put(-15,-4){\tiny 2.0}
\end{subfigure}
\end{tabular}
\caption{Convergence analysis of the Gibbs-Hamiltonian sampler. {\bf Left and middle panels:} Power spectrum correlation matrix ${\mathcal{R}}_{ij}$ of the first 1000 iterations of \textsc{Argo} with a mesh of 128$^3$.
Each entry of the matrix represents the correlation coefficient of the power spectra $P_i$ and $P_j$: ${\mathcal{R}}_{ij}=\frac{C_{ij}}{\sqrt{C_{ii}\,C_{jj}}}$, where $C_{ij}= \langle (P_i-\langle P_i \rangle ) (P_j-\langle P_j \rangle ) \rangle$ is the covariance matrix. {\bf Left panel:} correlation matrix for all modes of the power spectrum, {\bf middle panel:} correlation matrix for the lowest 30 modes, corresponding up to $k = 0.2\, h\, {\rm Mpc}^{-1}$. {\bf Right panel:} Potential scale reduction factor $\hat P$ of the \citet[][]{Gelman92} test  comparing the mean of variances of different chains with the variance of the different chain means.  
The cell number $i_{\rm c}$ of a $128^3$ mesh is plotted against the potential scale reduction factor ($\hat P-1$). 
Commonly a $\hat P-1$ of less then $0.1$ (blue line) is required to consider the chains to be converged at the target distribution. Here only two chains were compared, already showing that the majority of cells have converged. This result is already satisfactory, since including more chains will increase the statistics and reduce the potential scale reduction factor, eventually showing that all cells have converged.  }
 \label{fig:GR}
\end{figure*}

\begin{figure*} 
  \begin{tabular}{ccc}
   \includegraphics[width=6cm]{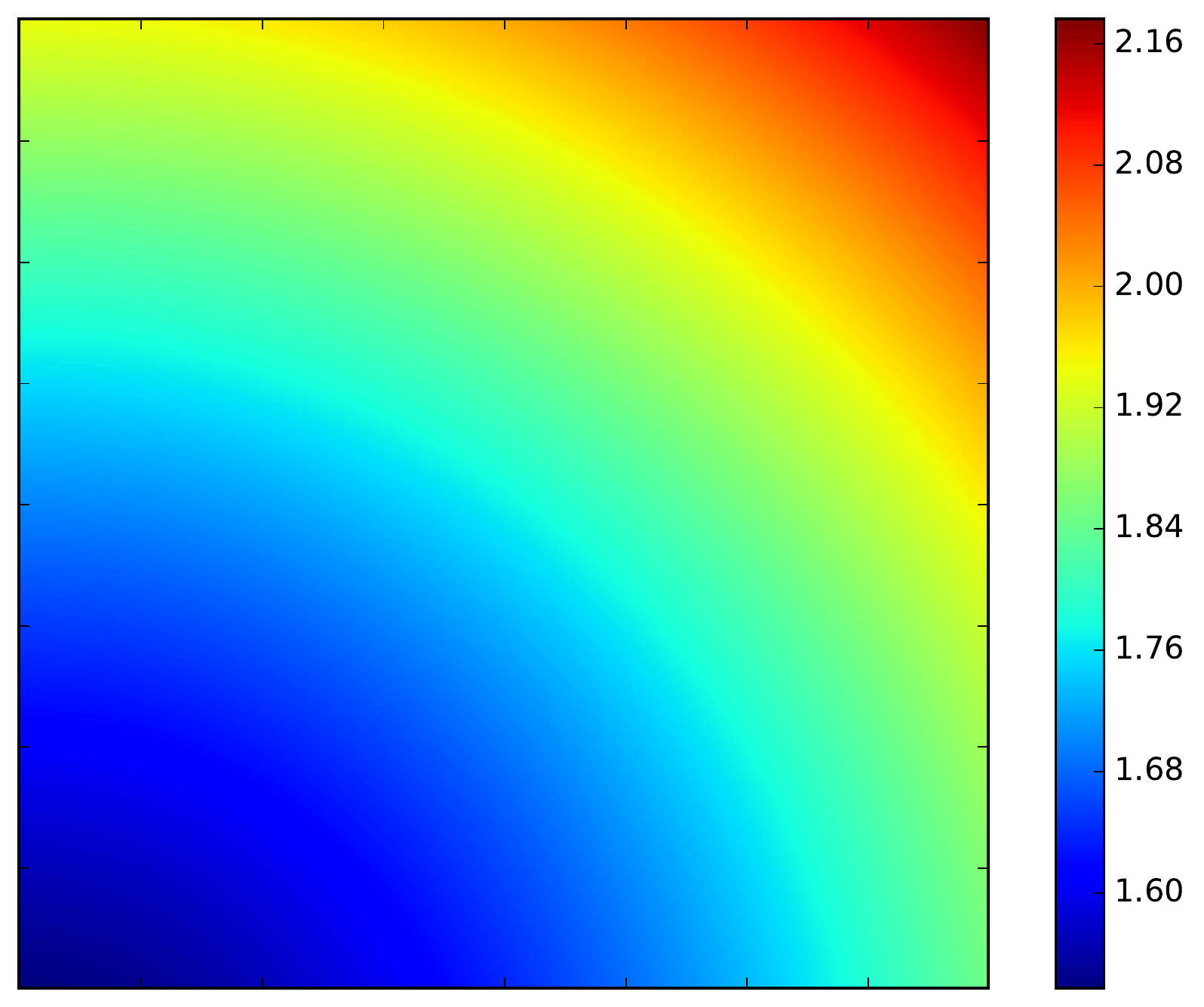}
   \includegraphics[width=6.25cm]{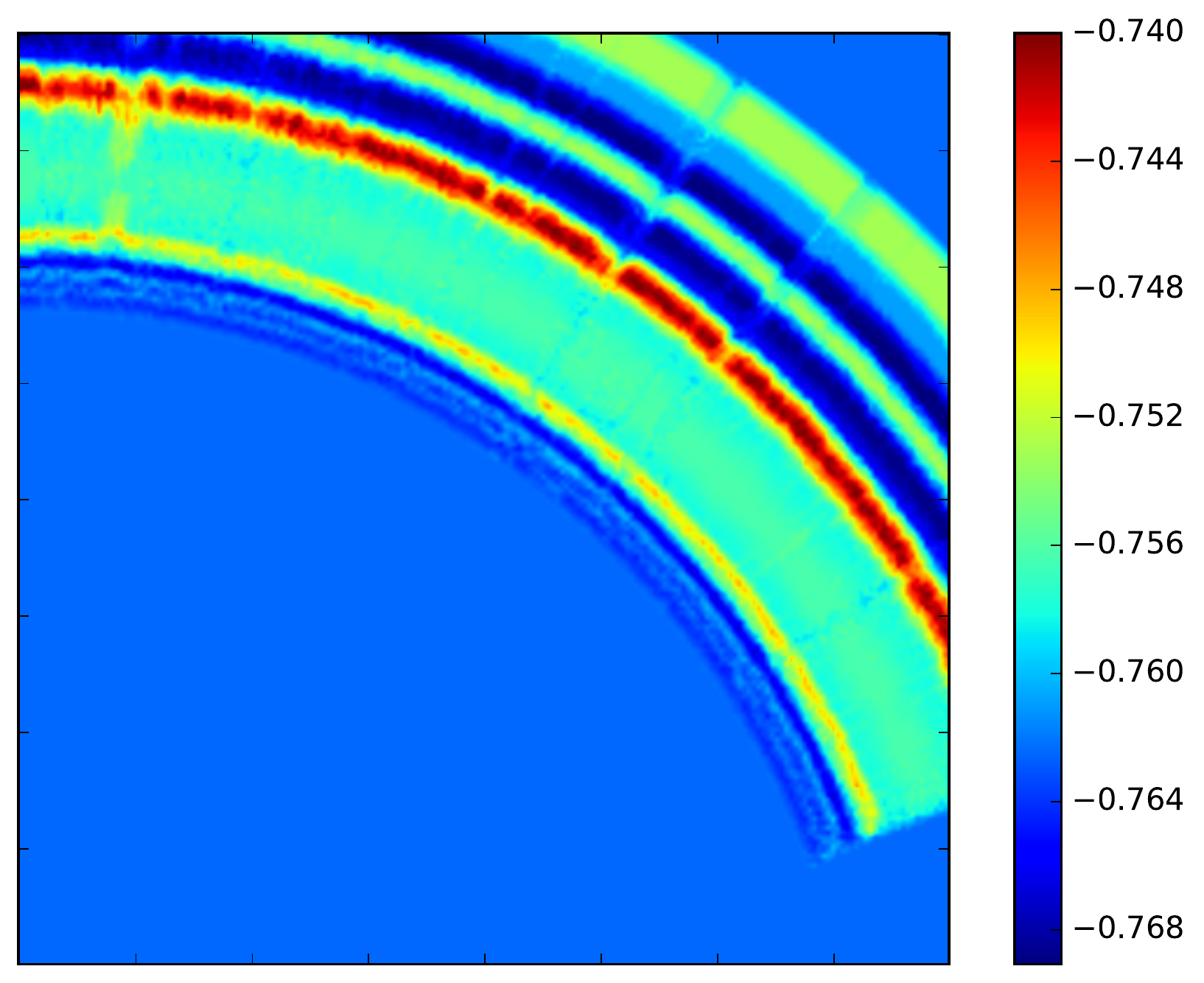}
   \includegraphics[width=6.1cm]{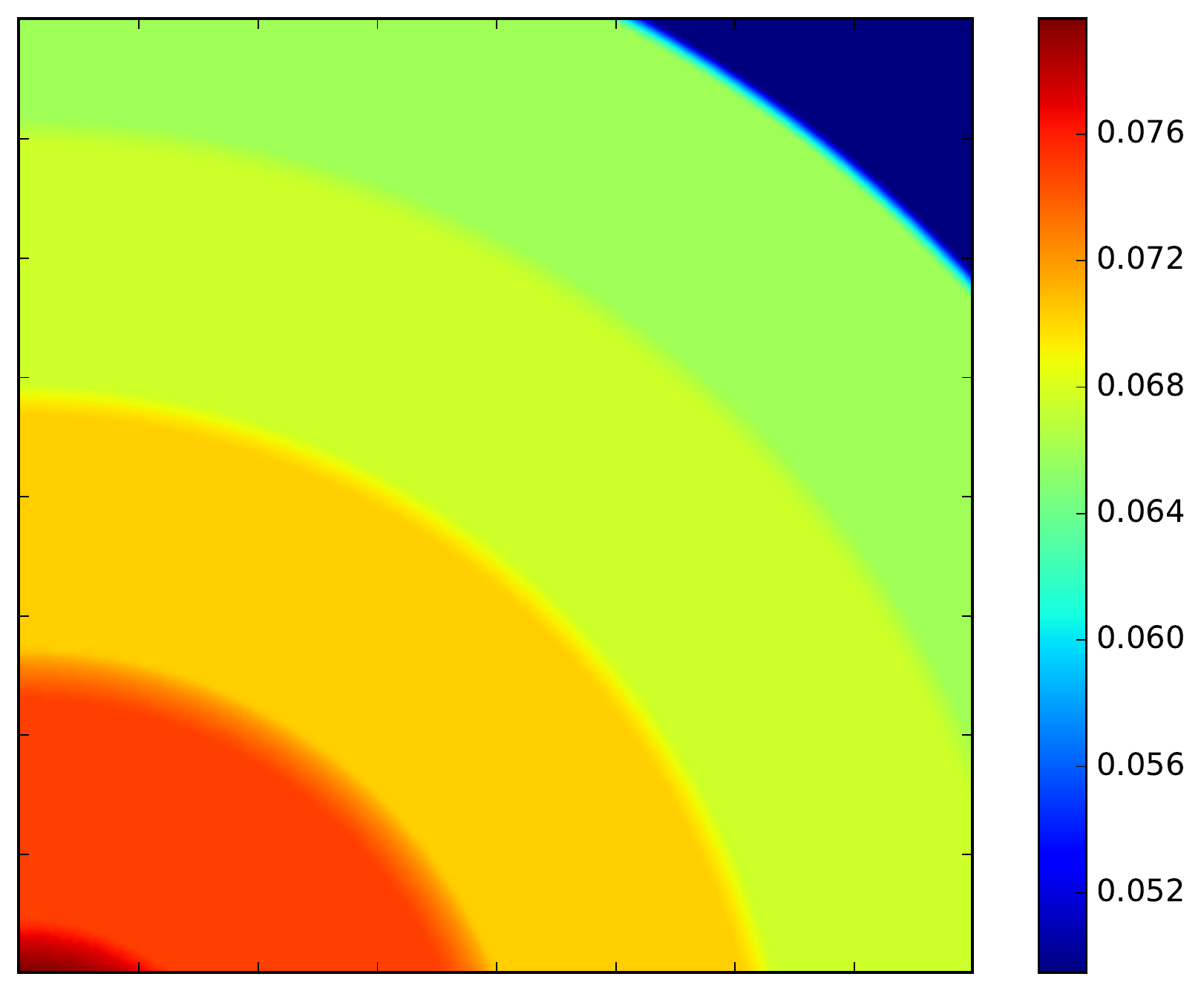}
   \end{tabular}
 \caption{Additional sampled quantities. Based on a light-cone mock catalogue in redshift space with $6.25\,h^{-1}$ Mpc resolution: slices of thickness $\sim$6 $h^{-1}$ Mpc in the $x-z$ plane of  the 3D cubical mesh of side 1250  $h^{-1}$ Mpc and $200^3$ cells for the following quantities: {\bf left panel:} the linear real space bias $b_{\rm L}$ multiplied with the nonlinear constant correction factor $f_{\rm b}=0.7$, {\bf middle panel:} the lognormal mean field $\mu$, and {\bf right panel:} the galaxy number density normalisation $\gamma$.}
 \label{fig:means}
\end{figure*}

\begin{figure*}
\begin{tabular}{cc}
\hspace{-0.3cm}
\begin{subfigure}{.3\textwidth}
\includegraphics[width=5.4cm]{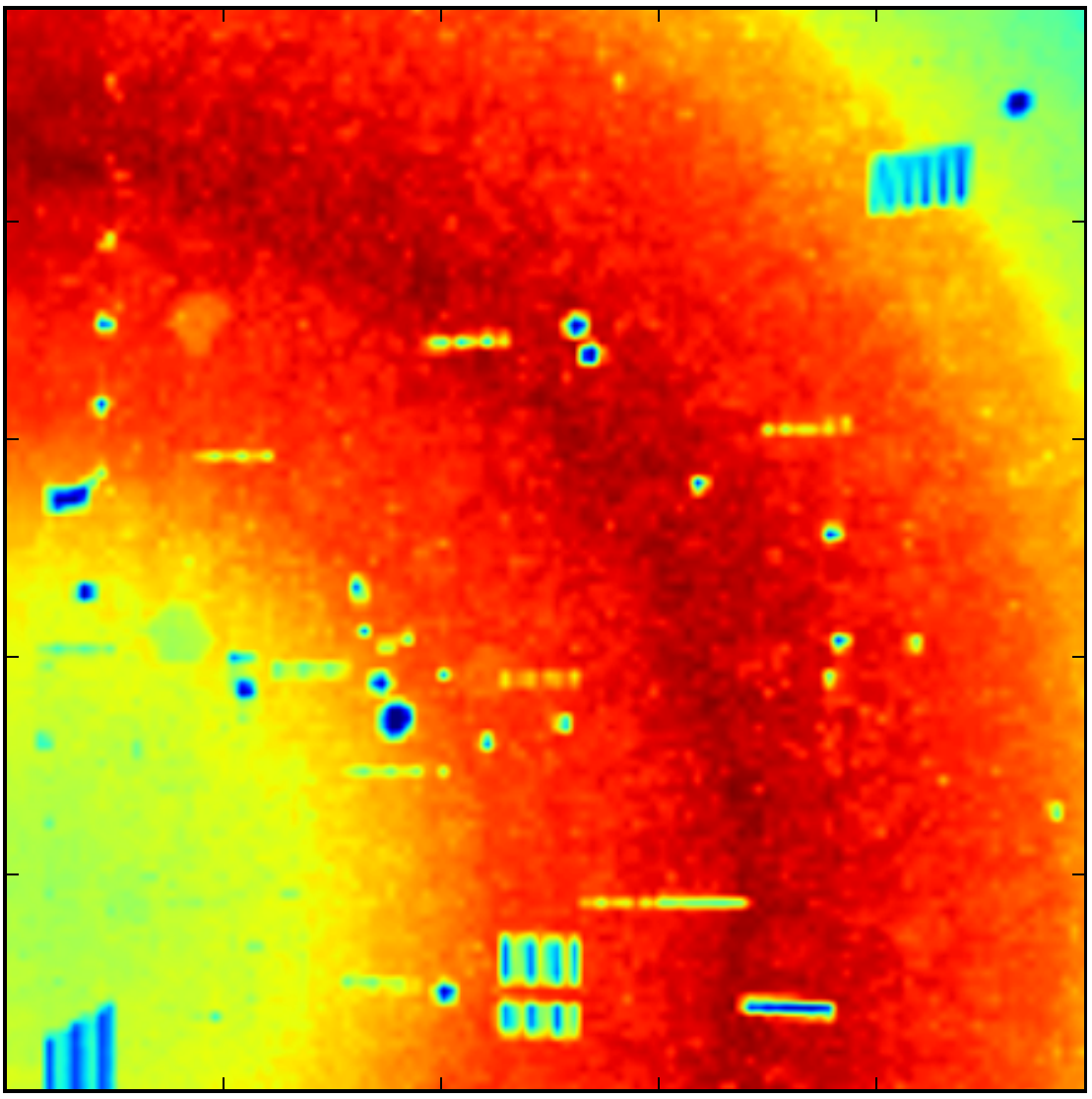}  
\end{subfigure}
\begin{subfigure}{.7\textwidth}
\includegraphics[width=.83cm]{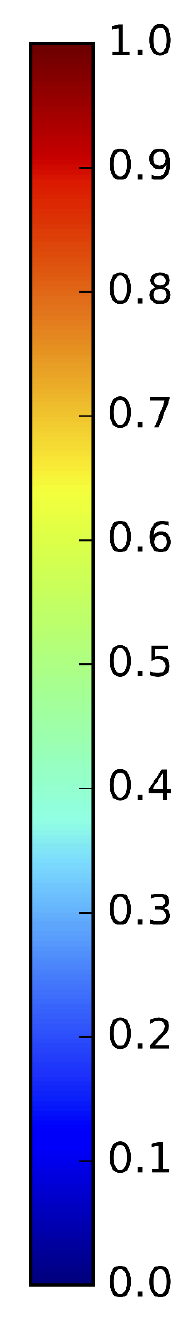}
 \hspace{-0.2cm}
  \includegraphics[width=7cm]{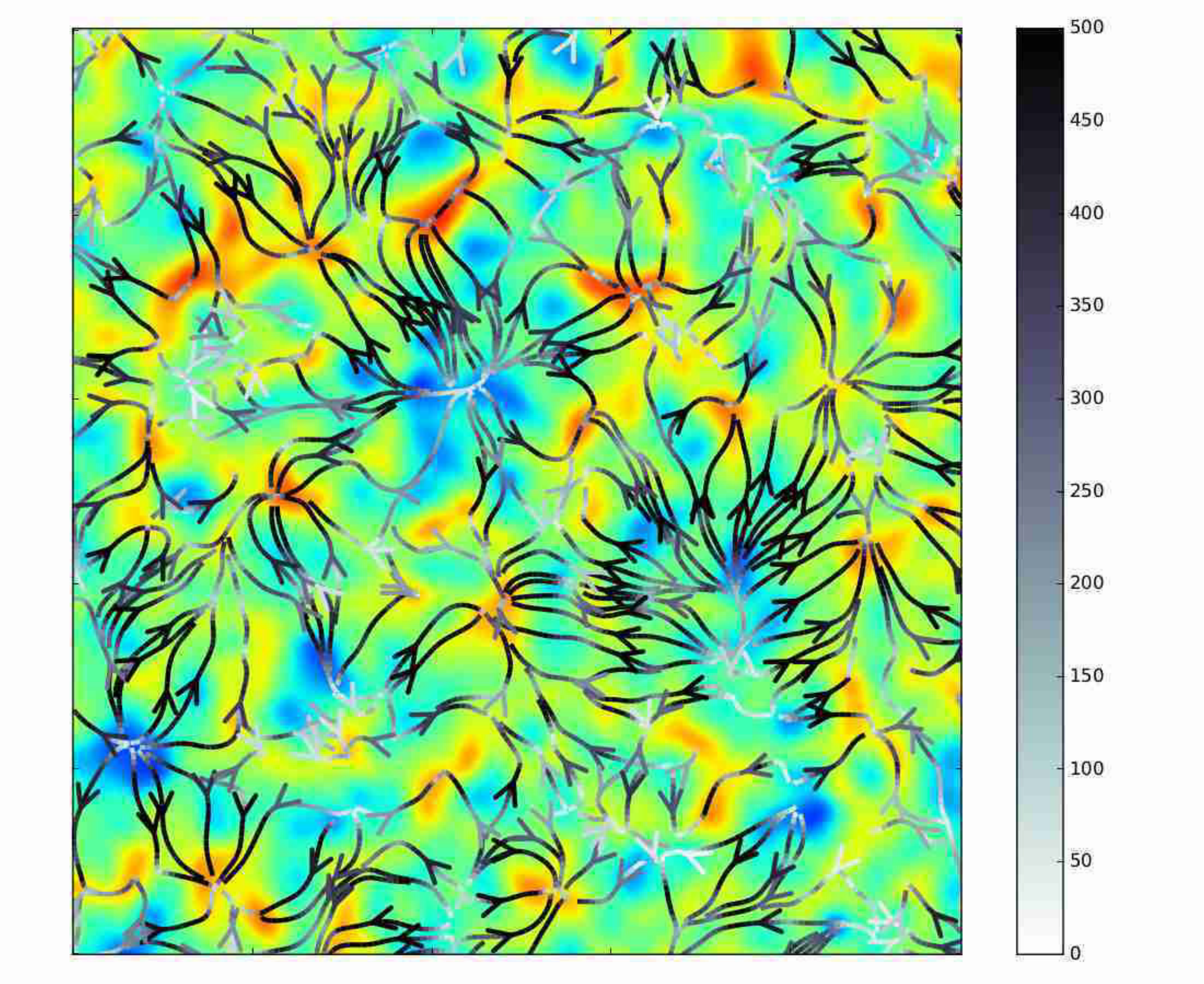}
\hspace{-1.5cm}
   \includegraphics[width=7cm]{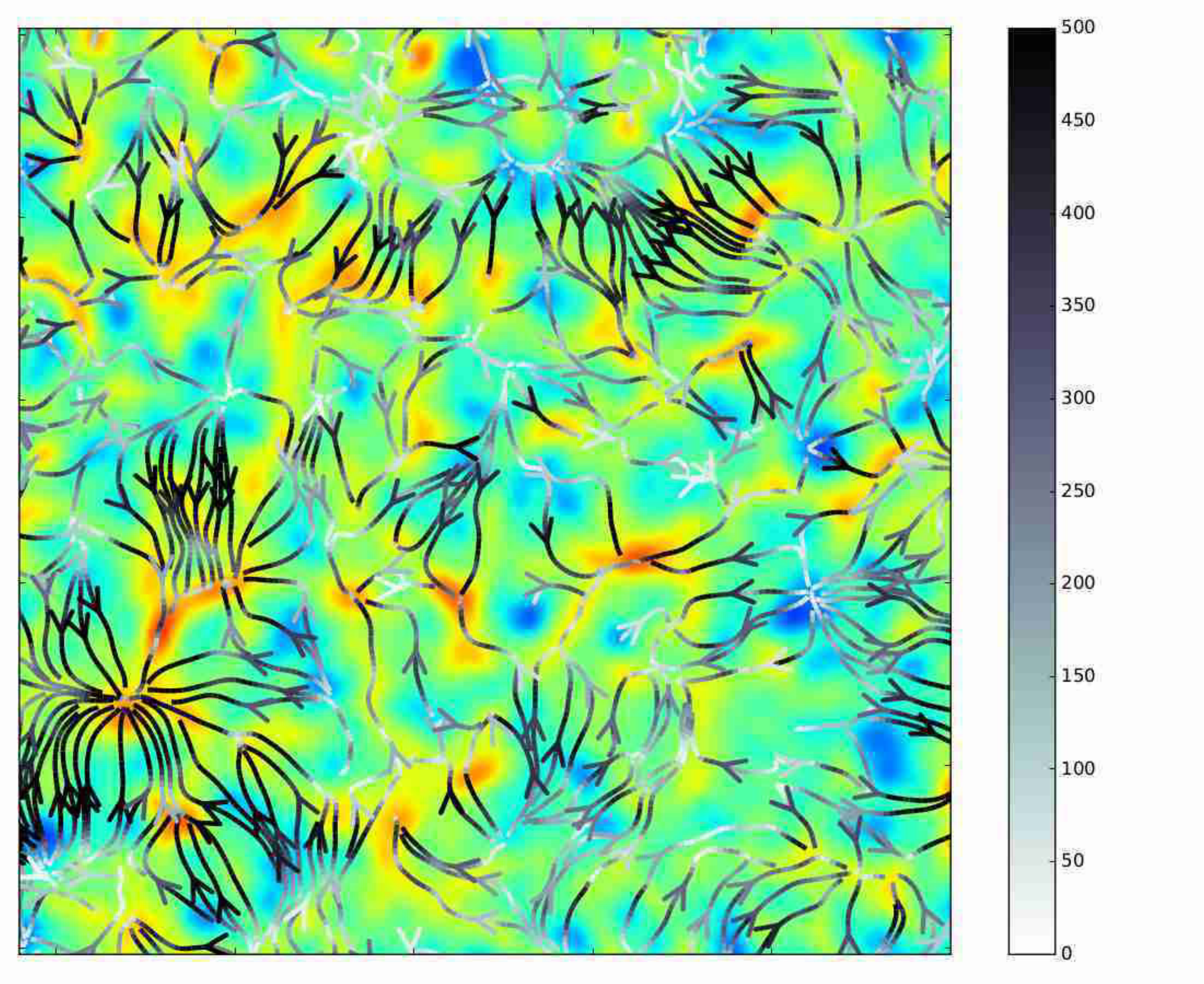}
\end{subfigure}
\end{tabular}                  
 \caption{Slices of thickness $\sim$30 $h^{-1}$ Mpc in the $x-y$ plane of  the 3D cubical mesh of side 1250  $h^{-1}$ Mpc and $200^3$ cells, showing a zoom-in region of 900 $h^{-1}$ Mpc side for visual purposes. {\bf Left panel:} the 3D completeness. Cosmic velocity fields with $6.25\,h^{-1}$ Mpc resolution with an additional Gaussian smoothing of the density and velocity field of $13\,h^{-1}$ Mpc smoothing radius  based on ({\bf middle panel:}) a light-cone mock catalogue in redshift space and on  {\bf right panel:} the BOSS DR12 data. The density of the stream lines  
 corresponds to the field strength of the flows, whereas the color of the stream lines indicates its velocity at a particular position. The colour code for the density field is red for high  and blue for low densities.  A more quantitive comparison is shown in the figures below and in \S \ref{sec:cw}.}
 \label{fig:vel}
\end{figure*}

\begin{figure*}
\begin{tabular}{ccc}
\hspace{-0.5cm}
\begin{subfigure}{.3\textwidth}
\includegraphics[width=5.45cm]{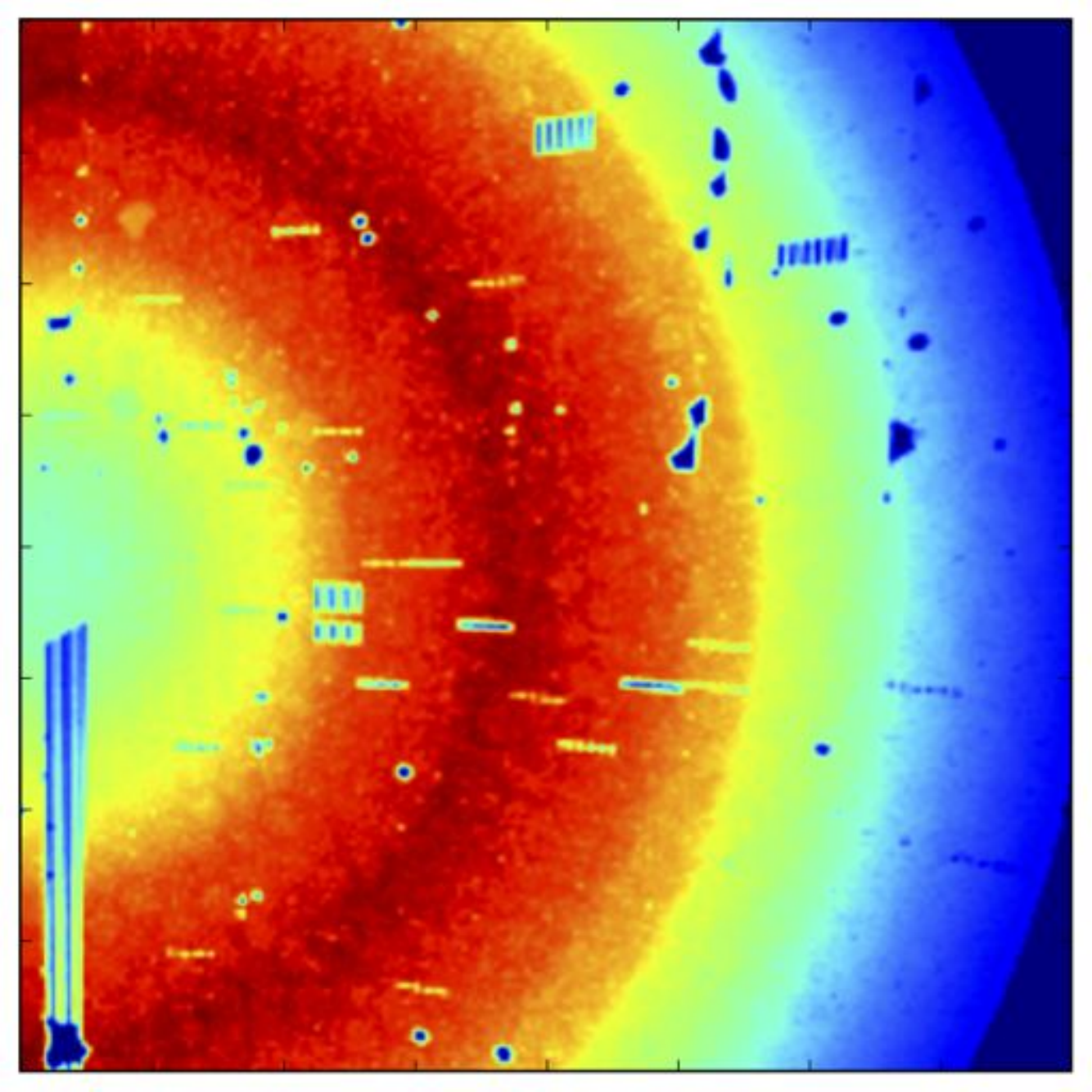} 
\vspace{-0.4cm}
\end{subfigure}
\begin{subfigure}{.1\textwidth}
\includegraphics[width=.84cm]{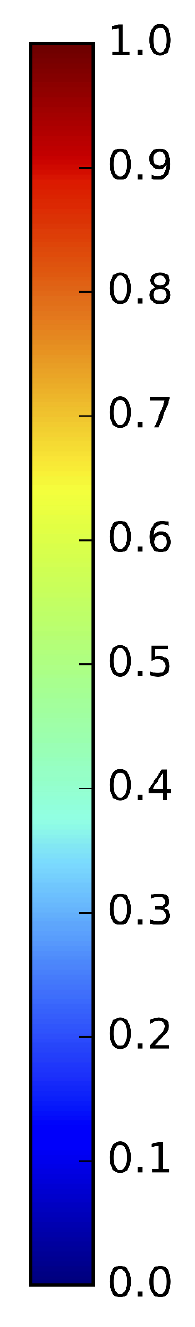}
\end{subfigure}
\hspace{-.75cm}
\begin{subfigure}{.6\textwidth}
\hspace{0.025cm}
   \includegraphics[width=5.6cm]{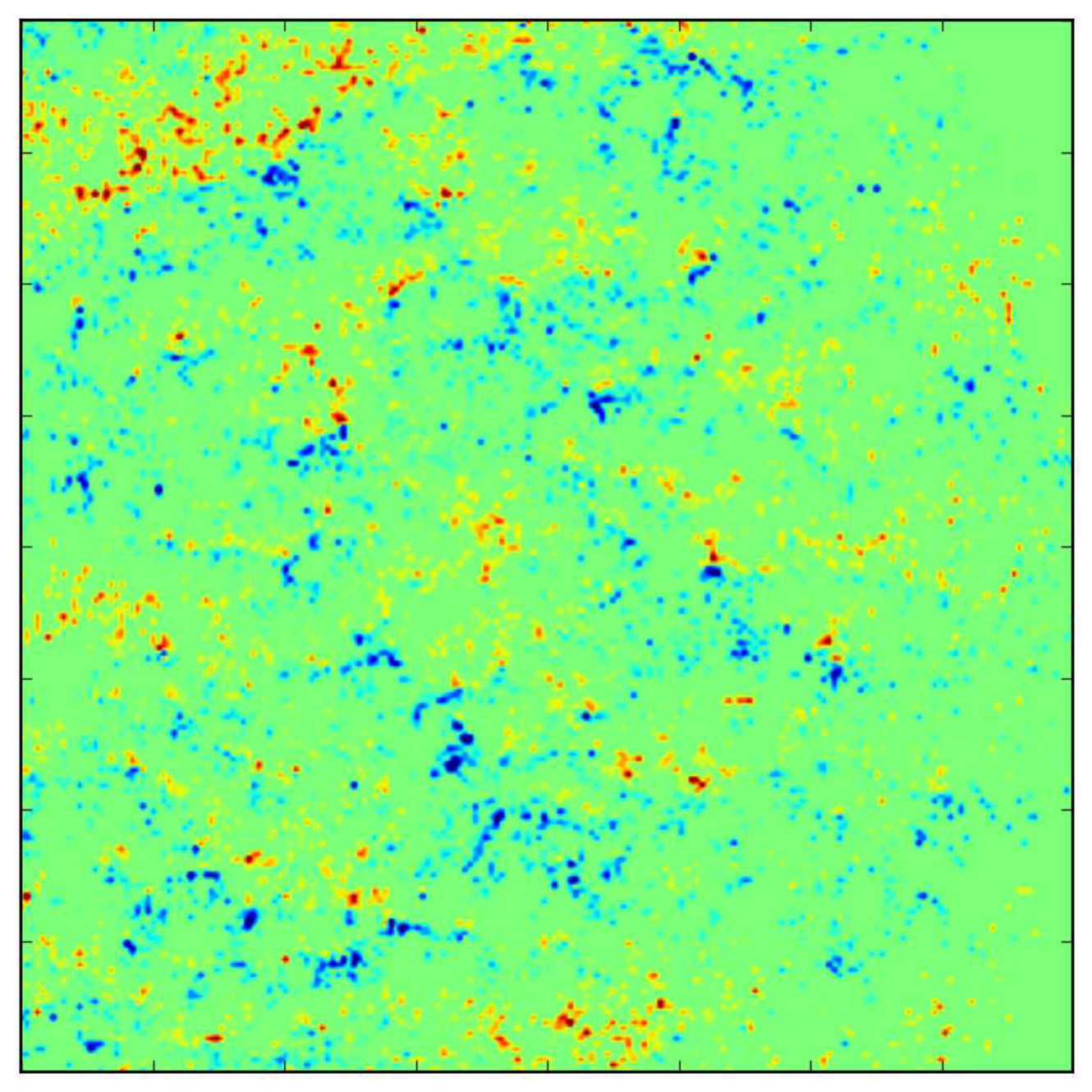}
\hspace{-0.2cm}
   \includegraphics[width=5.6cm]{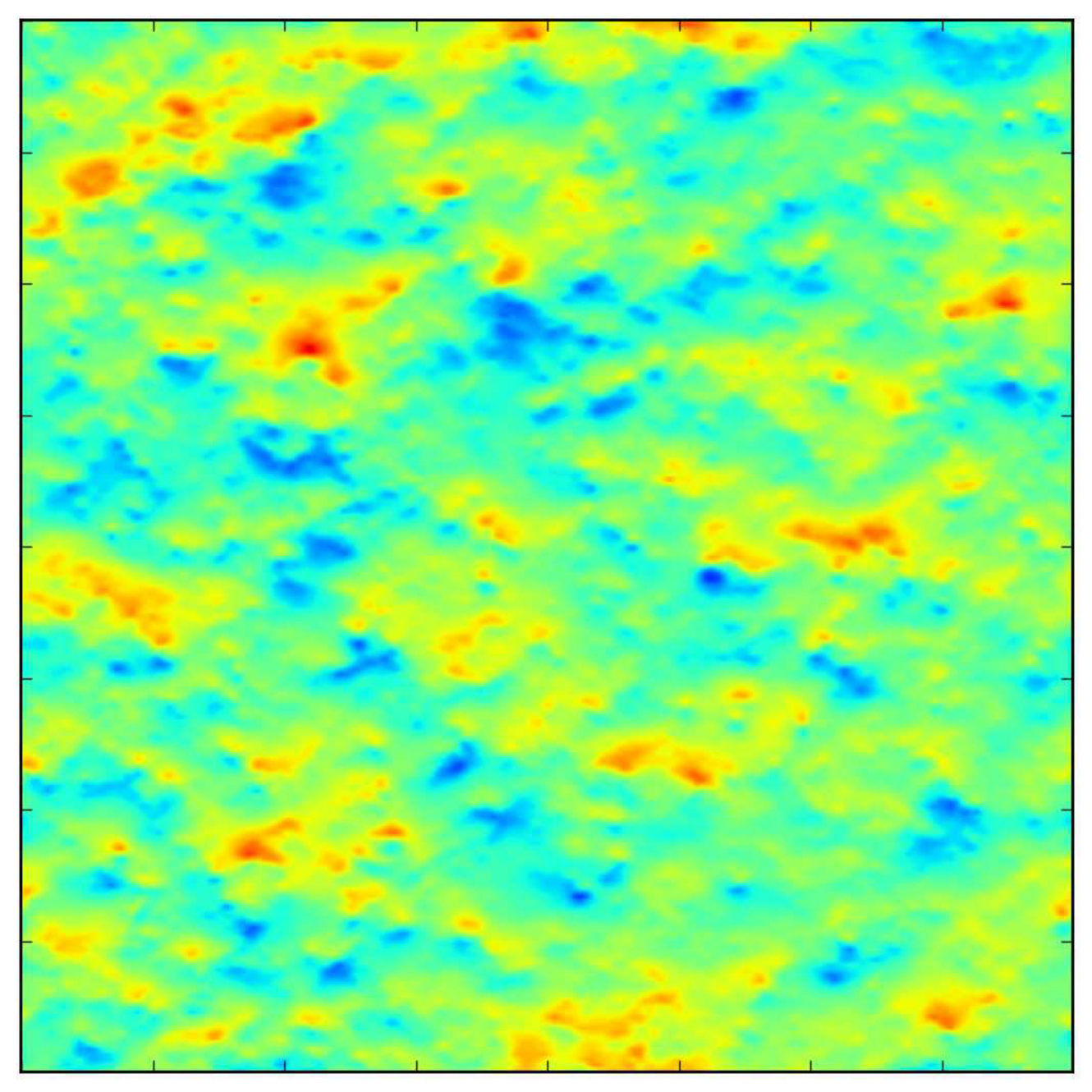}
\end{subfigure}%
\end{tabular}
 \caption{Based on a light-cone mock catalogue in redshift space with $6.25\,h^{-1}$ Mpc resolution and side 1250  $h^{-1}$ Mpc: slices in the $x-y$ plane of  ({\bf left panel:}) the 3D completeness, and the $x$-component of the velocity field  for {\bf middle panel:} the averaged mock galaxy velocities per cell, and   {\bf right panel:} one reconstructed velocity field sample with \textsc{argo} (compensating for completeness). The colour code for the density field is red for positive  and blue for negative peculiar velocities. A more quantitive comparison is shown in Fig.~\ref{fig:corrvel}.}
 \label{fig:velcomp}
\end{figure*}

\begin{figure*}  
\begin{tabular}{cc}
\hspace{-.5cm}
   \includegraphics[width=6.3cm]{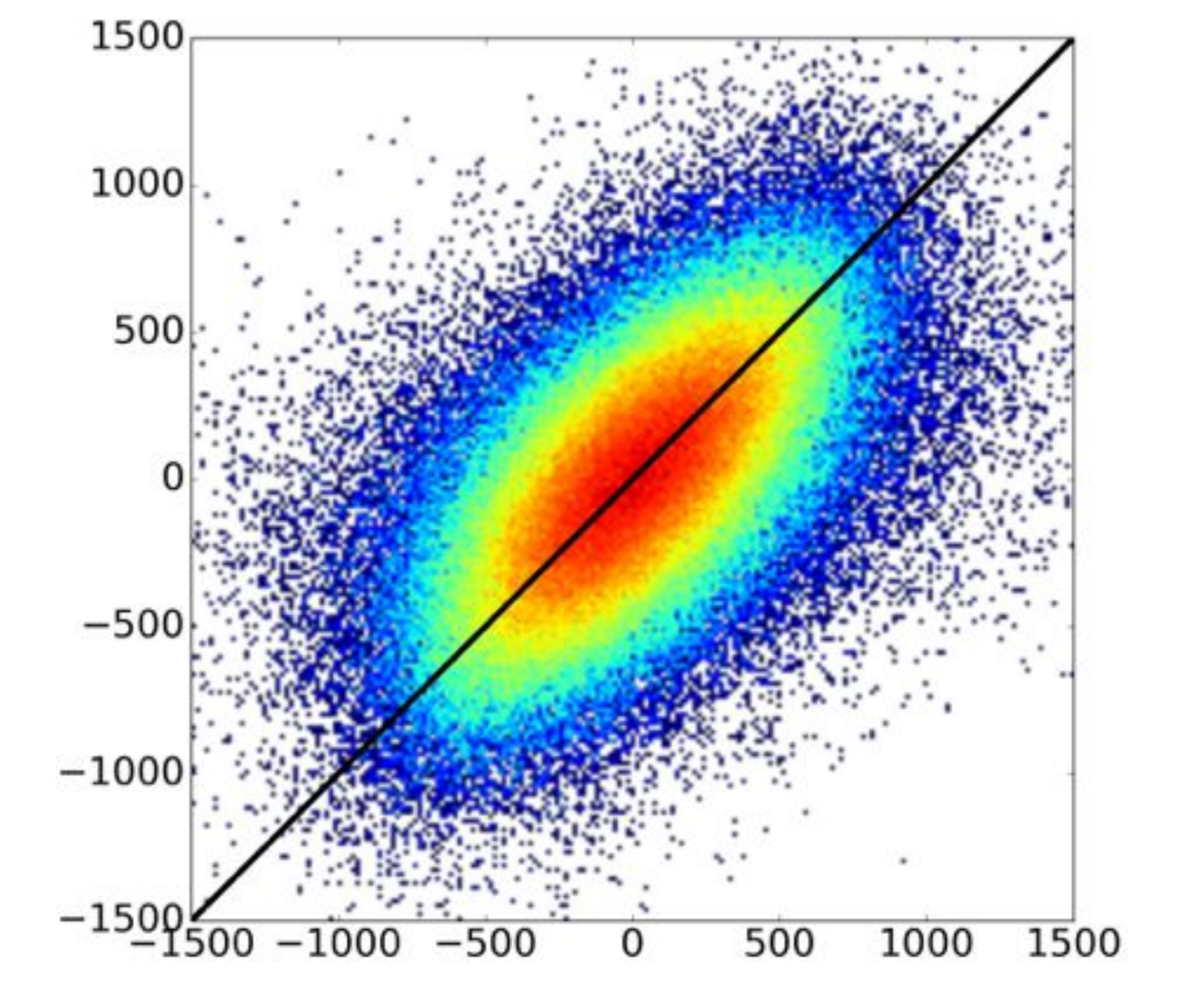}
\put(-140,130){\rotatebox[]{0}{\text{$r=0.61$}}}
   \put(-180,75){\rotatebox[]{90}{$v_{\rm \textsc{argo}}$ [km $s^{-1}$]}}
   \put(-115,-7){\rotatebox[]{0}{$v_{\rm mock}$} [km $s^{-1}$]}
\hspace{-.5cm}
   \includegraphics[width=6.3cm]{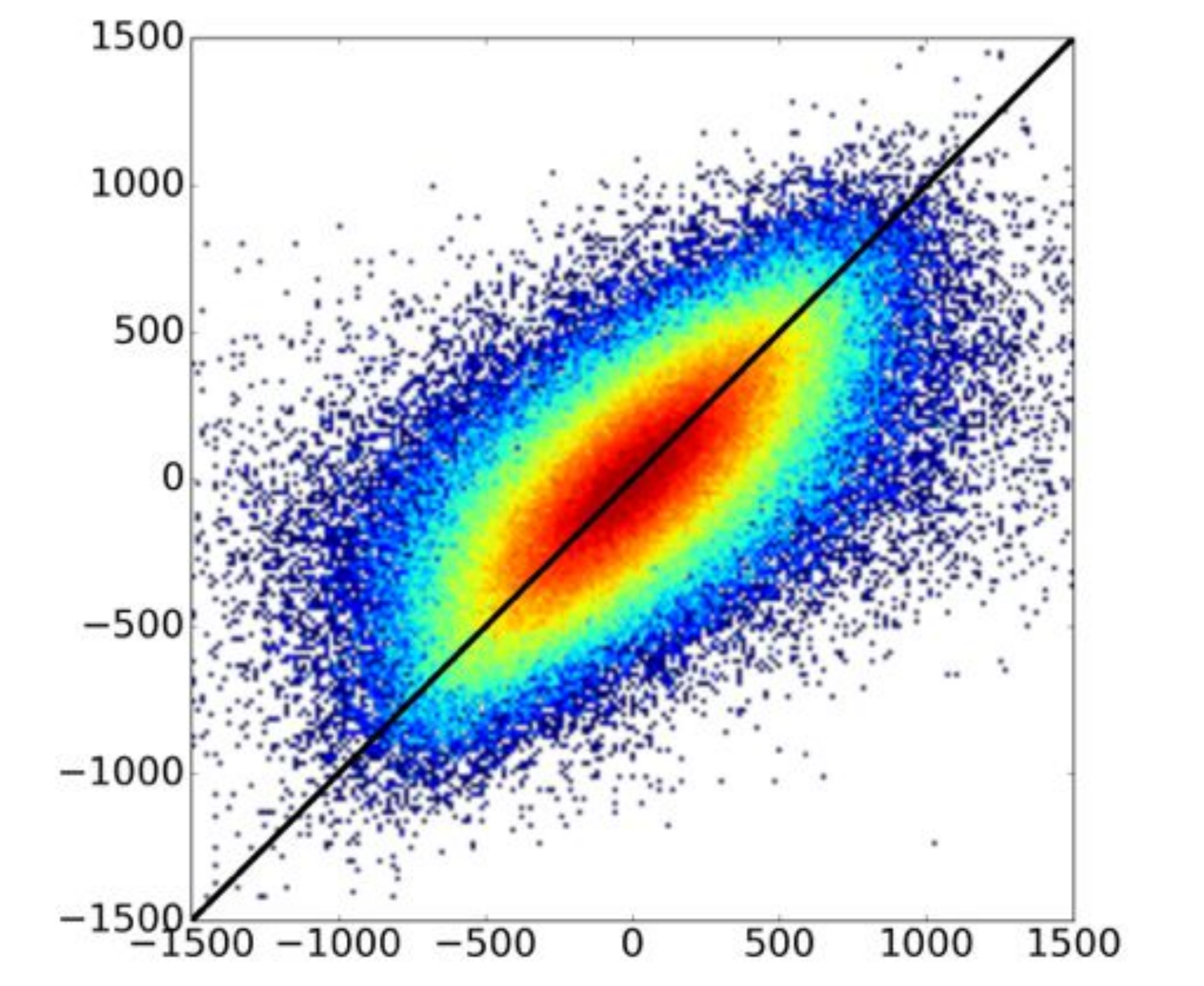}
\put(-140,130){\rotatebox[]{0}{\text{$r=0.68$}}}
   \put(-180,75){\rotatebox[]{90}{\text{$\langle v_{\rm \textsc{argo}}\rangle$} [km $s^{-1}$]}}
   \put(-115,-7){\rotatebox[]{0}{$v_{\rm mock}$} [km $s^{-1}$]}
\hspace{-.3cm}
   \includegraphics[width=.88cm]{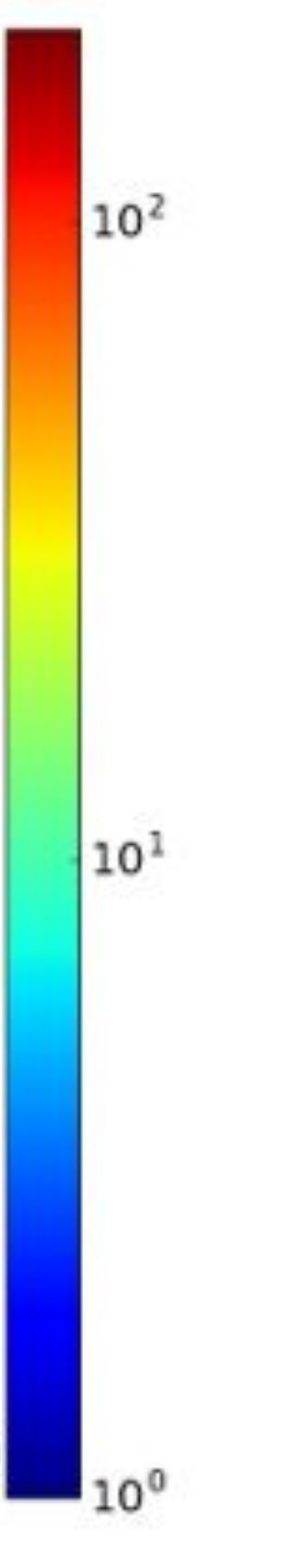}
\\
\hspace{-.5cm}
   \includegraphics[width=6.3cm]{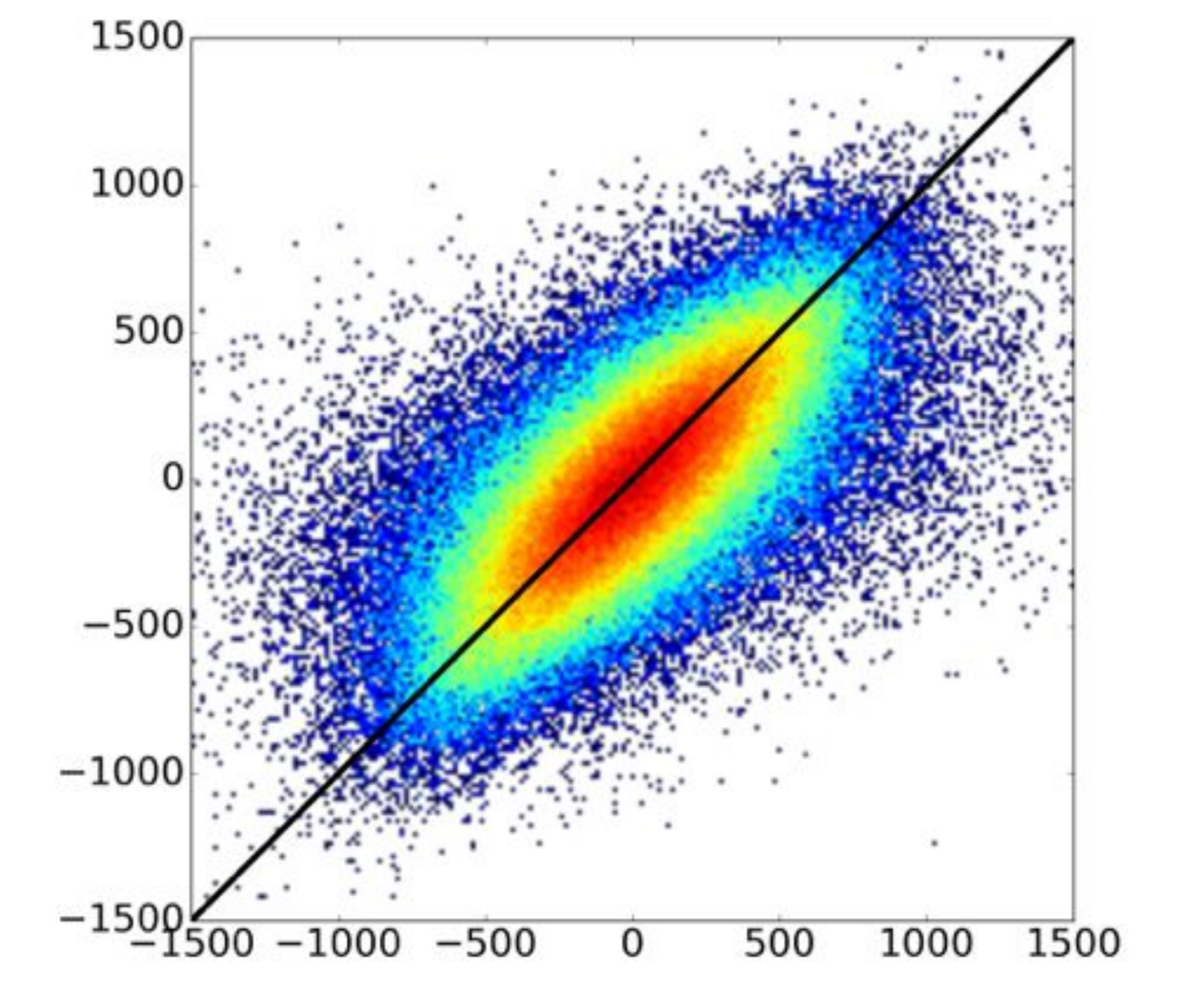}
\put(-140,130){\rotatebox[]{0}{\text{$r=0.7$}}}
   \put(-180,75){\rotatebox[]{90}{\text{$\langle v_{\rm \textsc{argo}}\rangle_{w>0.5}$} [km $s^{-1}$]}}
   \put(-115,-7){\rotatebox[]{0}{$v_{\rm mock}$} [km $s^{-1}$]}
\hspace{-.5cm}
   \includegraphics[width=6.3cm]{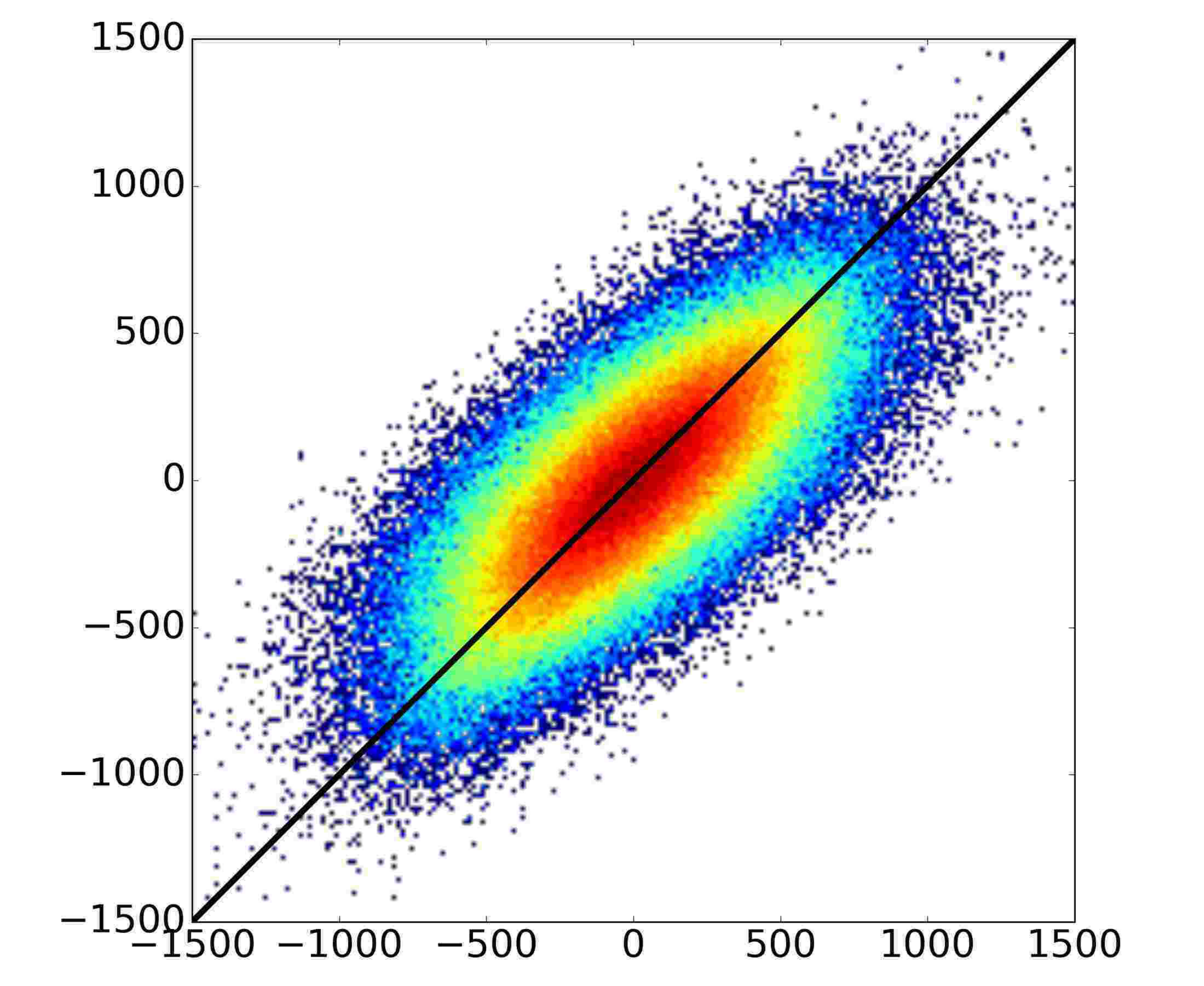}
\put(-140,130){\rotatebox[]{0}{\text{$r=0.76$}}}
   \put(-180,75){\rotatebox[]{90}{\text{$\langle v_{\rm \textsc{argo}}\rangle_{-3.5\% \,{\rm outliers}}$} [$s^{-1}$ km]}}
   \put(-115,-7){\rotatebox[]{0}{$v_{\rm mock}$} [km $s^{-1}$]}
\hspace{-.5cm}
   \includegraphics[width=6.3cm]{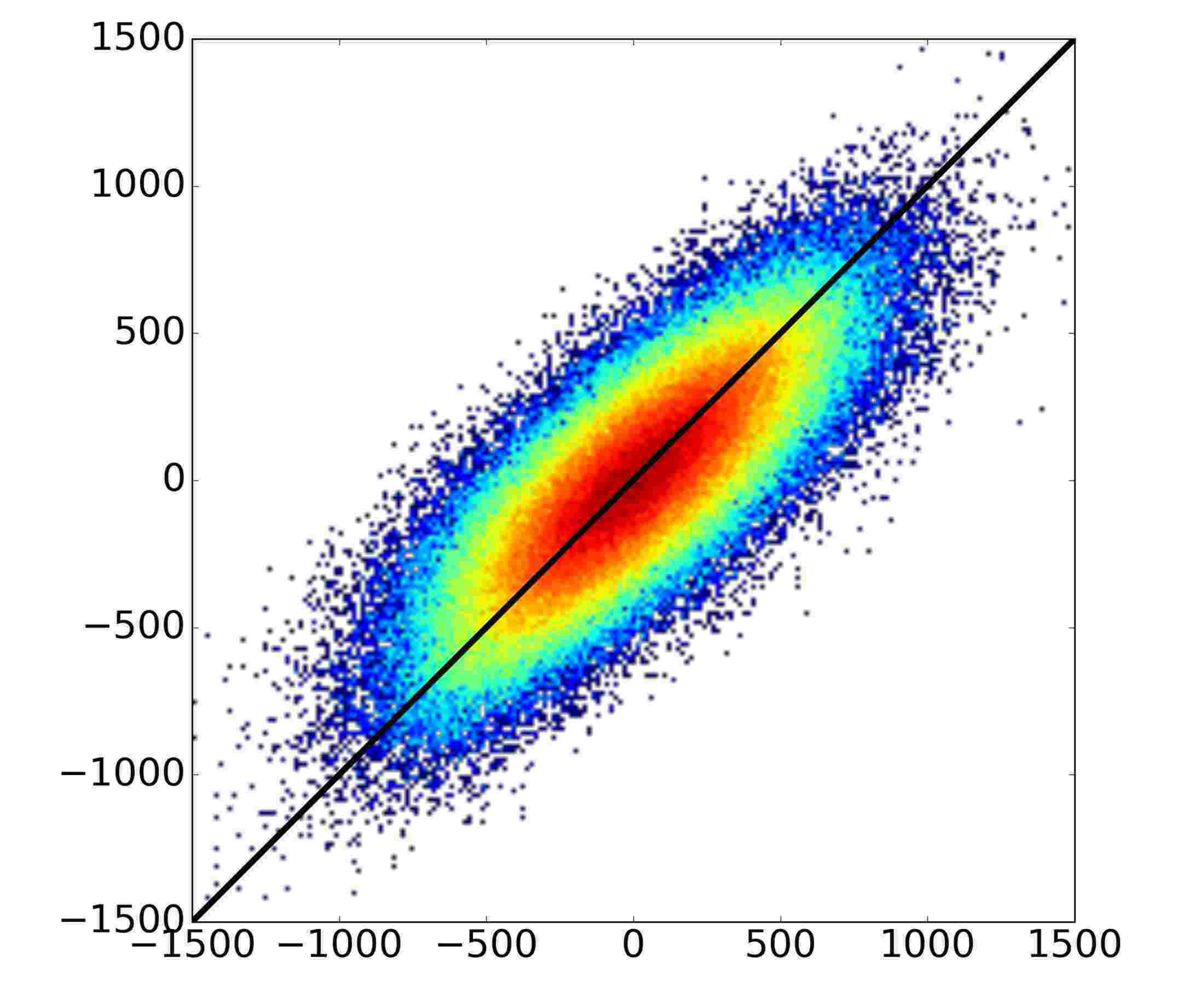}
\put(-140,130){\rotatebox[]{0}{\text{$r=0.8$}}}
   \put(-180,75){\rotatebox[]{90}{\text{$\langle v_{\rm \textsc{argo}}\rangle_{-10\% \,{\rm outliers}}$} [$s^{-1}$ km]}}
   \put(-115,-7){\rotatebox[]{0}{$v_{\rm mock}$} [km $s^{-1}$]}
\hspace{-.3cm}
   \includegraphics[width=.88cm]{CF_corr_vel_cb}
\end{tabular}
 	\caption{Velocity correlation taking one component of the velocity field for reconstructions with resolutions of $d_{\rm L}=6.25\,h^{-1}$ Mpc  with additional Gaussian smoothing of $r_{\rm S}=2\,h^{-1}$ Mpc. {\bf Upper left panel:} for one reconstructed sample, {\bf upper right panel:} for the mean over 6000 reconstructed samples, {\bf lower left panel:} same as upper right panel, but considering only galaxies with completeness  $w>0.5$ (for about 209000 galaxies, $\sim$82\% of the whole CMASS sample in the considered volume), {\bf lower middle panel:} same as upper right panel, but excluding galaxies for which the difference in the velocity reconstruction eceeds $|\mbi v|=700$ km $s^{-1}$ (i.e., excluding about 3.5\% of the sample), and {\bf lower right panel:} same as upper right panel, but excluding galaxies for which the difference in the velocity reconstruction eceeds $|\mbi v|=500$ km $s^{-1}$ (i.e., excluding about 10\% of the sample).}
 \label{fig:corrvel}
\end{figure*}

\begin{figure}  
   \includegraphics[width=8.5cm]{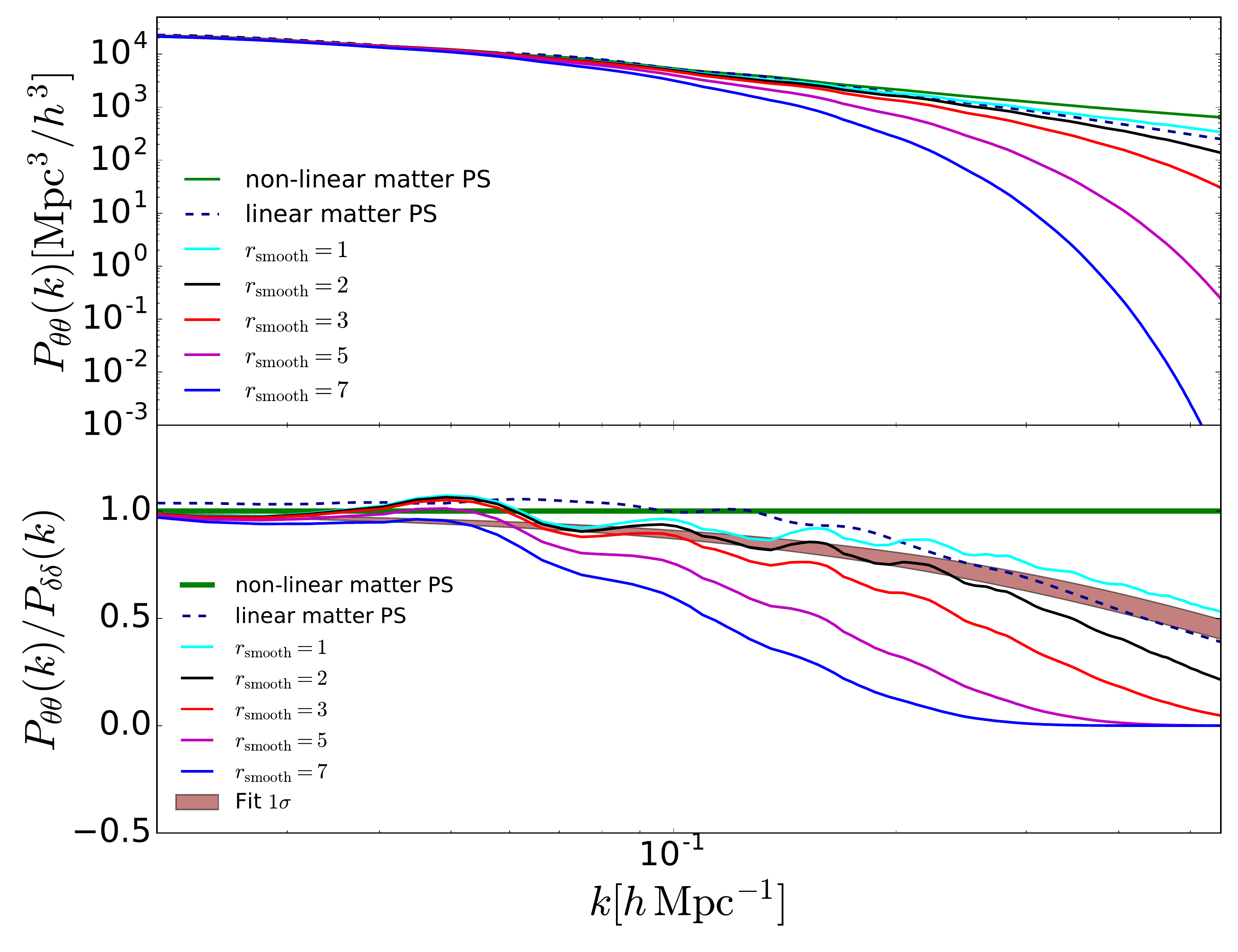}
 	\caption{{\bf Upper panel:} power spectrum of the scaled divergence of the peculiar velocity field for different smoothing scales for a typical realisation on a mesh of $200^3$ with resolution $d_{\rm L}=6.25\,h^{-1}$ Mpc. {\bf Lower panel:} ratio with respect to the nonlinear power spectrum from \citet[][]{Heitmann_et_al_2010}.  The shaded region represents the theoretical fit for the velocity divergence bias $b_{v}={\rm e}^{-(k/a)^b}$ by \citet[][]{Hahn2015} with the sigma region being computed based on the largest uncertainty found on the parameters $a$ and $b$. The wiggles are due to the more pronounced baryon acoustic oscillations in the mean theoretical power spectrum than in the particular realisation used in this plot.}
 \label{fig:pstheta}
\end{figure}

\begin{figure*}
 \begin{tabular}{cc}
\hspace{1cm}
\begin{subfigure}{.5\textwidth}
   \includegraphics[width=8.3cm]{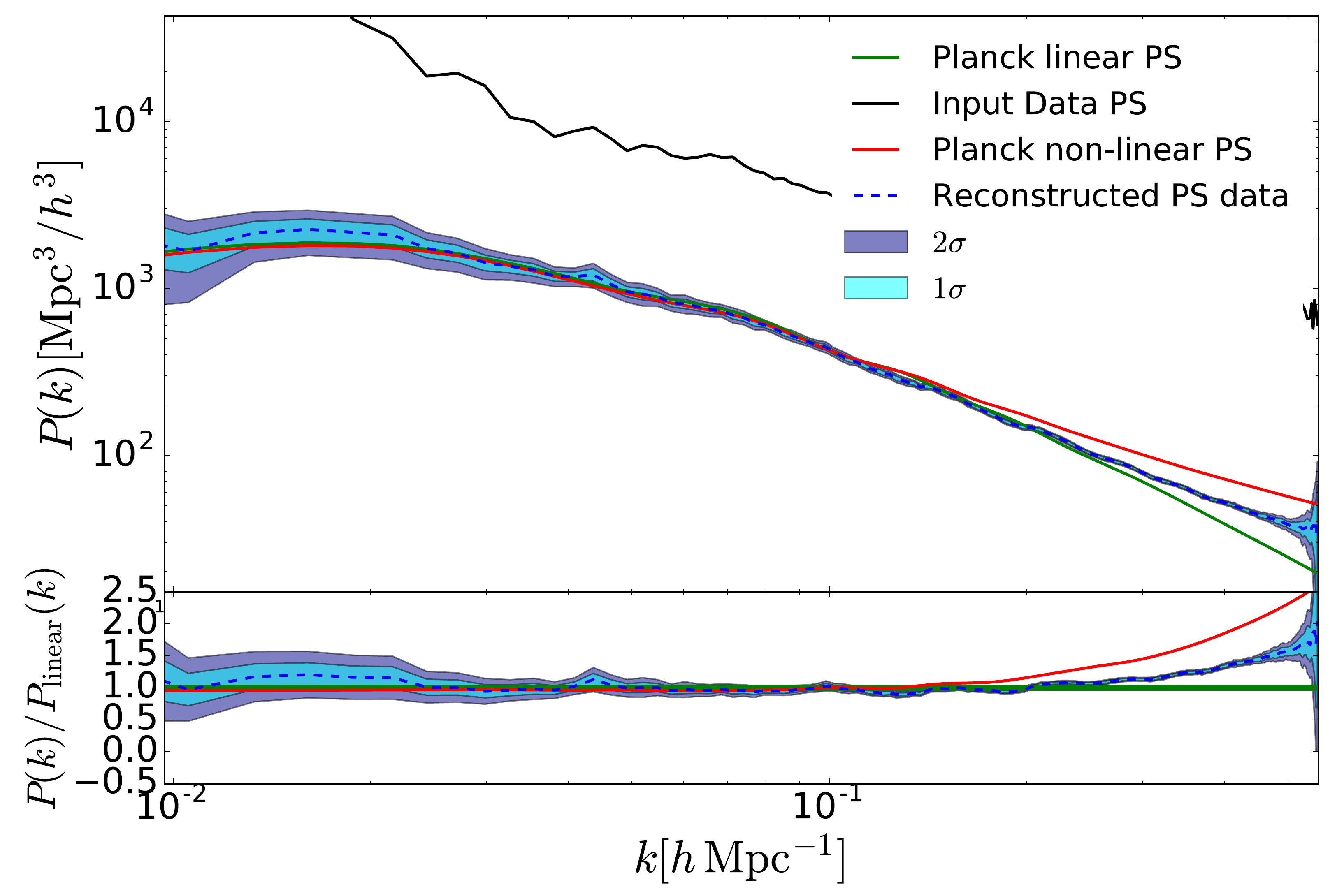}
\put(-200,80){${\rm L}=1250\,h^{-1}\,{\rm Mpc}$}
\put(-200,70){$d_{\rm L}=9.76\,h^{-1}\,{\rm Mpc}$}
\put(-200,60){\rotatebox[]{0}{\text{MOCKS BOSS DR12}}}
\vspace{-0.5cm}
\end{subfigure}%
\begin{subfigure}{.5\textwidth}
   \includegraphics[width=7.4cm]{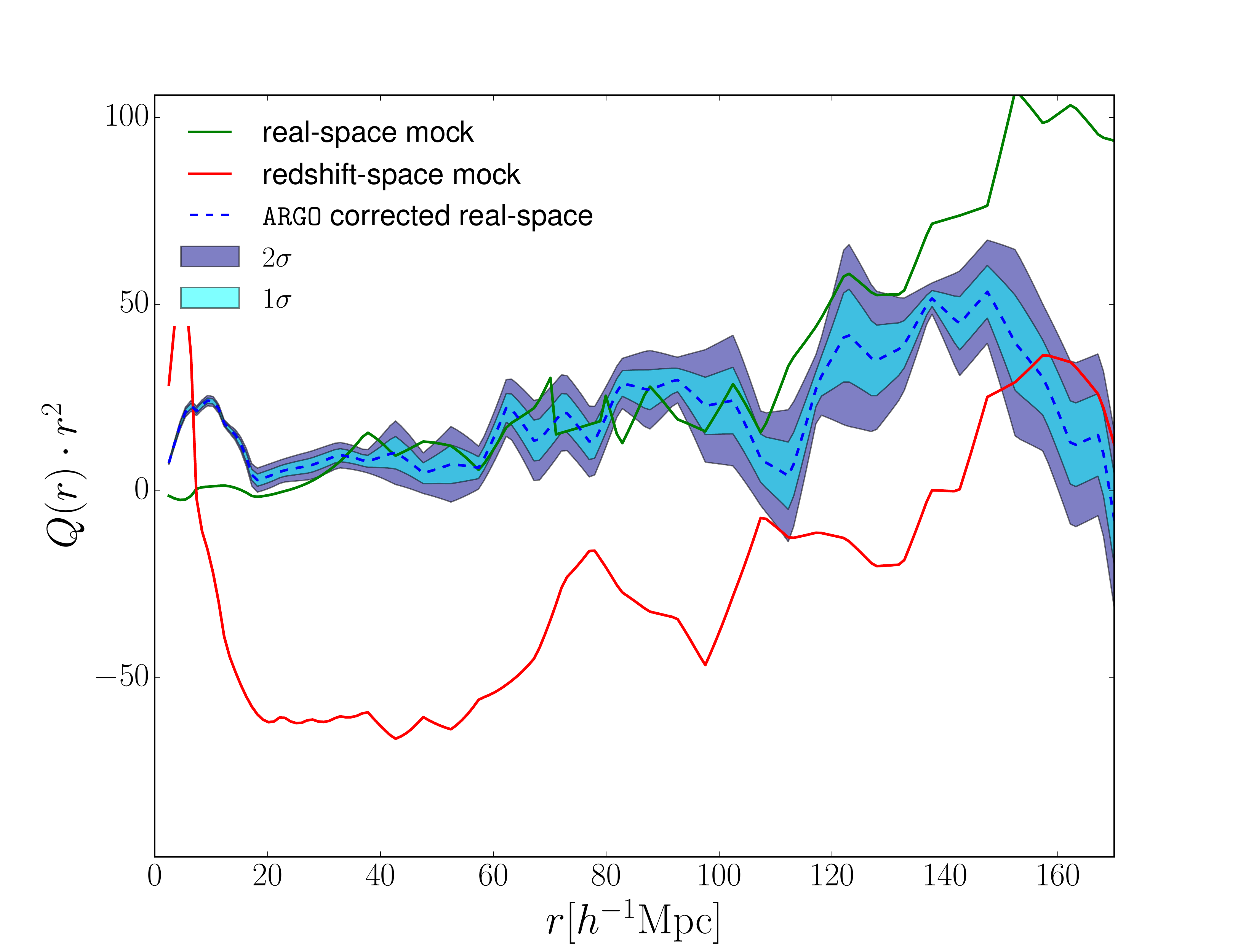}
\put(-180,40){${\rm L}=1250\,h^{-1}\,{\rm Mpc}$}
\put(-180,30){$d_{\rm L}=9.76\,h^{-1}\,{\rm Mpc}$}
\put(-180,20){\rotatebox[]{0}{\text{MOCKS BOSS DR12}}}
\end{subfigure}%
\vspace{-.55cm}
\\
\hspace{1cm}
\begin{subfigure}{.5\textwidth}
   \includegraphics[width=8.3cm]{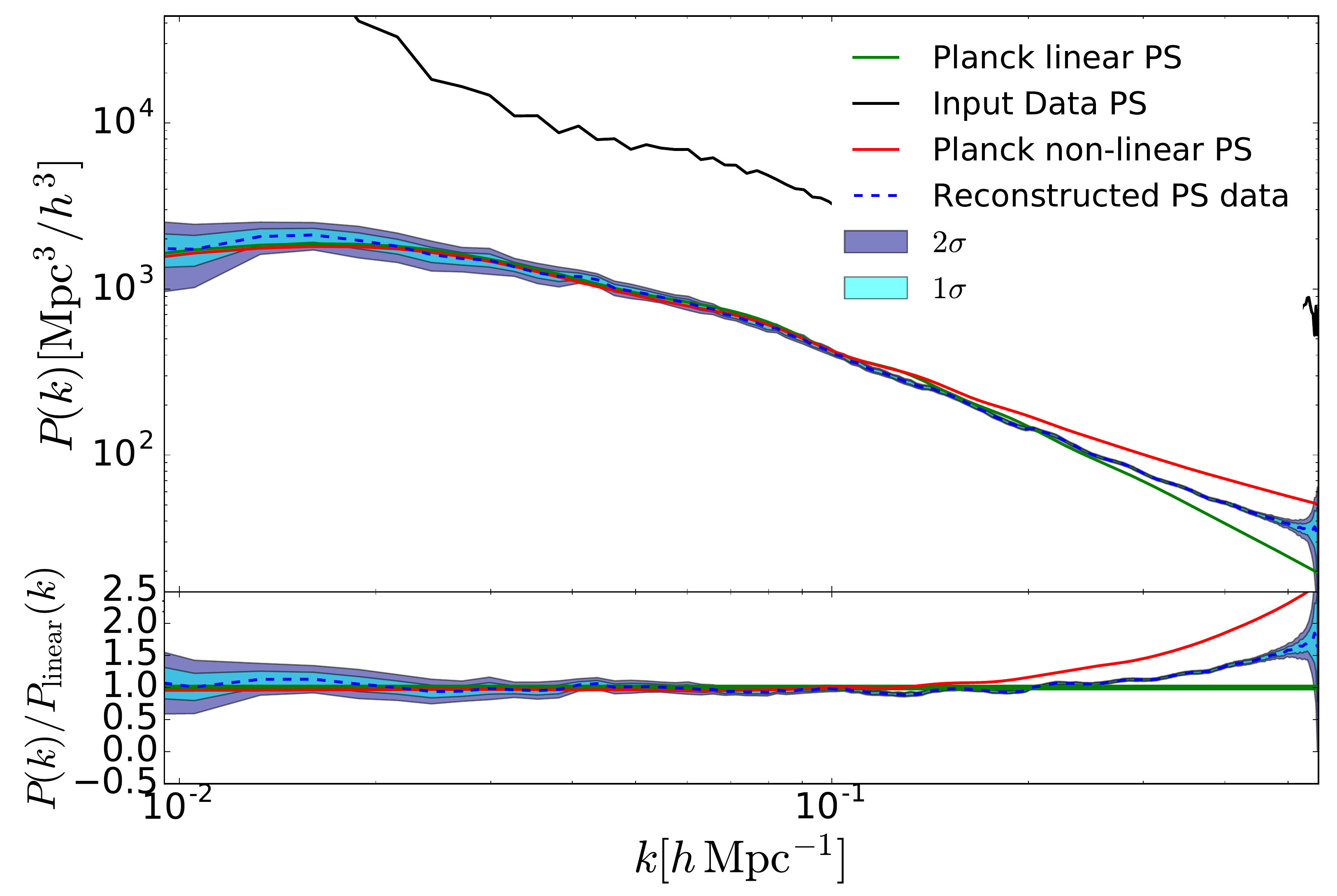}
\put(-200,80){${\rm L}=1250\,h^{-1}\,{\rm Mpc}$}
\put(-200,70){$d_{\rm L}=9.76\,h^{-1}\,{\rm Mpc}$}
\put(-200,60){\rotatebox[]{0}{\text{BOSS DR12}}}
\vspace{-0.5cm}
\end{subfigure}%
\begin{subfigure}{.5\textwidth}
   \includegraphics[width=7.4cm]{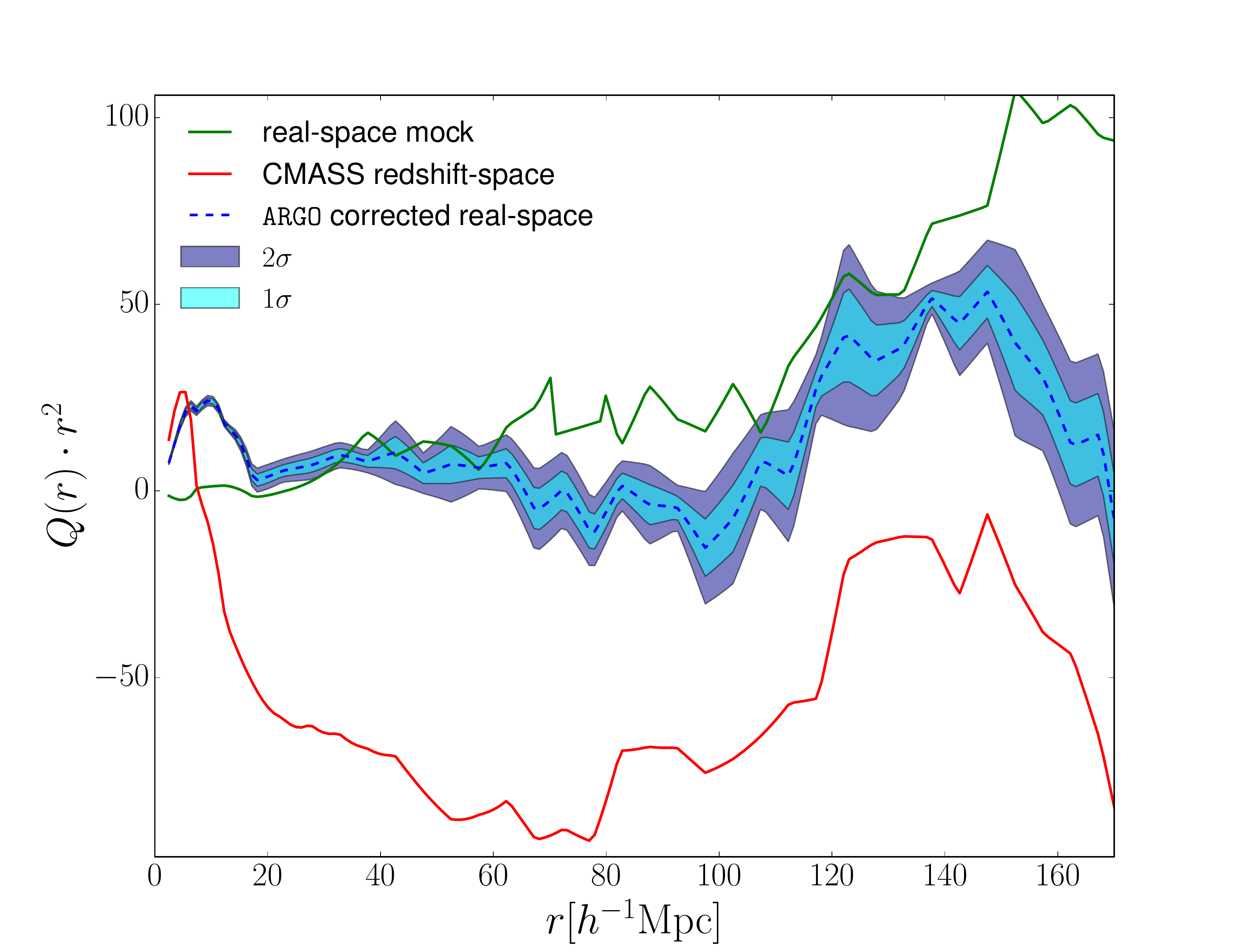}
\put(-180,40){${\rm L}=1250\,h^{-1}\,{\rm Mpc}$}
\put(-180,30){$d_{\rm L}=9.76\,h^{-1}\,{\rm Mpc}$}
\put(-180,20){\rotatebox[]{0}{\text{BOSS DR12}}}
\end{subfigure}%
\vspace{-.55cm}
\\
\hspace{1cm}
\begin{subfigure}{.5\textwidth}
   \includegraphics[width=8.3cm]{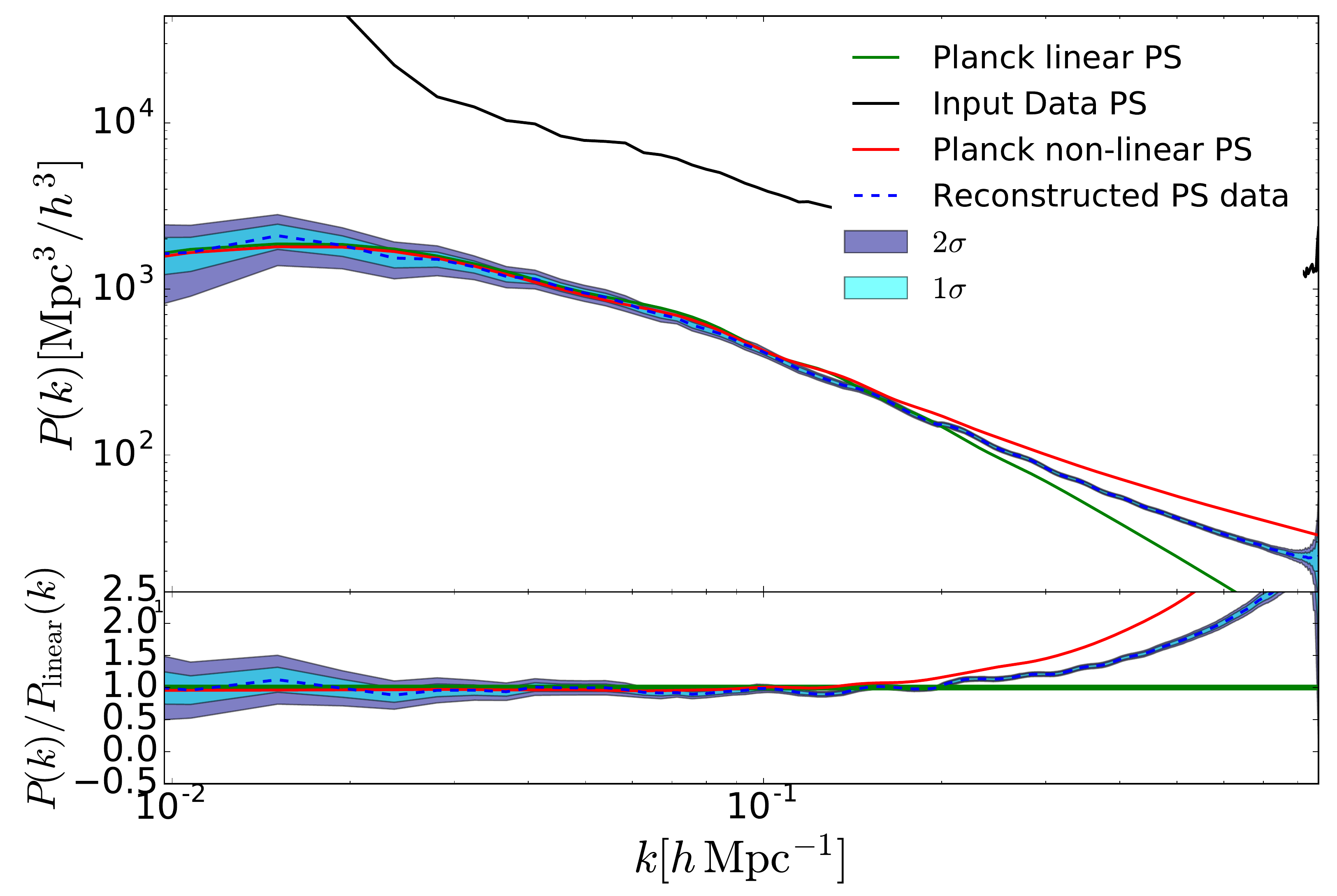}
\put(-200,80){${\rm L}=1250\,h^{-1}\,{\rm Mpc}$}
\put(-200,70){$d_{\rm L}=6.25\,h^{-1}\,{\rm Mpc}$}
\put(-200,60){\rotatebox[]{0}{\text{BOSS DR12}}}
\vspace{-0.5cm}
\end{subfigure}%
\begin{subfigure}{.5\textwidth}
   \includegraphics[width=7.4cm]{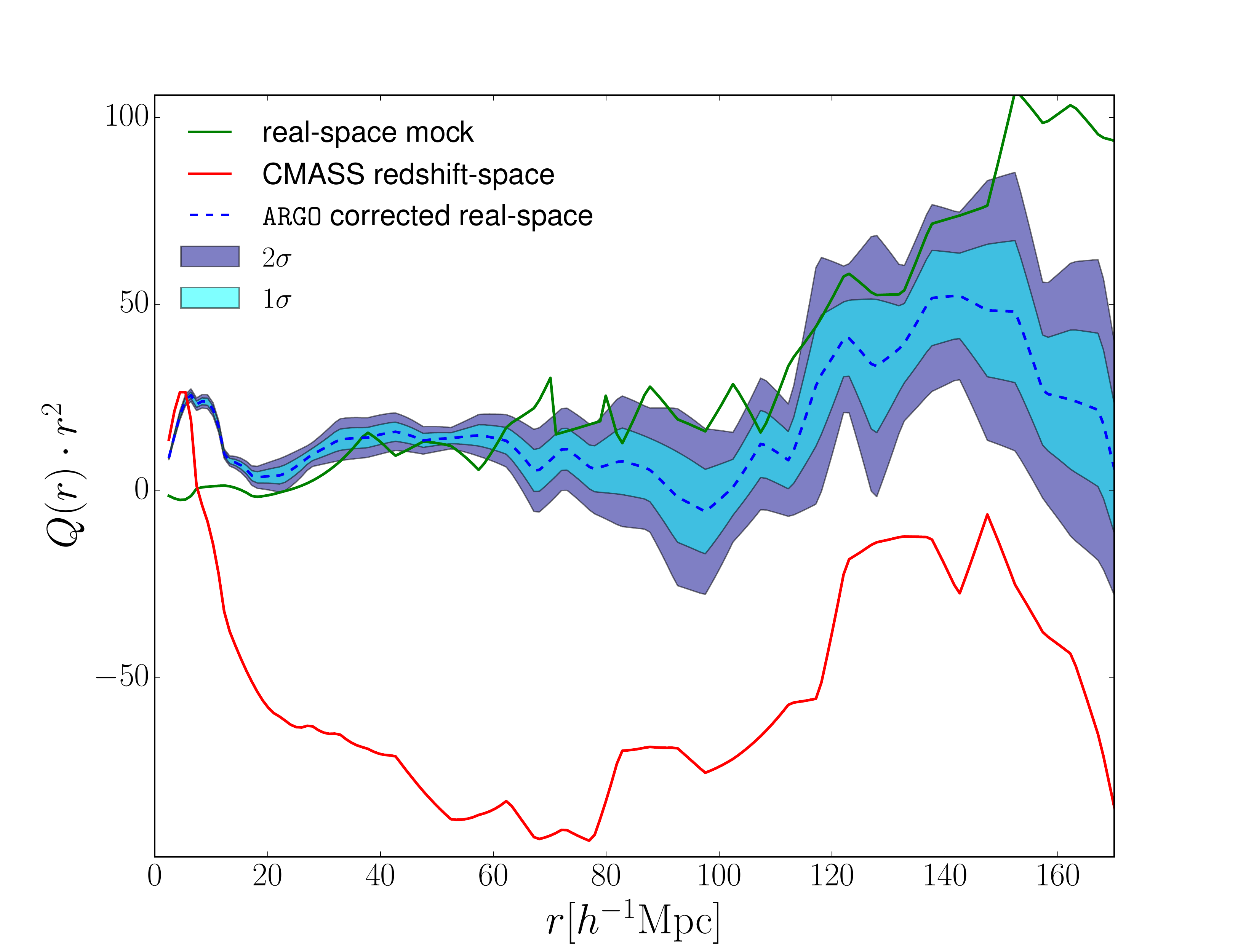}
\put(-180,40){${\rm L}=1250\,h^{-1}\,{\rm Mpc}$}
\put(-180,30){$d_{\rm L}=6.25\,h^{-1}\,{\rm Mpc}$}
\put(-180,20){\rotatebox[]{0}{\text{BOSS DR12}}}
\end{subfigure}%
 \end{tabular}
 \caption{{\bf Left panels:} power spectra of the reconstructed density fields $\delta(z_{\rm ref})$ on a mesh and {\bf right panels:} quadrupoles of the galaxy distribution $\{\mbi s^{\rm obs}\}$ and $\{\mbi r\}$  based on {\bf upper  panels}: a light-cone mock (including survey geometry) with $d_{\rm L}=9.76\,h^{-1}\,{\rm Mpc}$, {\bf middle  panels}: the BOSS DR12 data with $d_{\rm L}=9.76\,h^{-1}\,{\rm Mpc}$, and {\bf lower panels}: the BOSS DR12 on with $d_{\rm L}=6.25\,h^{-1}\,{\rm Mpc}$.   Power spectra show  the mean (dashed blue line) over 6000 samples with 1 and 2 $\sigma$ contours (light and dark blue shaded areas, respectively), as compared to the raw galaxy power spectrum (black solid line), the nonlinear  (red solid line), and the linear power spectrum (green solid line) assuming the fiducial cosmology. Quadrupole correlation functions show  the mean (dashed blue line) over 6000 samples (10 spaced samples in intervals from 500 iterations covering 4000 Gibbs-iterations for quadrupoles to reduce computations) with 1 and 2 $\sigma$ contours (light and dark blue shaded areas, respectively), as compared to the raw galaxy power spectrum (black solid line), and the corresponding computations for the catalogues in real (green line for mocks only) and redshift space (red line). }
\label{fig:pk}
\end{figure*}

\begin{figure*}
\begin{tabular}{cc}
\hspace{1cm}
\begin{subfigure}{.5\textwidth}
   \includegraphics[width=8.cm]{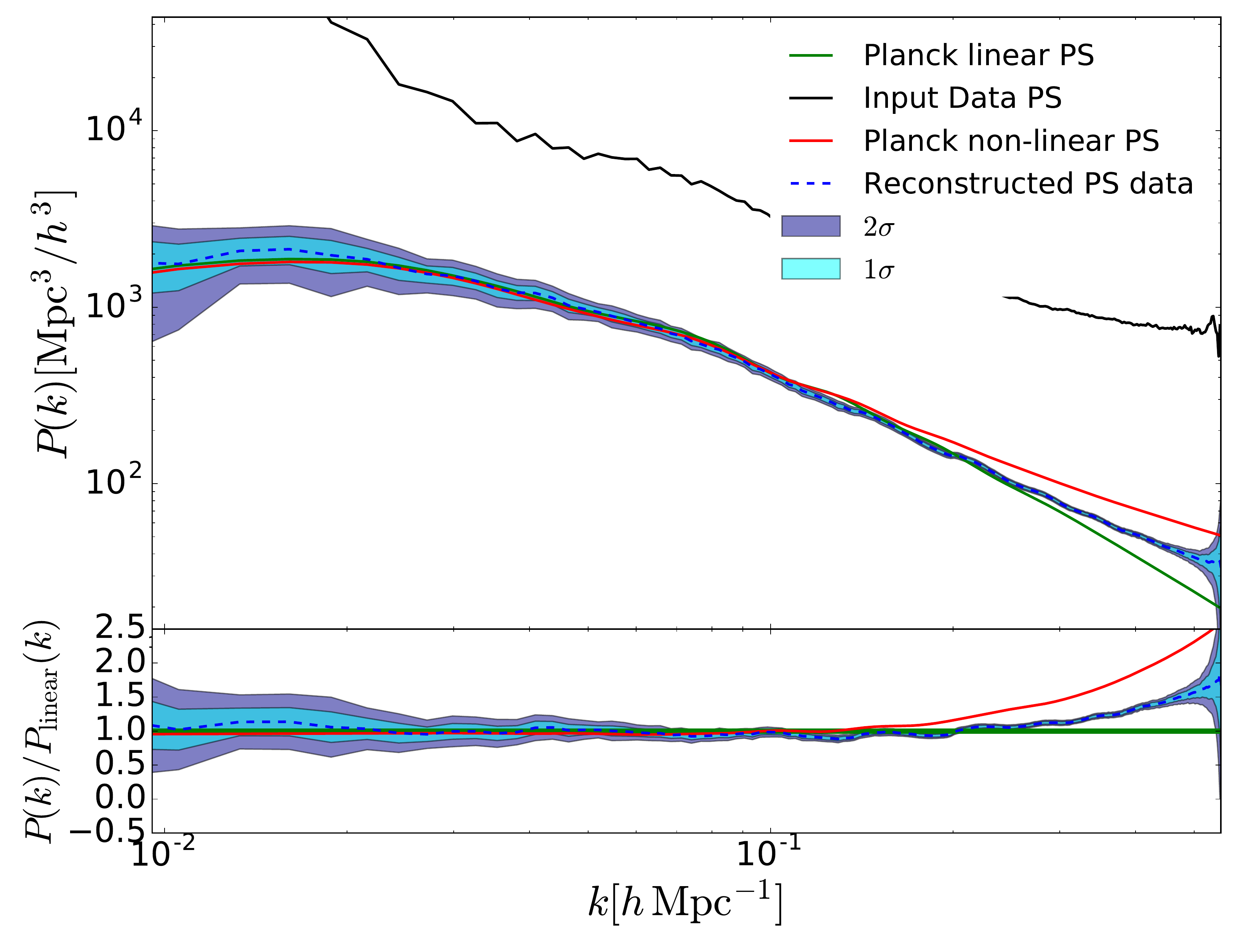}
\put(-195,100){${\rm L}=1250\,h^{-1}\,{\rm Mpc}$}
\put(-195,90){$d_{\rm L}=9.76\,h^{-1}\,{\rm Mpc}$}
\put(-195,80){\rotatebox[]{0}{\text{BOSS DR12}}}
\put(-195,70){with velocity dispersion}
\vspace{-0.5cm}
\end{subfigure}%
\begin{subfigure}{.5\textwidth}
\vspace{-.6cm}
   \includegraphics[width=7.7cm]{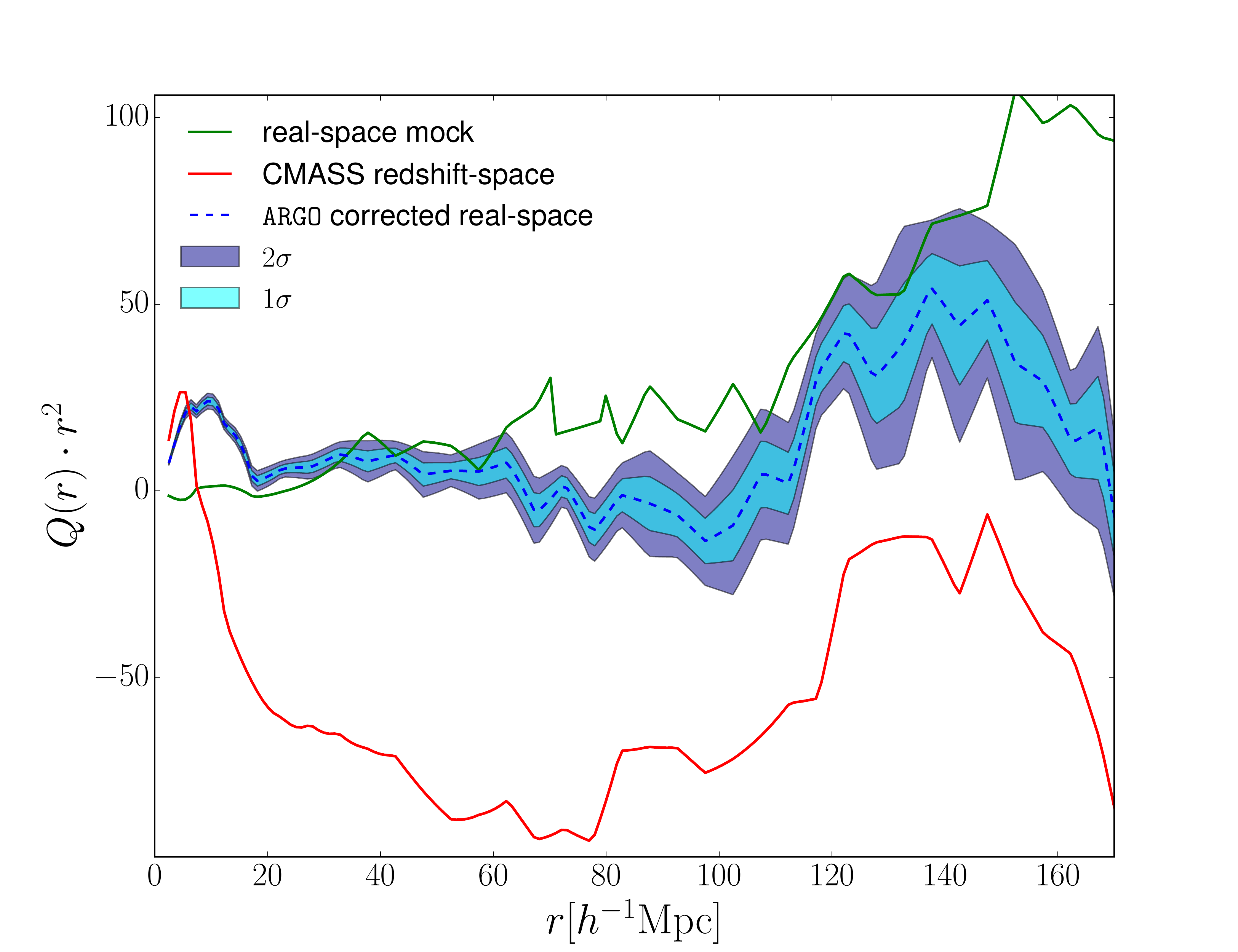}
\put(-185,55){${\rm L}=1250\,h^{-1}\,{\rm Mpc}$}
\put(-185,45){$d_{\rm L}=9.76\,h^{-1}\,{\rm Mpc}$}
\put(-185,35){\rotatebox[]{0}{\text{BOSS DR12}}}
\put(-185,25){with velocity dispersion}
\end{subfigure}%
\end{tabular}
\vspace{0.5cm}
 \caption{Same as Fig.~\ref{fig:pk} but including velocity dispersion. }
\label{fig:pkadd}
\end{figure*}

\begin{figure*}
\begin{tabular}{cc}
\hspace{-.13cm}
\includegraphics[width=6.6cm]{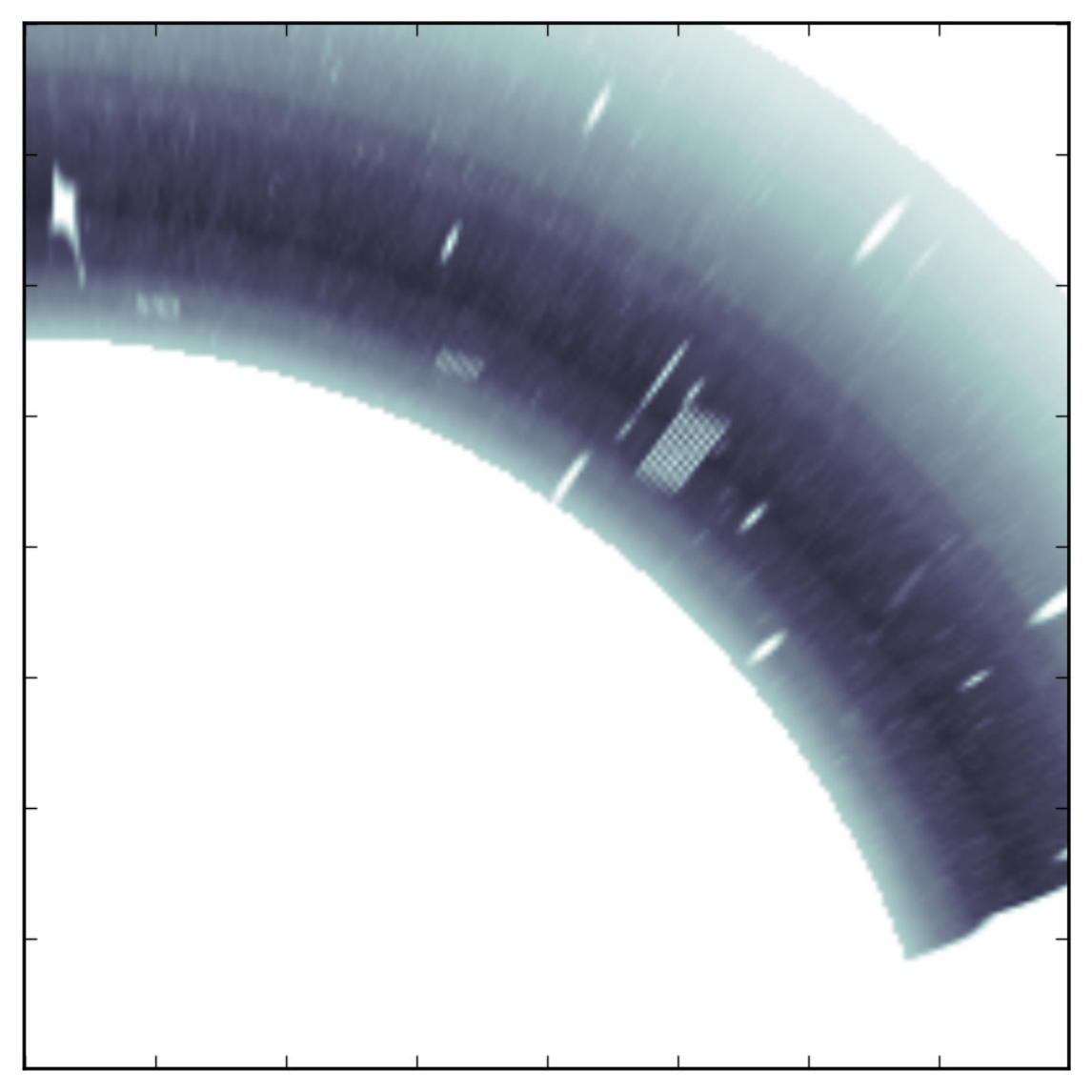}  
\hspace{-0.2cm}
\includegraphics[width=6.6cm]{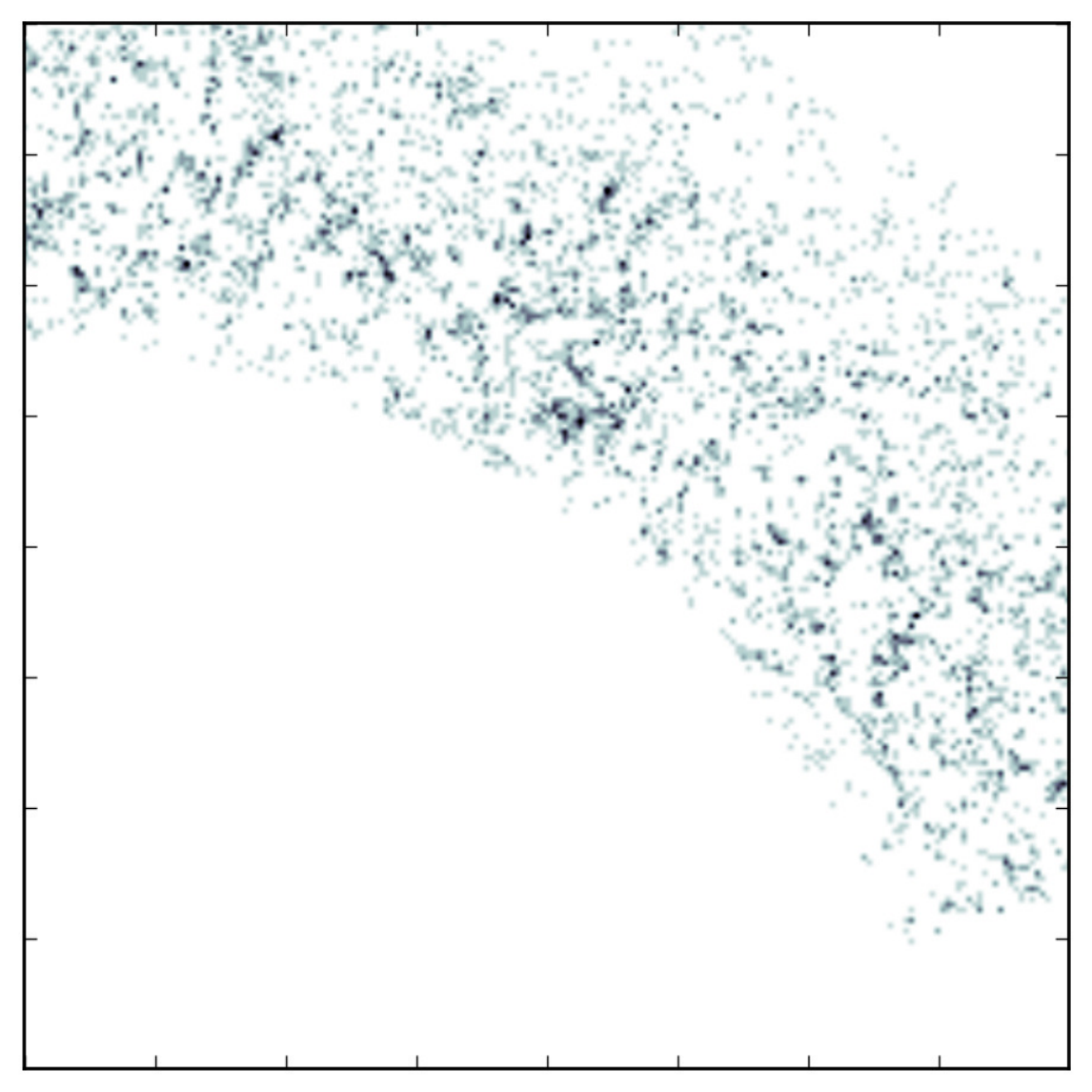} 
\includegraphics[width=1.03cm]{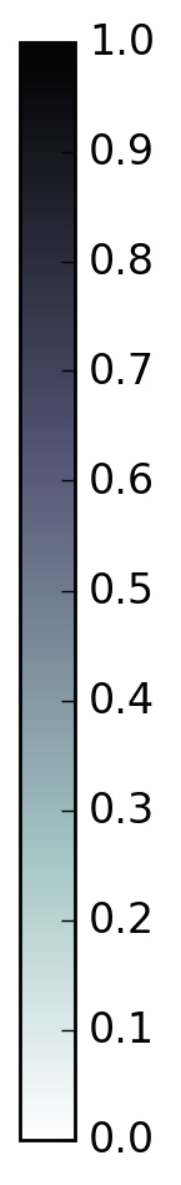}
\vspace{-0.1cm}
\\
\hspace{.1cm}
\includegraphics[width=6.6cm]{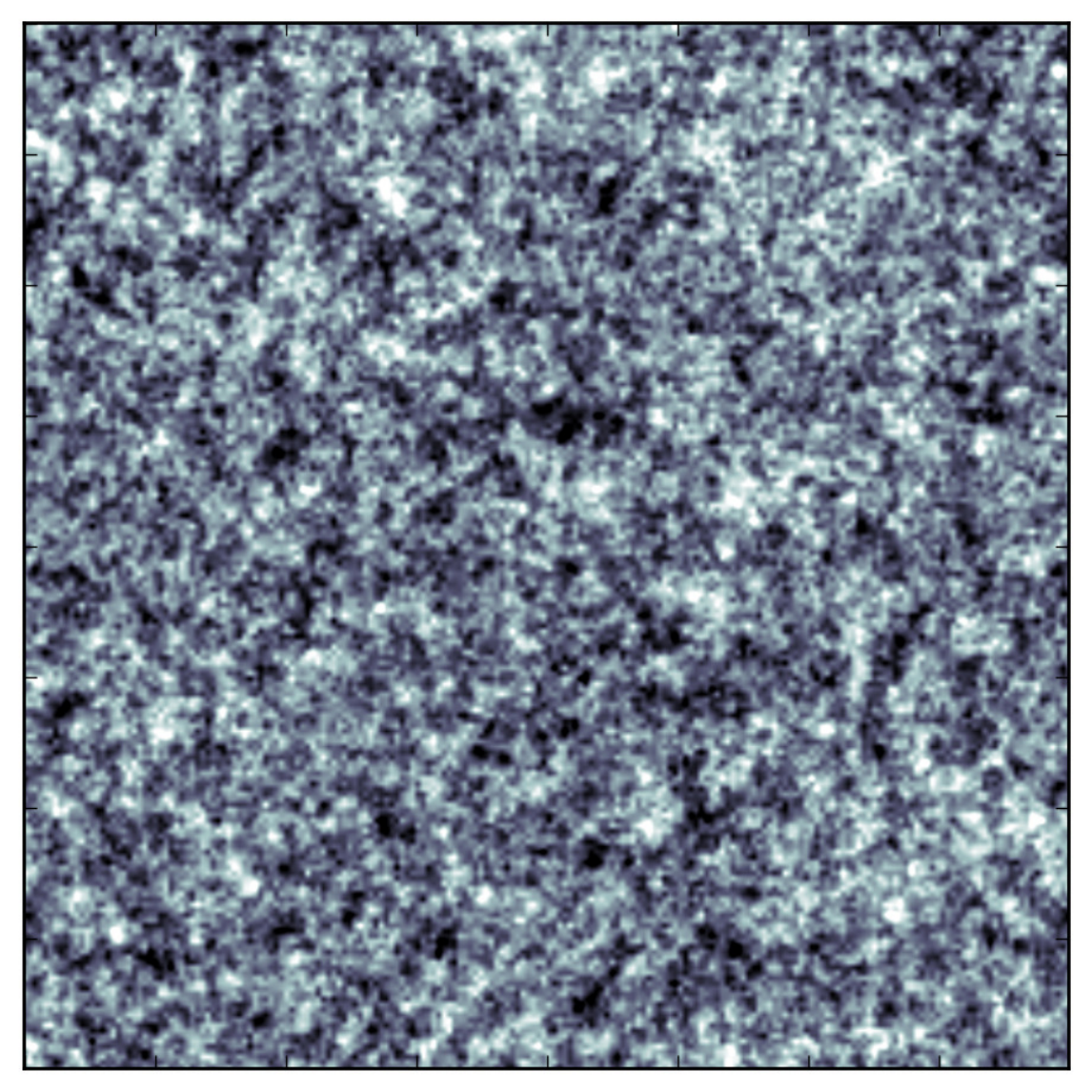} 
\hspace{-0.2cm}
\includegraphics[width=6.6cm]{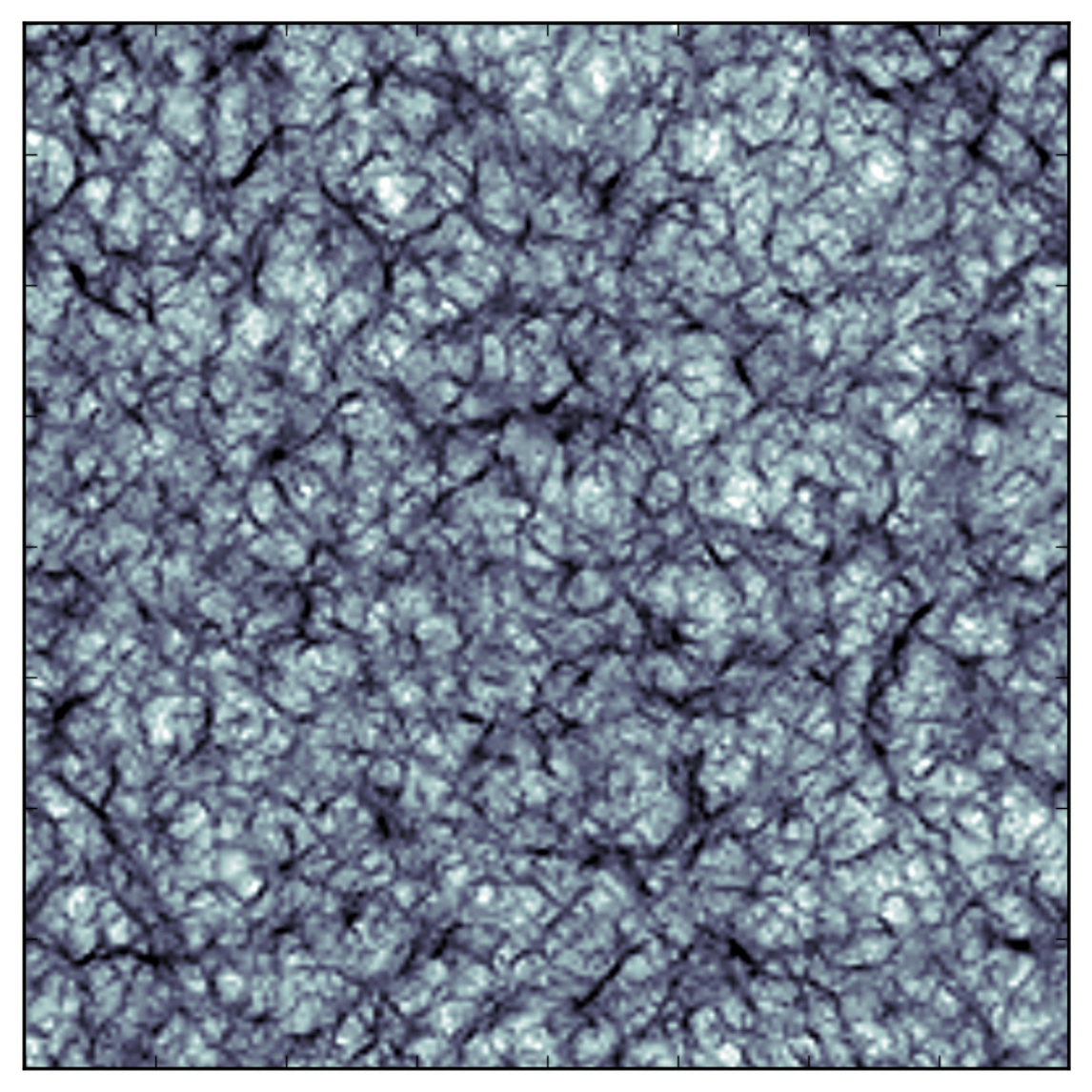} 
\includegraphics[width=1.233cm]{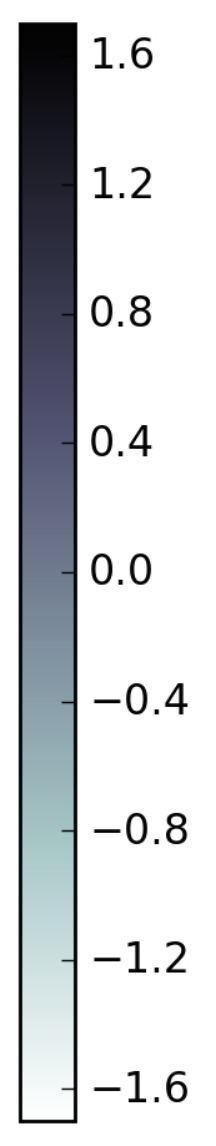}
\vspace{-0.1cm}
\\
\hspace{.0cm}
\includegraphics[width=6.6cm]{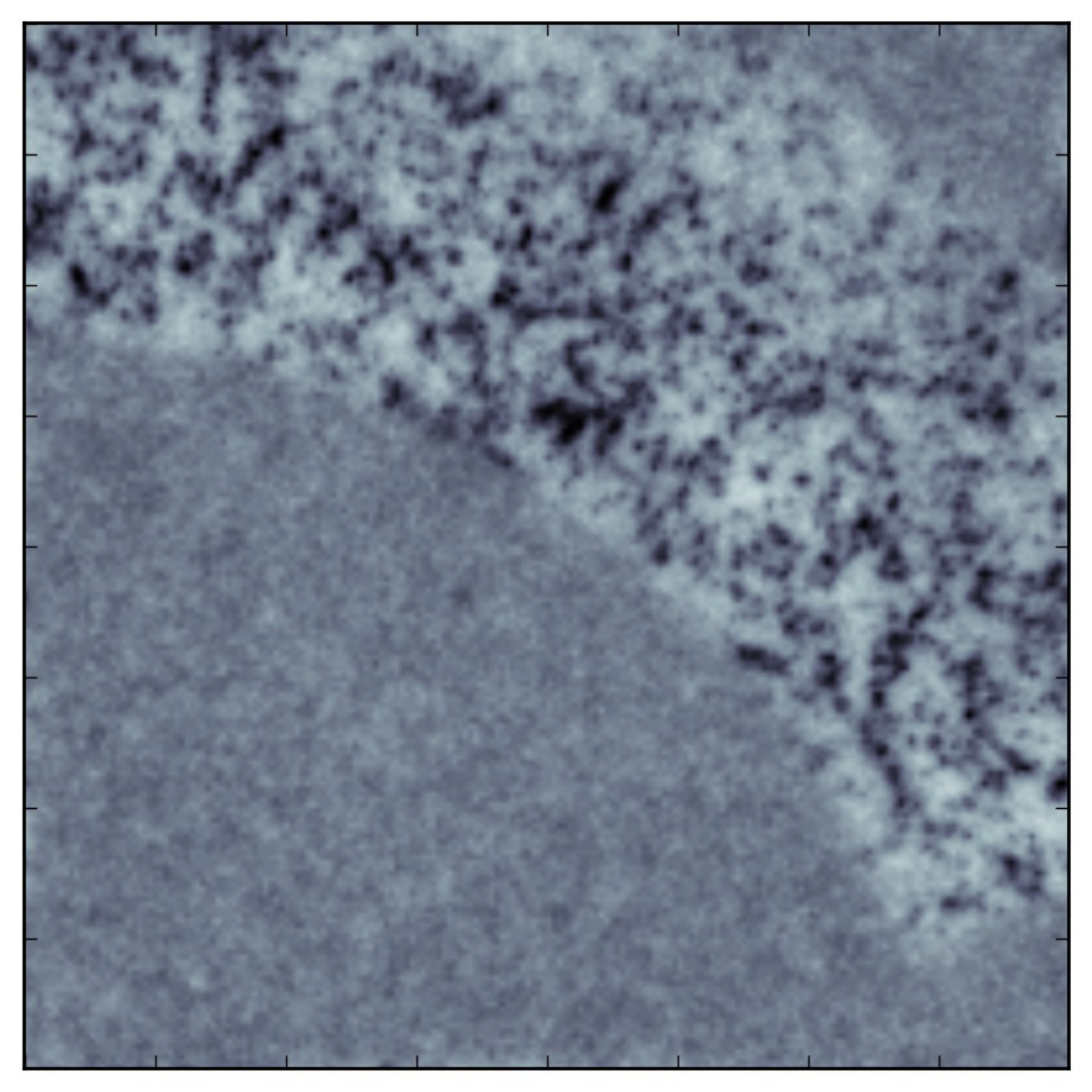} 
\hspace{-0.2cm}
\includegraphics[width=6.6cm]{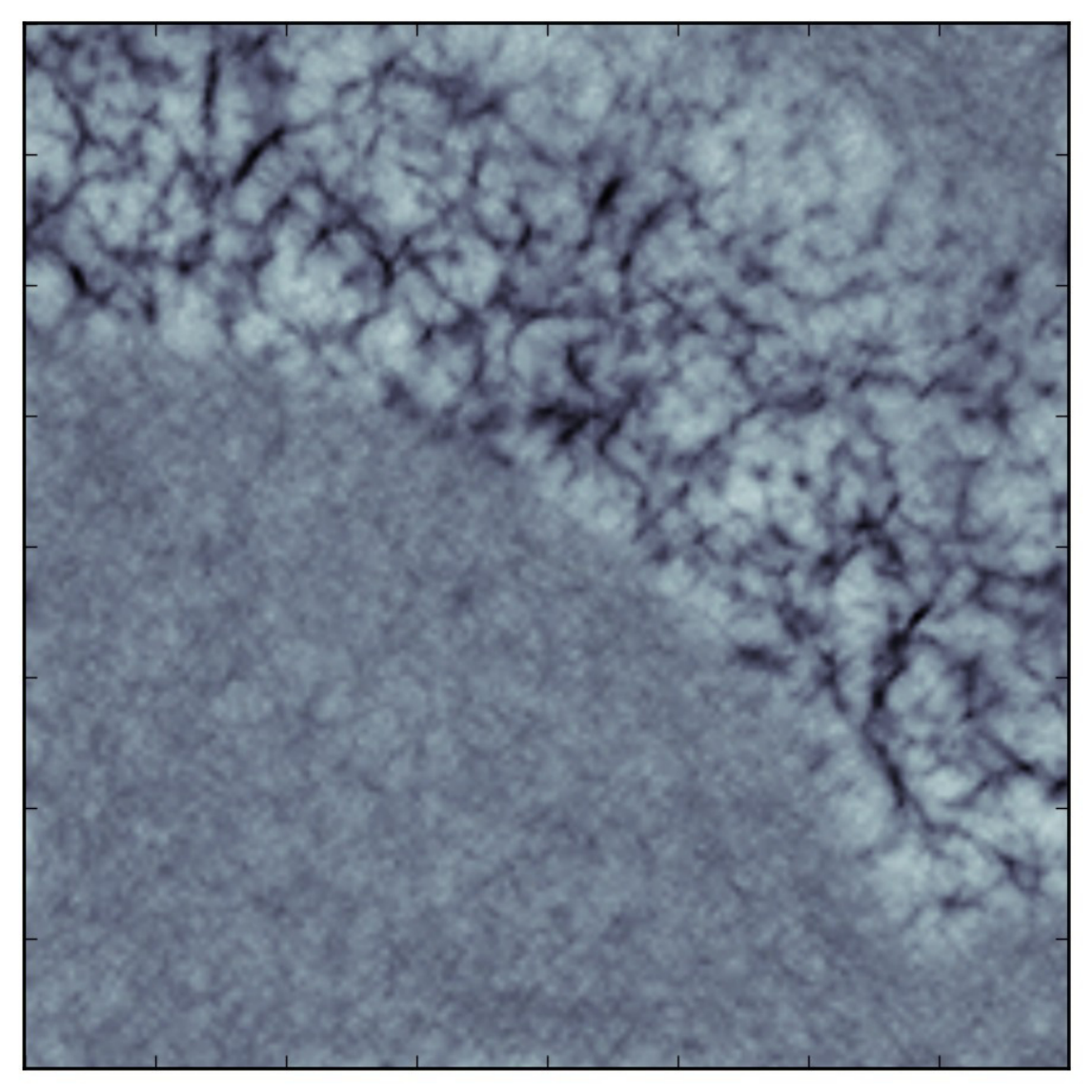} 
\includegraphics[width=1.23cm]{CF_slice_delta_cb}
\end{tabular}
\caption{\label{fig:slice2} Based on a light-cone mock catalogue in redshift space with $6.25\,h^{-1}$ Mpc resolution: slices of thickness $\sim$30 $h^{-1}$ Mpc in the $x-z$ plane of  the 3D cubical mesh of side 1250  $h^{-1}$ Mpc and $200^3$ cells for the following quantities: {\bf upper left panel:} the 3D completeness or window function multiplied with a factor of 0.8 for visualisation purposes, {\bf upper right panel:} the number counts per cell in real space, {\bf middle left panel:} one reconstructed linear logarithmic density sample of the lognormal-Poisson field, {\bf  middle right panel:} same as middle left panel for the Zeldovich transformed density,  {\bf  lower left panel:} the mean over the linear logarithmic density sample over 6000 reconstructions of the lognormal-Poisson field, and  {\bf  lower right panel:} same as lower left panel for the Zeldovich transformed density. }
\end{figure*}

\begin{figure} 
   \includegraphics[width=8cm]{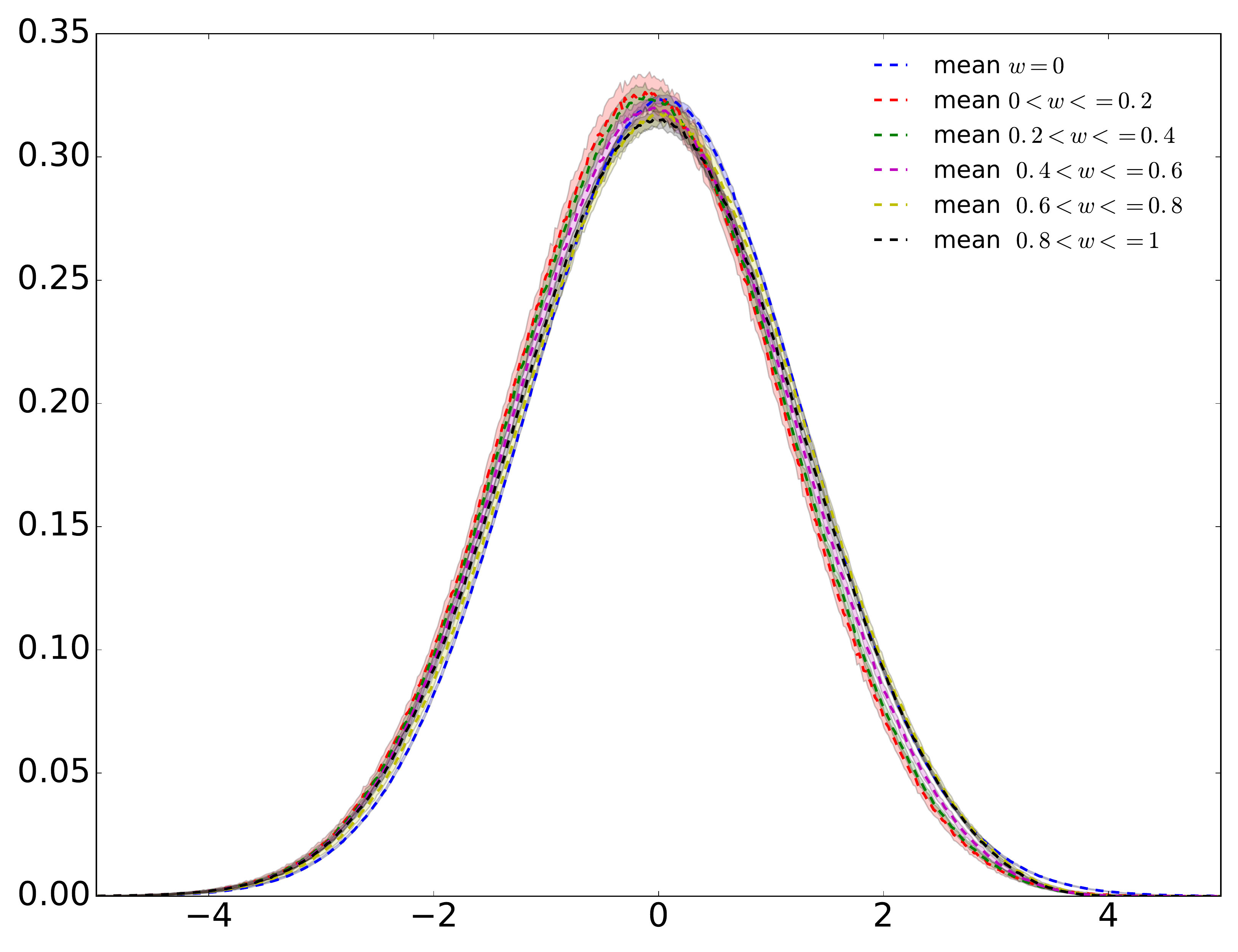}
\put(-235,75){\rotatebox[]{90}{PDF}}
\put(-112,-8){\rotatebox[]{0}{{$\delta_{\rm L}$}}}
 \caption{PDF of the matter statistics for different completeness values from 6000 reconstructions on a mesh of 200$^3$ and resolution $d_{\rm L}=6.25\,h^{-1}$ Mpc for the linear component reconstructed with the lognormal-Poisson model. The corresponding skewness range between $-10^{-4}$ and $-0.09$ with means being always smaller than $|\langle\delta_{\rm L}\rangle|<0.13$. The skewness is thus reduced by two orders of magnitude, as compared to a skewness of $\sim$7 corresponding to the galaxy overdensity on a mesh with a cell resolution of 10 $h^{-1}$ Mpc.}
 \label{fig:pdf}
\end{figure}

\begin{figure*}
\begin{tabular}{cc}
\includegraphics[width=6.6cm]{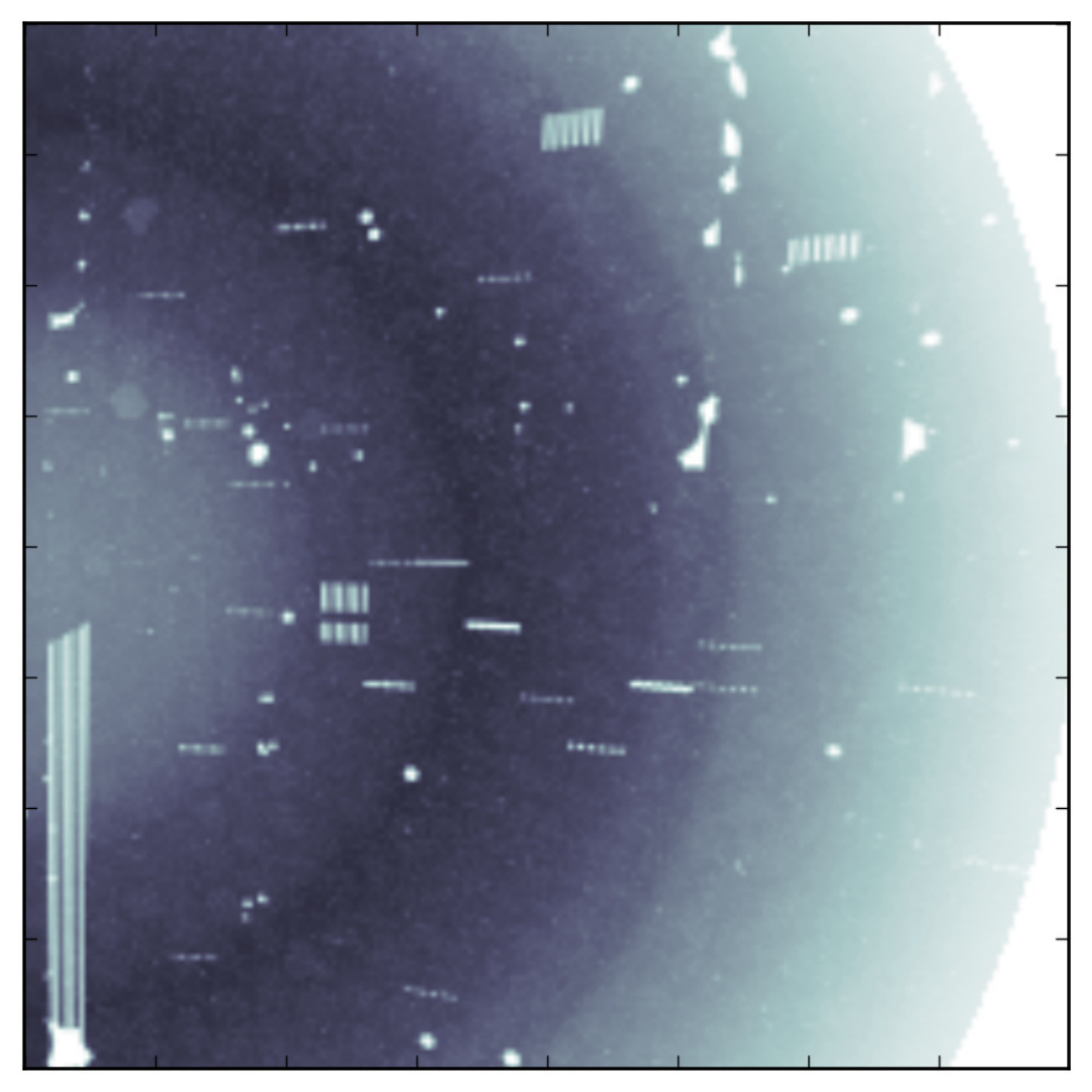}
\hspace{-0.2cm}
\includegraphics[width=6.6cm]{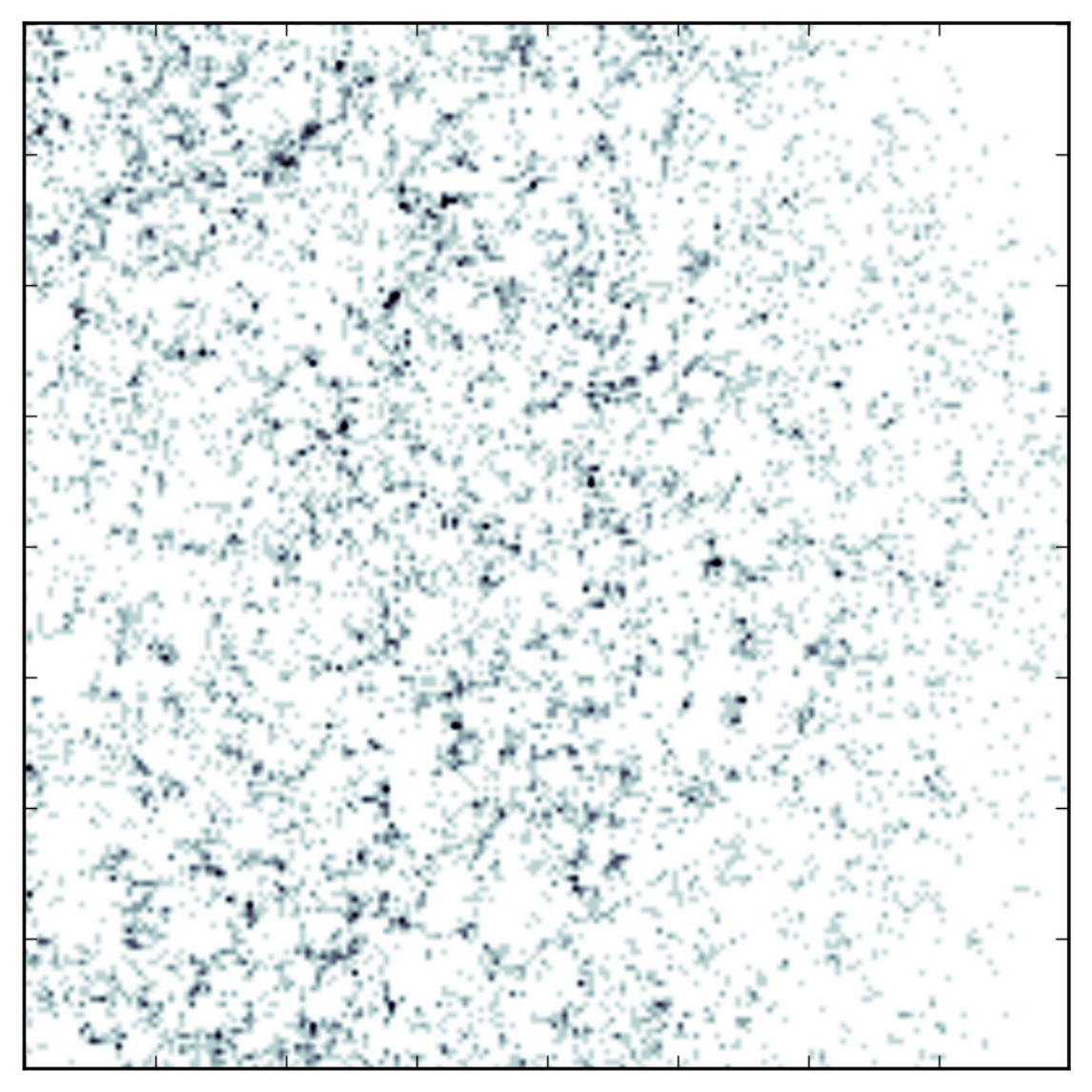}
\includegraphics[width=1.03cm]{CF_slice_comp_cb}
\vspace{-0.15cm}
\\  
\hspace{0.1cm}
\includegraphics[width=6.6cm]{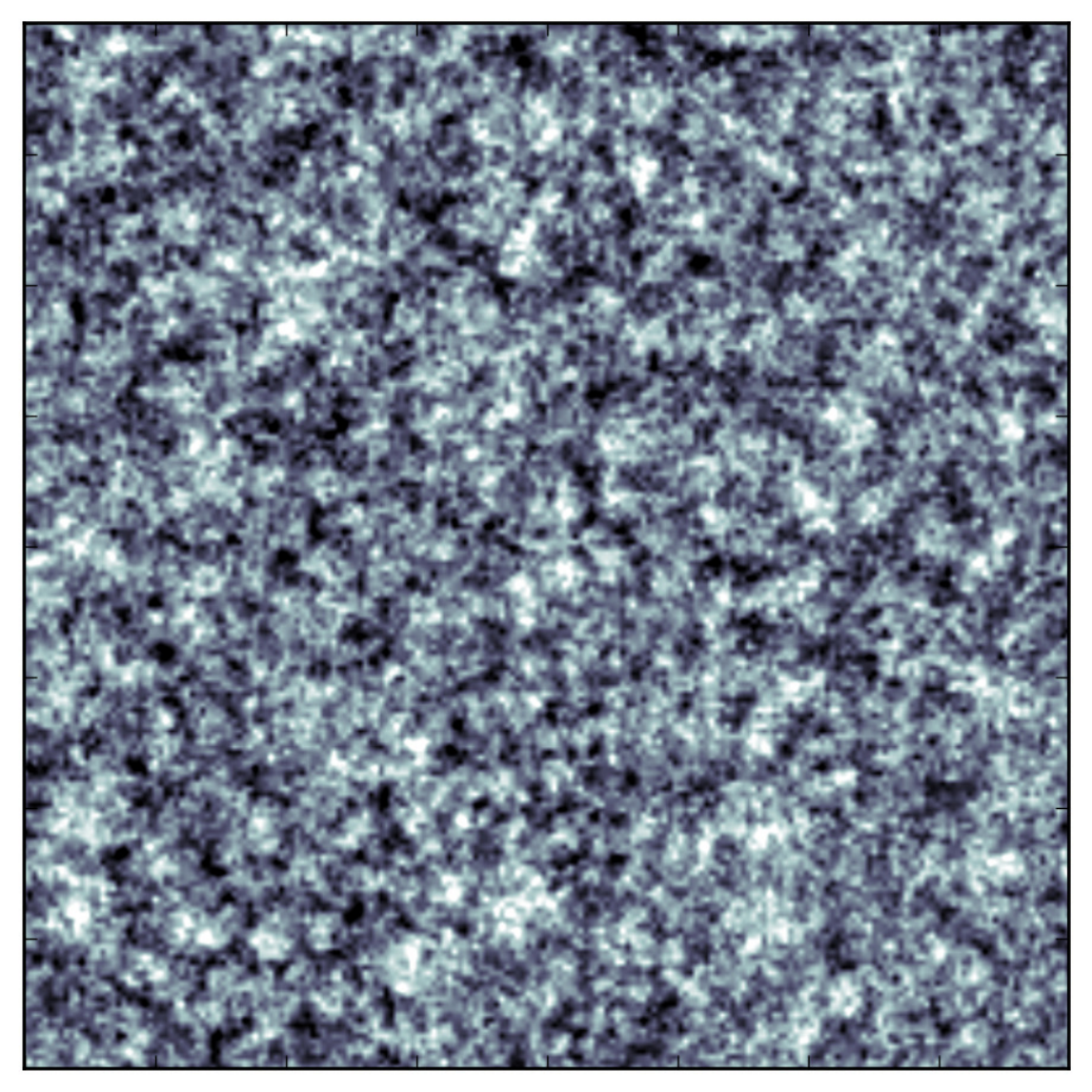}
%\hspace{-0.08cm}
\hspace{-0.18cm}
\includegraphics[width=6.6cm]{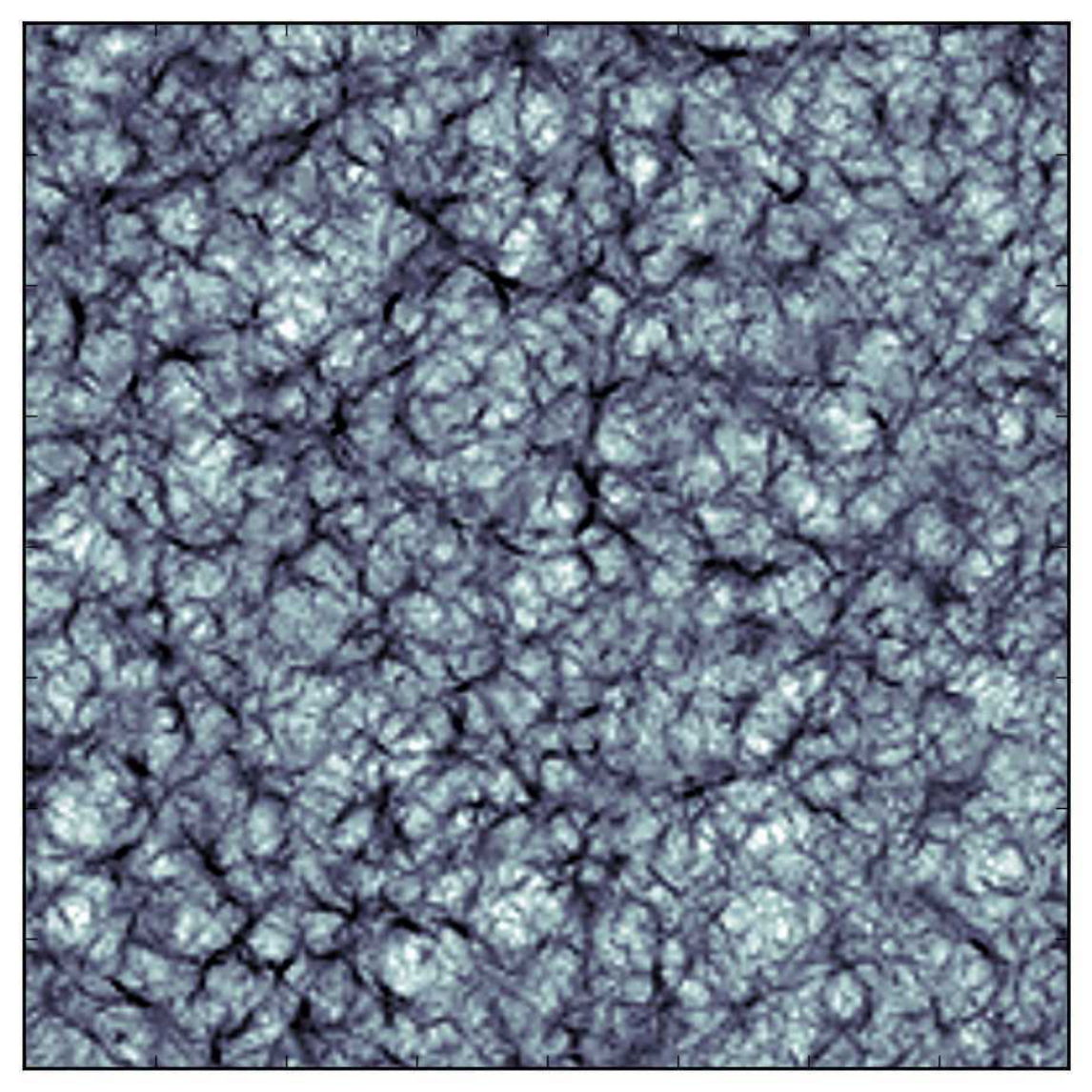}
\includegraphics[width=1.23cm]{CF_slice_delta_cb}
\vspace{-0.15cm}
\\
\hspace{0.01cm}
\hspace{0.01cm}
\includegraphics[width=6.6cm]{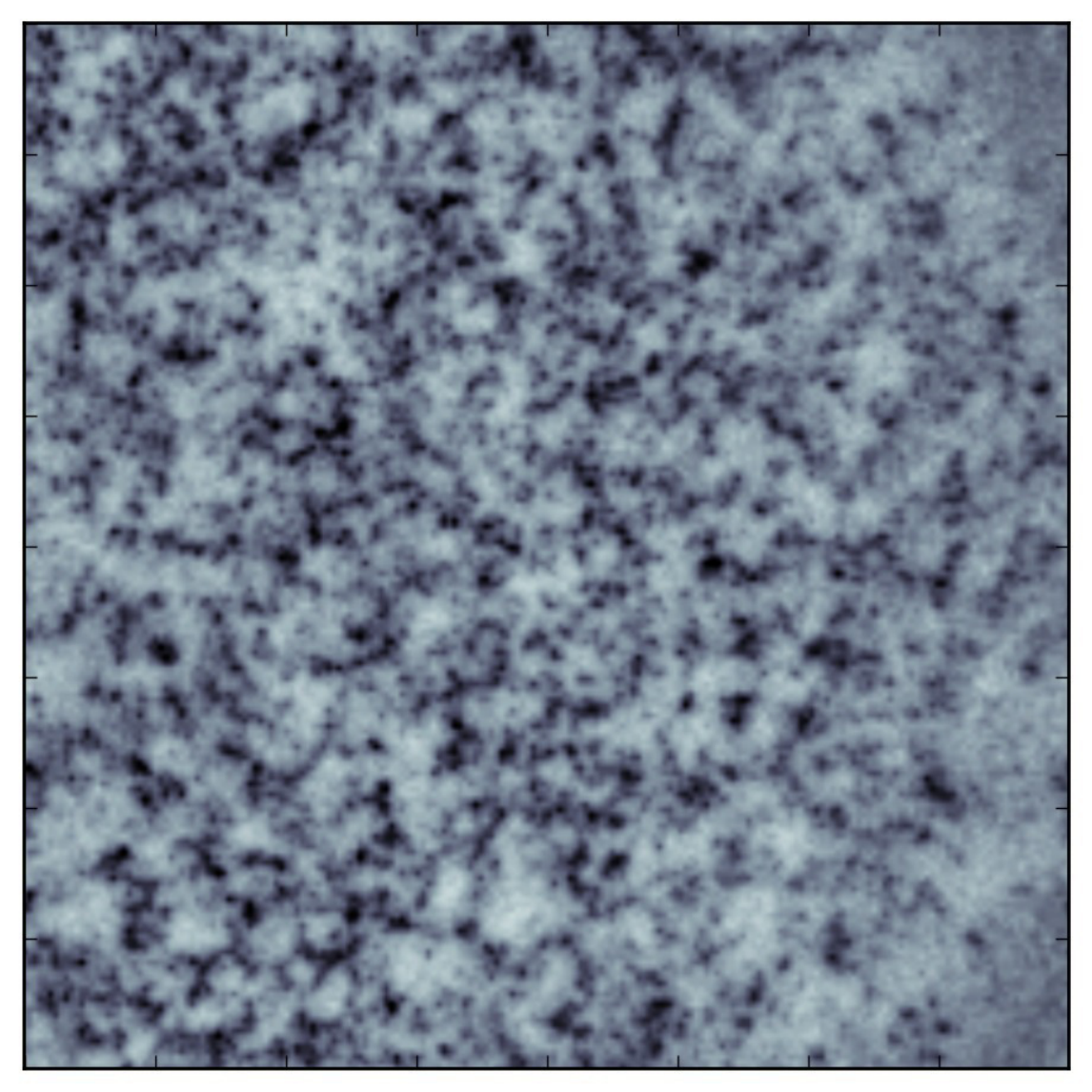} 
\hspace{-0.2cm}
\includegraphics[width=6.6cm]{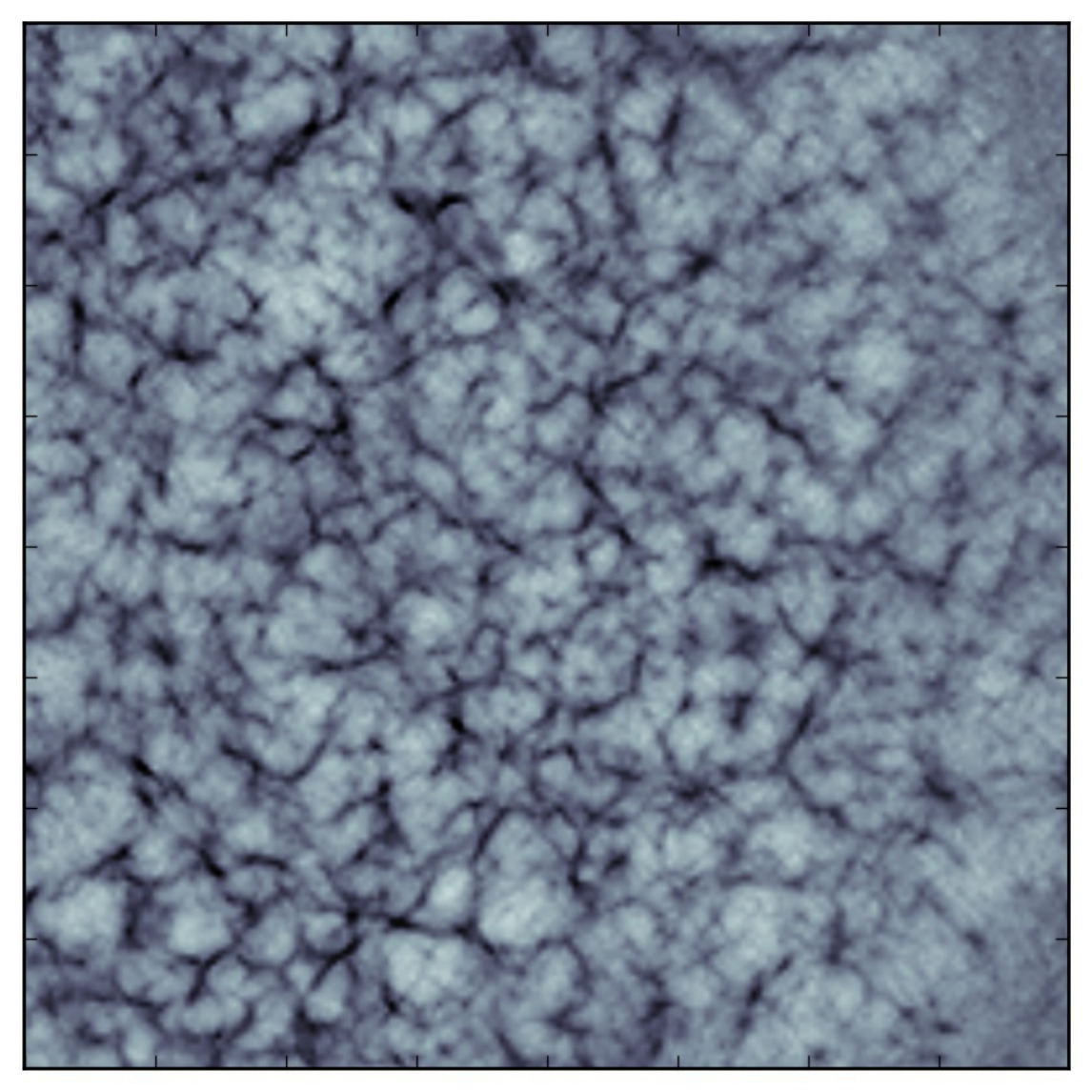} 
\hspace{-0.18cm}
\hspace{0.002cm}
\includegraphics[width=1.23cm]{CF_slice_delta_cb}
\vspace{-0.15cm}
\end{tabular}
\caption{\label{fig:slice3} Same as Fig.~\ref{fig:slice2}, but for the $x-y$ plane.}
\end{figure*}

\begin{figure*}
\begin{tabular}{cc}
\hspace{-1.06cm}
\includegraphics[width=7.72cm]{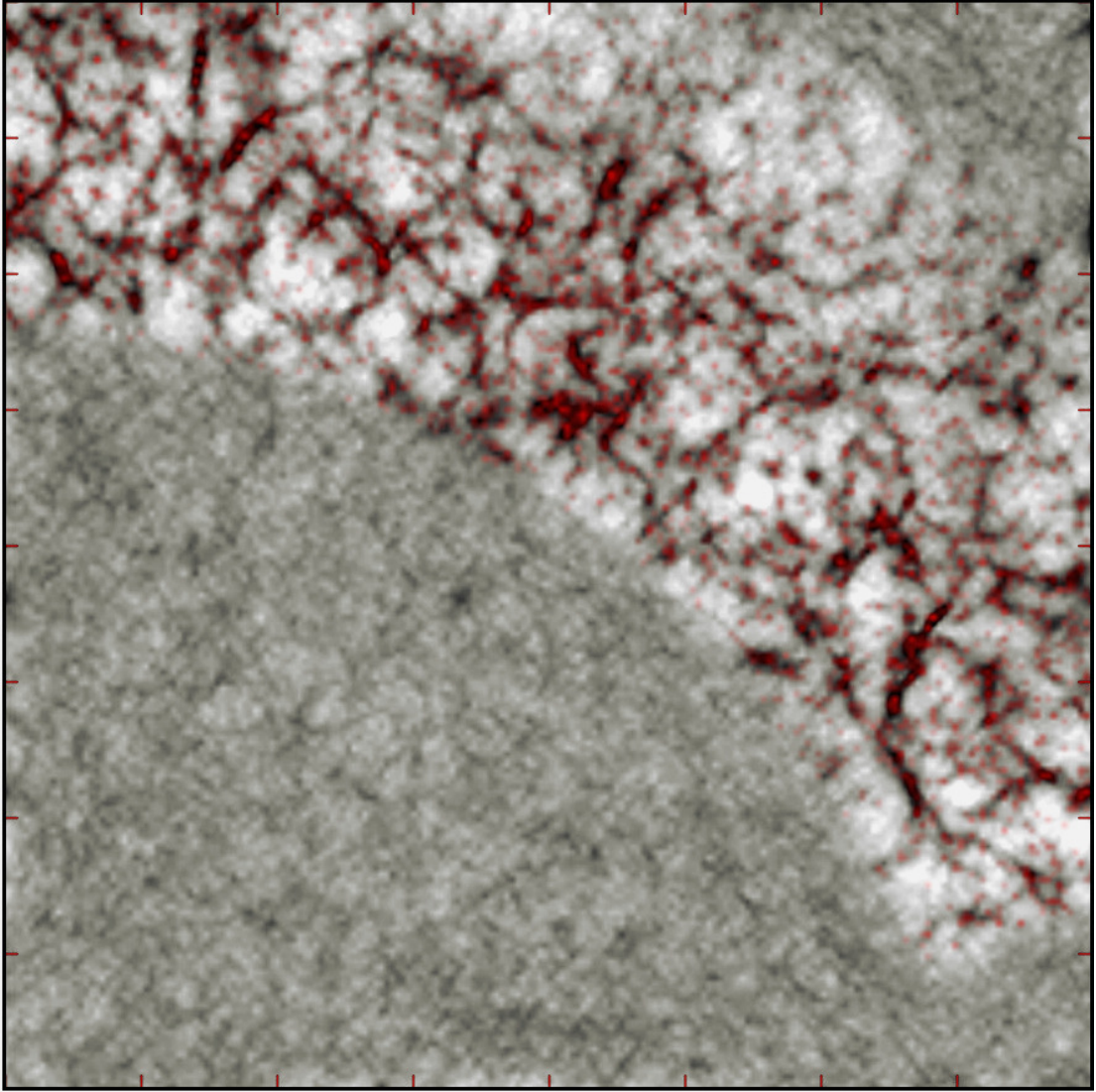} \hspace{0.35cm}
\includegraphics[width=7.7cm]{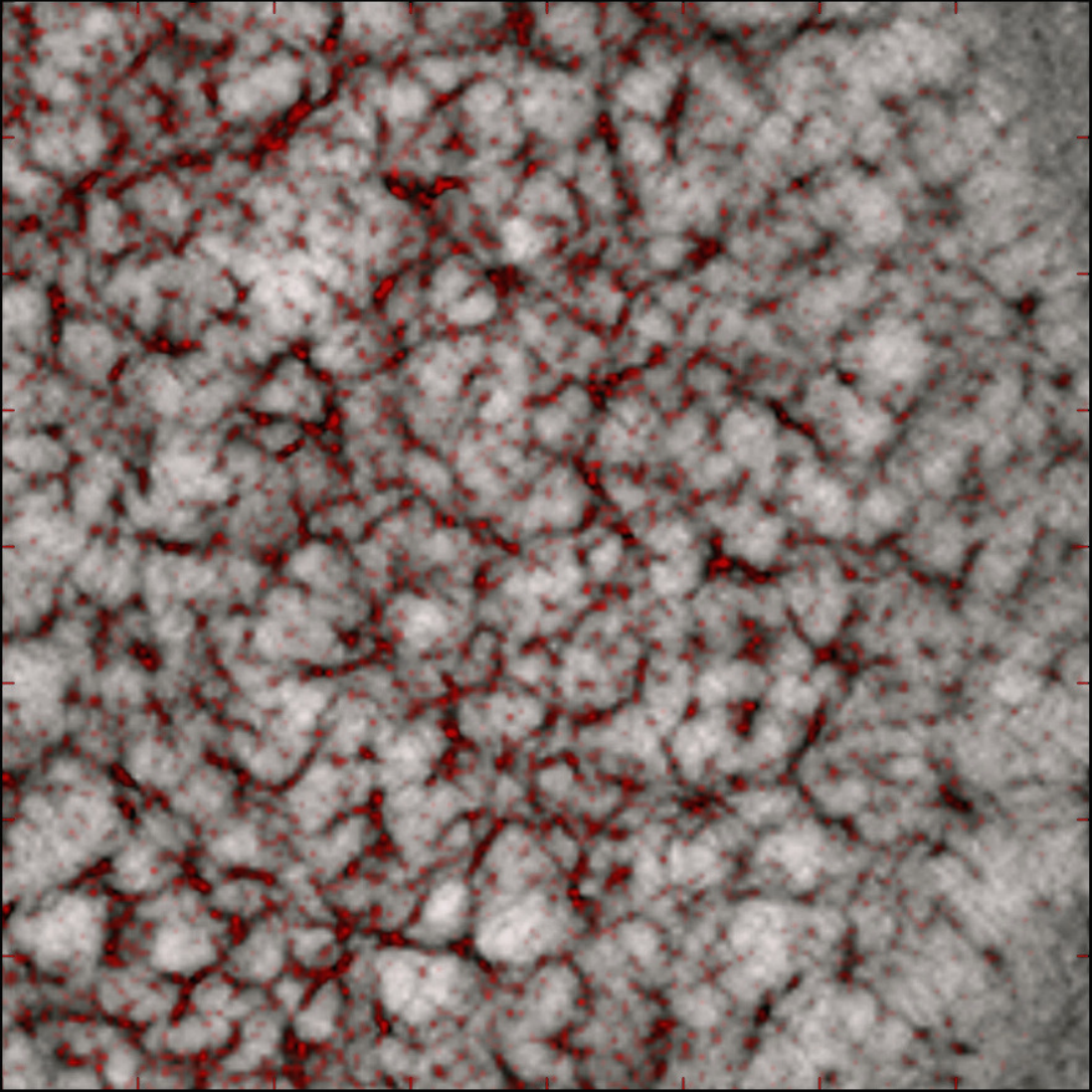} 
\end{tabular}
\caption{\label{fig:withgal} Slices of the ensemble averaged Zeldovich transformed density field shown in Figs.~\ref{fig:slice2} (left) and \ref{fig:slice3} (right) with another colour bar for visualisation purposes and the corresponding real space galaxy number count per cell overplotted in red.}
\end{figure*}

\begin{figure*}
\begin{tabular}{cc}
\includegraphics[width=9.9cm]{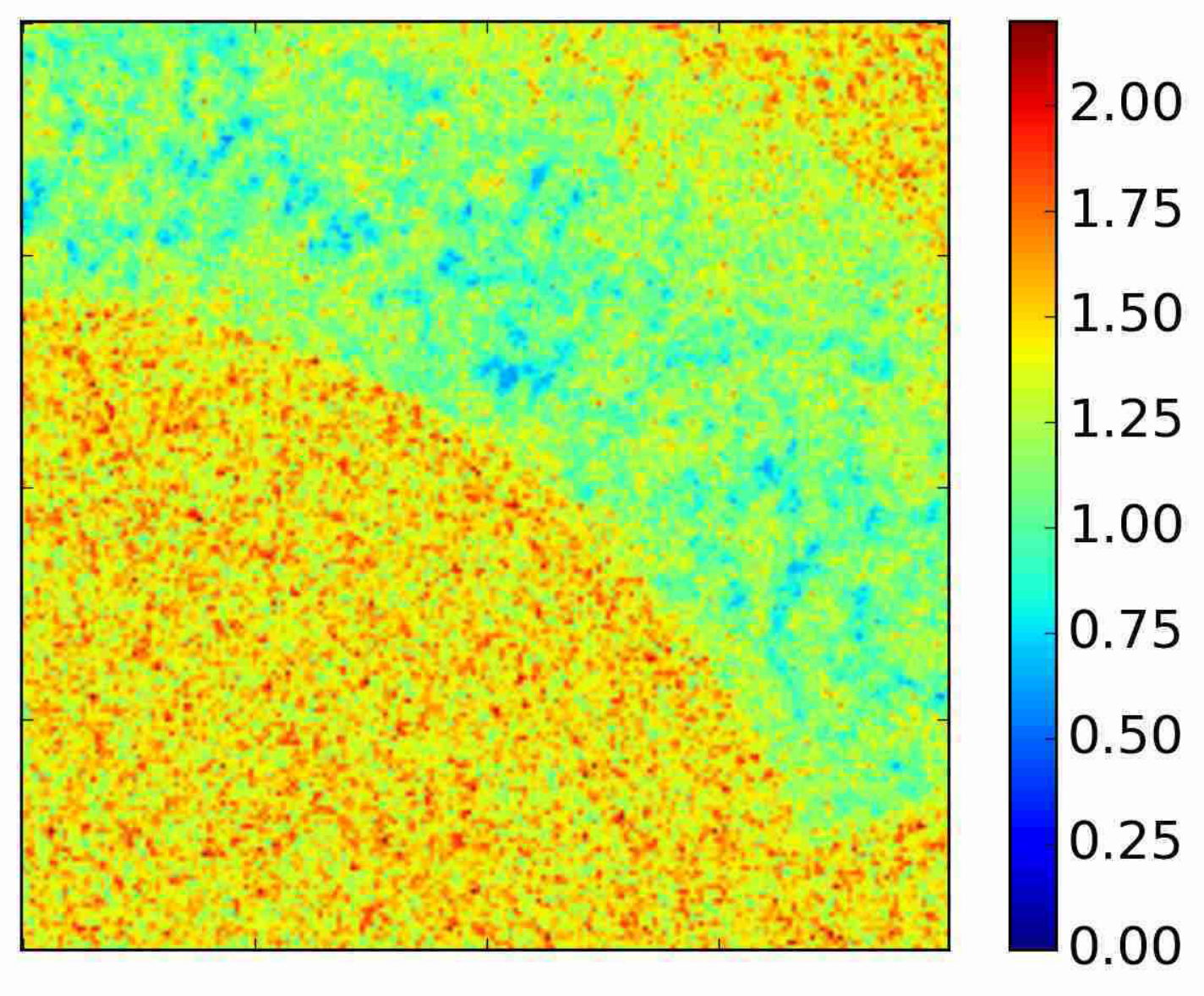}  
\hspace{-1.9cm}
\includegraphics[width=9.9cm]{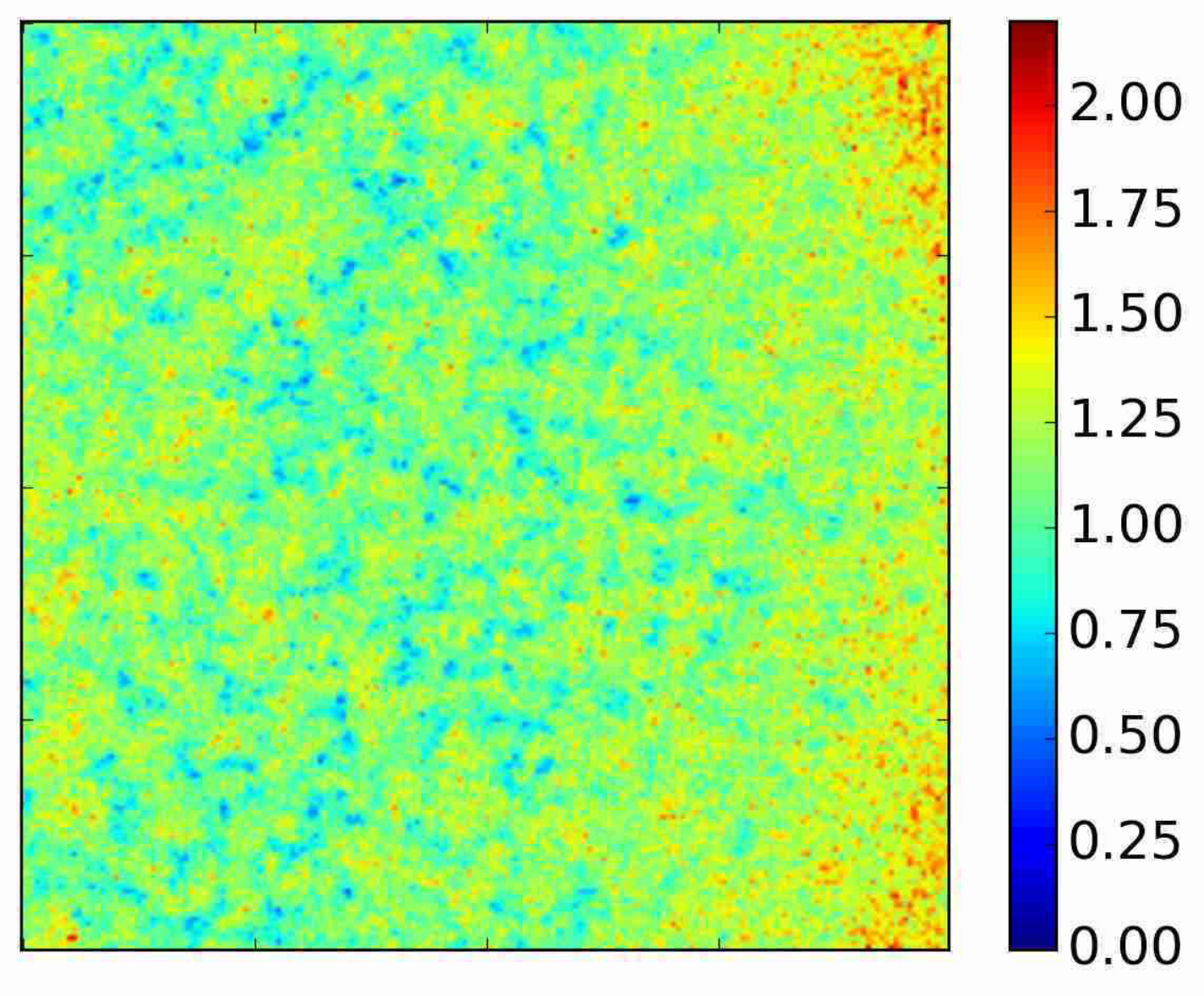}  
\end{tabular}
\caption{\label{fig:signal2noisecw} Slices of the  variance  corresponding to the  {\bf (left panel:)} $x-z$ plane, and {\bf (right panel:)}  $x-y$ plane shown in Figs.~\ref{fig:slice2} and \ref{fig:slice3}, respectively.  }
\end{figure*}

In this paper, we use $N$-body based mock galaxy catalogues constructed to match the clustering bias, survey mask, selection functions, and number densities of the BOSS DR12 CMASS galaxies. This permits us to test our method, as both real space and redshift space catalogues are known.
Finally, we apply our analysis method to the BOSS DR12 CMASS data. Let us describe the input galaxy catalogues below.

\subsection{BOSS DR12 galaxy catalogue}

This work uses data from the Data Release DR12 \citep[][]{AAA15} of the Baryon Oscillation Spectroscopic Survey (BOSS) \citep[][]{EWA11}. The BOSS survey uses the SDSS 2.5 meter telescope at Apache Point Observatory \citep[][]{Gunn-06} and the spectra are obtained using the double-armed BOSS spectrograph \citep[][]{2013AJ....146...32S}. The data are then reduced using the algorithms described in \citep[][]{BSA12}.  The target selection of the CMASS and LOWZ samples, together with the algorithms used to create large-scale structure catalogues (the \textsc{mksample} code), are presented in  \citet[][]{2016MNRAS.455.1553R}.

We restrict this analysis to the CMASS sample of luminous red galaxies (LRGs), which is a complete sample, nearly constant in mass and volume limited between the redshifts $0.43\le z \le 0.7$ (see \citet{Andersonboss} for details of the targeting strategy).

\subsection{Mock galaxy catalogues in real and redshift space}
\label{sec:mocks}

The mock galaxy catalogues used in this study were presented in \citet[][]{sergio15}, and are extracted from one of the \textsc{BigMultiDark} simulations\footnote{\url{http://www.multidark.org/MultiDark/}} \citep[][]{BigMD}, which was performed using \textsc{gadget-2} \citep{Springel2005} 
with $3,840^3$ particles on a volume of $(2.5\,h^{-1}{\rm Gpc}$ $)^3$ assuming $\Lambda$CDM Planck cosmology with \{$\Omega_{\rm M}=0.307115, \Omega_{\rm b}=0.048206,\sigma_8=0.8288,n_s=0.9611$\}, and a Hubble constant ($H_0=100\,h\,\kmsmpc$) given by  $h=0.6777$.  Haloes were defined based on  the Bound Density Maxima (BDM) halo finder \citep{Klypin1997}.

They have been constructed based on the Halo Abundance Matching (HAM) technique to connect haloes to galaxies \citep{Kravtsov04,Neyrinck04,Tasitsiomi04,Vale04,Conroy06,Kim08,Guo10,Wetzel10,Trujillo11,Nuza13}.

At first order HAM assumes a one-to-one correspondence between the 
luminosity or 
stellar and dynamical masses: 
 galaxies with more stars are assigned
to more massive haloes or subhaloes. 
The luminosity in a red-band is sometimes used instead of stellar mass.
 There should be some degree of stochasticity in the relation between stellar and dynamical
masses due to deviations in the merger history, angular momentum,
halo concentration, and even observational errors \citep{Tasitsiomi04,Behroozi10,Trujillo11,Leauthaud11}. 
 Therefore, we include a scatter in that relation necessary to accurately  fit the clustering of the BOSS data \citep[][]{sergio15}.

\section{Results}
\label{sec:results}

In this section we present the results obtained with the \textsc{argo} code including the cosmic evolution treatment described in Sec.~\ref{sec:method} on the mock galaxy catalogues and finally on the data.
Let us first describe the preparation of the data.

\subsection{Preparation of the data}

As explained in Sec.~\ref{sec:method} our method requires the galaxy number counts on a mesh. Therefore we need first to assume a fiducial cosmology (the same as the mock catalogues described in Sec.~\ref{sec:mocks}), and transform angular coordinates (right ascension $\alpha$ and declination $\delta$) and redshifts into comoving Cartesian coordinates $x,y,z$
\ba
 x &=& r \cos{\alpha}\cos{\delta} \nonumber \\
 y &=& r \sin{\alpha}\cos{\delta} \nonumber \\
 z &=& r \sin{\delta}\nonumber \,,
\ea
with the comoving distance given by
\ba
 r &=&\displaystyle \frac{H_0}{c} \int\limits_0^z  \frac{dz'}{\sqrt{\Omega_{\mathrm{M}}(1+z')^3+\Omega_\Lambda}} \,.
\ea
With these transformations we can then grid the galaxies on a mesh and obtain the galaxy number count per cell $\mbi N_{{\rm G}}$.
In paticular, we consider in our analysis cubical volumes of L=1250 $h^{-1}$ Mpc side length with 128$^3$ and 200$^3$ cells {\color{black} (and cubical volumes of L=3200 $h^{-1}$ Mpc side length with 512$^3$ cells, see appendix \ref{sec:app})}, and with the lower left corner of the box at
\ba 
x_{\rm llc} &=& -1500 \, h^{-1} \rm{Mpc} \nonumber \\
y_{\rm llc} &=& \hspace{0.15cm}-650 \, h^{-1} \rm{Mpc} \nonumber \\
z_{\rm llc} &=&  \hspace{.74cm}0 \, h^{-1} \rm{Mpc} \, . \nonumber
\ea

\subsubsection{Completeness: angular mask and radial selection function}

Furthermore, our data model requires the completeness in each cell $\mbi w$ (see Eq.~\ref{eq:lambda}).
The first ingredient in our three dimensional completeness is the angular mask of a position in the sky. The mask is provided as polygons with equal completeness  \citep[see left panel of Fig.~\ref{fig:mask}, and the \textsc{mangle} software package,][]{Hamilton-Tegmark-04,Swanson08}.
As a first step in the 3D completeness calculation we project the angular mask to 3D by throwing large numbers of sight lines evaluating the sky mask with \textsc{mangle}. The result of such a projection is shown on the right panel of  Fig.~\ref{fig:mask}.
Next we need to define the radial selection function from the number density distribution as a function of redshift
\ba
f(r) \propto \frac{1}{r^2}\frac{\Delta N_{\rm G}}{\Delta r}\,.
\ea
normalised to one. 
In principle the radial selection function should be evaulated in real space to avoid the so-called "Kaiser rocket" effect \citep{Kaiser-87,nusser2014}. This is only possible when the real space positions are reconstructed, as we do here. 
Obtaining the real space radial selection function can be expressed as an additional Gibbs-sampling step for iteration $j+1$ 
\ba
{f(r)^{j+1}}&\curvearrowleft& \mathcal P_{f}\left(f(r)|\{r^{j}\}\right)\,{,}\label{eq:f}
\ea
for the set of recovered galaxy distances in the previous iteration $\{r^{j}\}$.
Once we have the radial selection function we can multiply it with the 3D projected angular mask to get the 3D completeness. 
The radial selection functions as provided by the CMASS galaxy catalogue in redshift space and the reconstructed real space one are shown in Fig.~\ref{fig:fr_argo}. The agreement between both is very good, being compatible within 2-$\sigma$ throughout almost the entire redshift range. However, we see some tiny differences at distances where the selection function suffers strong gradients at the smallest distances, indicating that this approach could become important if such extreme cases happen more often.

\subsection{Application to galaxy catalogues}

In this section we present results from first testing the  method on light-cone mocks resembling the BOSS CMASS survey geometry,  radial selection function, and galaxy bias for which both the galaxy in real space and in redshift space are available; and second applying the same method to the BOSS DR12 data.
We explore the scales between 6 and 10 $h^{-1}$ Mpc. In particular, we consider grids with $128^3$ and $200^3$ cells with cubical volumes of L=1250 $h^{-1}$ Mpc side.
We need to obtain 12000 density and peculiar velocity samples using 8 cores for a mesh of $128^3$ ($200^3$) cells about 200 (550) CPU hours with about $<$1 ($\sim$3) minute(s)  per Gibbs-iteration given the survey geometry in this case study (less than 40\% of the volume is covered with data). The memory requirements are 170 (870) Mb, respectively.
{\color{black} An additional set of reconstructions considering volumes of 3200 $h^{-1}$ Mpc and 512$^3$ cells has been done (see appendix \ref{sec:app}).}
We have chosen 10 redshift and completeness bins for the range in which the CMASS data are defined in our study $0.43<z<0.7$ to sample the renormalisation of the lognormal fields (Eq.~\ref{eq:mu3}), and the normalisation of the number densities in the power law bias model (Eq.~\ref{eq:gamma}). A too fine resolution in redshift and completeness would  introduce too much stochasticity in the derived $\mu$ and $\gamma$ constants. We consider, however, 10 redshift bins to recover the peculiar velocity field at different redshifts  (see Sec.~\ref{sec:vel}). Here we do not take more redshift bins to save computational costs. In fact such a redshift spacing of 0.0225 is enough to model the cosmic evolution of CMASS galaxies \citep[see][]{sergio15,2016MNRAS.456.4156K}.
The power spectrum correlation matrix shown in Fig.~\ref{fig:GR} demonstrates that after less than 200 iterations the chain converges to power spectra which are highly correlated. The correlation is less strong if one considers only the first 30 bins up to $k\sim0.03\,h\,{\rm Mpc}^{-1}$, since cosmic variance due to empty  and low completeness regions in the volume dominates those scales. Nevertheless, even on those scales we expect to have high correlations between power spectra of different iterations after convergence due to the constrained phases by the data.
It is therefore safe to disregard the first 1000 iterations of the chains until the power spectra have converged and use a total of 6000 iterations for our analysis for each setup (meshes of $128^3$ and $200^3$ for mocks and observations). We further demonstrate that we succeed in sampling from the posterior distribution function  estimated through the \citet[][]{Gelman92} test as shown on the right panel of Fig.~\ref{fig:GR}  \citep[for details, see appendix in ][]{Ata15}.
The linear real space bias $b_{\rm L}$, one typical sample of the lognormal mean field $\mu=\langle\log(1+\delta)\rangle$, and of the galaxy density normalisation $\gamma$, are shown in  Fig.~\ref{fig:means}. We find that it is crucial to sample the bias and the mean fields on at least 5 bins to get accurate density reconstructions free of radial selection biases. However, the reconstructions  are robust against different redshift  bins in $\gamma$. 
We find that the theoretical prediction for the mean field $\mu=-\sigma^2/2\simeq-0.760$ for resolutions of $6.25\,h^{-1}$ Mpc is compatible within 4\% with our numerical sampling result.
The technique presented in this work permits us to get peculiar velocity fields which are compensated for the survey geometry and selection functions. This can be qualitatively appreciated in  Figs.~\ref{fig:vel} and \ref{fig:velcomp}.
On a quantitative level, we find that the velocities are highly correlated and {\color{black} approximately} unbiased with the true velocities  (see Fig~\ref{fig:corrvel}) for the case in which the density fields was smoothed with a Gaussian kernel with radius of $r_{\rm S}=2\,h^{-1}$ Mpc for a resolution of $d_{\rm L}=6.25\,h^{-1}$ Mpc. We note that the maximum a posteriori (MAP) solution, such as Wiener-filtering, will yield biased results, although for Wiener-filtering the variance can be separately added to the MAP solution and such a bias is known \citep[][]{zaroubi}.  The statistical correlation coefficient we find is about 0.7 including about 10\% of satellite galaxies with virial motions, which is what one finds for CMASS galaxies. {\color{black} We have checked this result testing boundary effects and cosmic variance by considering the full volume covered by the CMASS sample (see appendix \ref{sec:app}). We find very similar results to the sub-volume reconstructions, which at most decrease the statistical correlation coefficient to about 0.69.} This correlation can be considerably improved by excluding these satellite galaxies from the analysis. As a proxy we consider two cases. One excluding galaxies for which the velocity difference between true and reconstructed exceeds 500 and 700 km $s^{-1}$. The first one removes $\sim$10\% of the galaxies, and the second one $\sim$3.5\%. Since not all satellite galaxies will be outliers the answer will be probably closer to the latter case, rasing the statistical correlation coefficient to about $r=0.75$, {\color{black} which is a priori a considerable improvement with respect to previous methods  \citep[see, e.g.,][though a proper comparison between methods remains to be done based on the same mocks]{Schaan2015,2016A&A...586A.140P}}. Although we are using only linear theory here, our method includes a couple of ingredients which can explain this improvement, such as being a self-consistent (iterative) method, yielding linearised density fields, for which the pixel window has been exactly solved (the counts in cells, i.e., the nearest grid point, are treated through the full Poisson likelihood), and nonlinear bias has been taken into account.
 The smoothing scale could be considered another parameter of our model. However, it can be derived from the velocity divergence power spectrum $P_{\theta\theta}$ with $\theta\equiv-\frac{1}{fHa}\nabla\cdot\mbi v$ prior to running any Markov chain, {\color{black} as it has been done here}.  In particular, one expects $P_{\theta\theta}$ to converge towards the linear power spectrum in the transition to the nonlinear regime at about $k\sim0.15-0.2\,h$ Mpc$^{-1}$ \citep[][]{Jennings2012,Hahn2015}.
 This is expected as the velocity divergence is closer to the Gaussian field than the gravitationally evolved density field \citep[see, e.g.,][]{kitaura_vel}. In fact while the density is enhanced in the potential wells, virialisation prevents galaxies from getting larger and larger velocities. As a consequence, the power spectrum of the velocity divergence is close to the linear density field in the quasi-linear regime, eventually being even more suppressed at high $k$ values.
 Fig.~\ref{fig:pstheta} shows that such an agreement down to scales of $k\sim0.2\,h$ Mpc$^{-1}$ is indeed achieved for smoothing scales of about $r_{\rm S}=2\,h^{-1}$ Mpc. In fact for a smoothing scale $r_{\rm S}$ between 1 and 2 $h^{-1}$ Mpc one can potentially obtain unbiased results beyond $k=0.5\,h$ Mpc$^{-1}$.   While our chains with $128^3$ were run with velocities derived from density fields smoothed with $r_{\rm S}=7\,h^{-1}$ Mpc, our reconstructions with $200^3$ were run using $r_{\rm S}=2\,h^{-1}$ Mpc. This variety of smoothing scales serves us to test the robustness of the velocity reconstructions depending on this parameter.
In fact  we manage to recover the monopoles in real space down to scales of  about $k\sim0.2\,h$ Mpc$^{-1}$  (see left panels in Fig.~\ref{fig:pk} and \ref{fig:pkadd}, for the lognormal-Poisson).
We have checked that the theoretical prediction from renormalised perturbation theory for the bias correction parameter $f_{\rm b}$ can be sampled as a free parameter yielding compatible results, $f_{\rm b}=0.70 \pm 0.05$ vs $f_{\rm b}=0.66 \pm 0.1$  from theory when considering the first 30 bins in the power spectrum, i.e., $k\lsim0.03\,h\,{\rm Mpc}^{-1}$ . 
Given the volume we consider in this work of $(1250\,h^{-1}\,{\rm Mpc})^3$ we expect cosmic variance to cause deviations from zero in the quadrupoles. Therefore, we show the quadrupole of the real space mock galaxy catalogue as a reference. The upper right panel in Fig.~\ref{fig:pk} demonstrates that we cover the real space quadrupole down to scales of about $r\sim20\,h^{-1}$ Mpc. Deviations on large scales ($\gsim120\,h^{-1}$ Mpc) between the recovered and the true quadrupoles are due to the large empty volume which pushes the solution to be closer to zero than in the actual mock catalogue. In fact, we showed in a previous paper that one can recover with this method the quadrupole features of the  particular realisation when considering complete volumes \citep[][]{Kitaura:2015bca}.
The results are consistent when comparing lower to higher resolution reconstructions (middle to lower right panels). However, we see that the uncertainty (shaded regions) in the quadrupole increases in the higher resolution case. This is expected as the coarser grid smooths the peculiar velocities and tends to underestimate them. 
In addition, we have  run a reconstruction chain including velocity dispersion, showing that this will also enhance the error bars in the quadrupole, however yielding the same qualitative results as without that term (see lower panels in Fig.~\ref{fig:pkadd}). We observe a slightly enhanced uncertainty in the monopole and quadrupole on large scales. A proper treatment of the velocity dispersion requires, however, at least a density dependent dispersion term, or even looking at the tidal field Eigenvalues \citep[see][]{Kitaura:2015bca}. This is however, computationally more expensive and requires a number of additional parameters. We thus leave such an effort for later work.
The accuracy of the quadrupole reconstruction presented in this paper seems to be superior than in some of the standard BAO reconstruction techniques \citep[][]{Vargas15,Burden15}, see in particular, right panel in Fig.~8 in \citet[][]{2016MNRAS.456.4156K} showing the quadrupole after BAO reconstruction for a set of mock Multidark-\textsc{patchy} BOSS DR12 CMASS catalogues very similar to the ones used here. We note that while the monopoles of the dark matter field are trivially computed from the reconstructed samples on complete meshes, the computation of the quadrupoles of the galaxies needs more computational efforts to account for survey geometry and radial selection functions \citep[see, e.g.,][]{AAB14}. 
Fig.~\ref{fig:slice2} shows slices in the $x-z$ plane of the galaxy number counts, the completeness, and the reconstructed density fields. One can clearly recognise prominent features in the data in the reconstructed density fields. It is remarkable however, how these features appear balanced without selection function effects, in such reconstructions. Only when one computes the mean over many realisations, one can see that larger significance in the reconstructions correlates with higher completeness values. The vanishing structures in unobserved regions further demonstrates the success in sampling from the posterior distribution function. Fig.~\ref{fig:pdf} shows that the lognormal fields are indeed reasonably Gaussian distributed in terms of the univariate probablity distribution function.  In fact the absolute skewness is reduced from about 6.4 to less than 0.03 with means being always smaller than $|\langle\delta_{\rm L}\rangle|<0.13$ for different completeness regions. As we will analyse below the 3pt statistics does, however, not correspond to a Gaussian field.

\subsection{The cosmic web from lognormal-Poisson reconstructions}
\label{sec:cw}

So far we have been reconstructing the linear component of the density field in Eulerian space at a reference redshift within the lognormal approximation.
We can, however, get an estimate of the nonlinear cosmic web by performing structure formation within a comoving framework, i.e., without including the displacement of structures, as our reconstructed linear density fields already reside at the final Eulerian coordinates. One can use cosmological perturbation theory to make such a mapping \citep[see][]{kitlin}. We will rely here on the classical \citet[][]{1970A&A.....5...84Z} framework. By demanding mass conservation from Lagrangian to Eulerian space $\rho(\mbi q)\dd\mbi q=\rho(\mbi r)\dd\mbi r$, we get an equation for the cosmic evolved density field within  comoving coordinates: $1+\delta^{\rm PT}(\mbi q)={\mat J}^{-1}$ (with the supercript standing for perturbation theory), where $\mat J$ is the Jacobian matrix often called the tensor of deformation: ${\cal D}_{ij}\equiv\delta^{\rm K}_{ij}+\Psi_{i,j}(\mbi q, z)$.
By doing the proper diagonalisation one finds that the comoving evolved density field can be written as: 
\be
\delta^{\rm PT}(\mbi q, z)=\frac{1}{(1-D(z)\lambda_1(\mbi q))(1-D(z)\lambda_2(\mbi q))(1-D(z)\lambda_3(\mbi q))}-1\,,
\label{eq:zeld1}
\ee
where $\lambda_i$ are the Eigenvalues of the deformation tensor with $\lambda_1\geq\lambda_2\geq\lambda_3$.
This framework is helpful to gain insight over the formation of the cosmic web  \citep[see][]{hahn}. In fact we could use the reconstructed velocity field to compute the shear tensor and study the cosmic web \citep[][]{1996Natur.380..603B}. We will however, focus on  the largest Eigenvalue denoting the direction of first collapse to form the filamentary cosmic web. We can Taylor expand the previous equation within the Eulerian framework yielding
\be
\delta^{\rm PT}(\mbi r, z)\simeq D(z)\lambda_1(\mbi r)+\lambda^+(\mbi r, z)\,,
\label{eq:zeld}
\ee
with $\lambda^+$ being the higher order contributions including the rest of Eigenvalues, which can be approximated by $\lambda^+(\mbi r, z)\simeq-\langle D(z)\lambda_1(\mbi r)\rangle$. This expression avoids the problem of formation of caustics, as present in Eq.~\ref{eq:zeld1}. We have tested other expansions  including the rest of Eigenvalues, however, with less success in describing the nonlinear cosmic web. The operation of retaining the information of the largest Eigenvalue can also be interpreted, as filtering out the noisy part of the Gaussian field. This technique could potentially be useful to effectively enhance the cosmic web of a low resolution simulation for mock catalogue production. We leave a more thorough investigation of other possible comoving structure formation descriptions for later work. 
Since this theory is based on the Gaussian density field, we will compute the Eigenvalues  based on the linear component of the density field $\delta_{\rm L}$. In particular, we will compute them from the gravitational potential $\phi_{\rm L}\equiv\nabla^{-2}\delta_{\rm L}$, solving the Poisson equation with the inverse Laplacian operator in Fourier space, to obtain the correspoding tidal field tensor.
By applying Eq.~\ref{eq:zeld} we thus get the linear component of the gravitationally evolved density field in Eulerian space, which we will denote as $\delta^{\rm PT}_{\rm L}(\mbi r)$. We now can compute the nonlinear component by doing the transformation $\delta^{\rm PT}(\mbi r)=\exp(\delta^{\rm PT}_{\rm L}(\mbi r)+\mu(\delta^{\rm PT}_{\rm L}(\mbi r))-1$, having the physical meaninful property of yielding positive definite density fields. To ensure that this field shares the same power spectrum, as the lognormal reconstructed density field $\delta(\mbi r)=\exp(\delta_{\rm L}(\mbi r)+\mu(\delta_{\rm L}(\mbi r)))-1$, we apply in Fourier space 
\be
\hat{\delta}^{\rm PT,{\rm f}}_{\rm L}(\mbi k)=\sqrt{P^{\rm trans}(k)}\frac{\hat{\delta}^{\rm PT}_{\rm L}(\mbi k)}{\sqrt{\langle|\hat{\delta}^{\rm PT}_{\rm L}(\mbi k)|^2\rangle_{\Delta k}}}\,,
\ee
where the nonlinear transformed  power spectrum $P^{\rm trans}(k)$ is found iteratively.  The ratio between the target power spectrum and the one obtained at a given iteration is multiplied to $P^{\rm trans}(k)$ from the previous iteration until the nonlinear power spectra averaged in $\Delta k$-shells coincide $\langle|\hat{\delta}^{\rm PT,{\rm f}}(\mbi k)|^2\rangle_{\Delta k}\simeq\langle|\hat{\delta}(\mbi k)|^2\rangle_{\Delta k}$ (i.e., the power spectrum from the nonlinear transformed lognormal density field), in a given $k$-range within a given accuracy. As a starting guess of $P^{\rm trans}(k)$ we take $\langle|\hat{\delta}_{\rm L}(\mbi k)|^2\rangle_{\Delta k}$ (i.e., the power spectrum from the linear lognormal density field).  In practice, less than 15 iterations are necessary to be accurate within better than 1\% up to at least 70\% of the Nyquist frequency using about 100 $\Delta k$-bins for meshes of $200^3$ cells on cubical volumes of 1250 $h^{-1}$ Mpc side, requiring less than 100$s$  on 8 cores. This operation is justified, as we are dealing with the Gaussian component of the density field, permitting us to define a pseudo white noise ${\hat{\delta}^{\rm PT}_{\rm L}(\mbi k)}/\sqrt{\langle|\hat{\delta}^{\rm PT}_{\rm L}(\mbi k)|^2\rangle_{\Delta k}}$, which allows modifications of the two point statistics.  In fact, the PDF of $\delta^{\rm PT}_{\rm L}$  is very Gaussian. %, as one can see in Fig.~\ref{fig:pdf}.  
This calculation is parameter free, and does not require any further input than the lognormal field (and the window function to compute the completeness dependent renormalised mean fields).
Effectively, these transformations retain the two-point statistics, while improving the three-point statistics of the lognormal field, hereby extracting the cosmic web structure of the density field, which is  diluted in the lognormal reconstructions {\color{black} \citep[for a similar concept see][]{Leclercq2013}}. We note, that the distribution of peaks even prior to the nonlinear tidal field transformation do not correspond to a random lognormal realisation, as they are based on the galaxy distribution within the posterior sampling analysis, which already suffered displacements due to the action of gravity.
The results of this study are shown in Figs.~\ref{fig:slice2}, \ref{fig:slice3}, \ref{fig:withgal}, and \ref{fig:signal2noisecw}. One can see how the closely Gaussian logarithmic-density field (lower left panels in Figs.~\ref{fig:slice2} and \ref{fig:slice3}) is transformed into a density field depicting the cosmic web (lower right panels in Figs.~\ref{fig:slice2} and \ref{fig:slice3}), which is in good agreement with the distribution of galaxies (see Fig.~\ref{fig:withgal}).
The ensemble average plots shown in the lower right panels of Figs.~\ref{fig:slice2} and \ref{fig:slice3} demonstrate the robustness of the reconstructed filamentary network. The  variance plots confirm as expected that the uncertainty on the density is larger in the voids and in the unobserved regions (see lower panels in Fig.~\ref{fig:signal2noisecw}). In fact, the variance depicts the negative of the filamentary network.
These density maps can be used for environmental studies \citep[see, e.g.,][]{Nuza14}.
They could be used as a reference for future applications including reconstructions of the initial conditions \citep[see e.g.][]{kit2mrs,Jasche2013,Kitaura2013,Wang2013,HKG13,Wang2014}.

\section{Summary and conclusions}
\label{sec:conc}

In this work, we have presented a Bayesian phase-space (density and velocity) reconstruction of the cosmic large-scale matter density and velocity field from the SDSS-III Baryon Oscillations Spectroscopic Survey Data Release 12 (BOSS DR12) CMASS  galaxy clustering catalogue.
We have demonstrated that very simple models can yield accurate results on scales larger than $k\sim0.2\,h$ Mpc$^{-1}$.

In particular we have used a set of simple assumptions. Let us list them here
\begin{itemize}
\item the statistical distribution of galaxies is described by the lognormal-Poisson model, 
\item linear theory relates the peculiar velocity field to the  density field,
\item the volume is a fair sample, i.e. ensemble averages are equal to volume averages,
\item cosmic evolution is modelled within linear theory with  redshift dependent growth factors, growth rates, and bias,
 \item a power law bias, based on the linear bias multiplied by a correction factor, which can be derived from renormalised perturbation theory, relates the galaxy expected number counts to the underlying density field. 
\end{itemize}
This has permitted us to reduce the number of parameters and derive them consistently from the data, with a given smoothing scale and a particular $\Lambda$CDM cosmological parameter set.

We have included a number of novel aspects in the \textsc{argo} code extending it to account for cosmic evolution in the linear regime. In particular, the Gibbs-scheme samples
\begin{itemize}
\item the density fields with a lognormal-Poisson model,
\item the mean fields of the lognormal renormalised priors for different completeness values,
\item the number density normalisation at different redshift bins,
\item the real space positions of galaxies from the reconstructed peculiar velocity fields,
\item and the real space radial selection function from the reconstructed real space positions of galaxies (accounting for the ``Kaiser-rocket'' effect).
\end{itemize}

Our results show that we can get unbiased dark matter power spectra up to $k\sim0.2\,h$ Mpc$^{-1}$, and unbiased isotropic quadrupoles down to scales of about 20  $h^{-1}$ Mpc, being far superior to redshift space distortion corrections based on traditional BAO reconstruction techniques which start to deviate at scales below 60 $h^{-1}$ Mpc. 

As a test case study we also analyse deviations of Poissonity in the likelihood, showing that the power in the monopole and the scatter in the quadrupoles is increased towards small scales.

 The agreement between the reconstructions with mocks and BOSS data is remarkable. In fact, the identical algorithm with the same set-up and parameters were used for both mocks and observations. This confirms that the cosmological parameters used in this study are already close to the true ones, the systematics are well under control, and gives further support to $\Lambda$CDM at least on scales of about $0.01\lsim k\lsim0.2\,h\,{\rm Mpc}^{-1}$.   

We also found that the reconstructed velocities have a statistical correlation coefficient compared to the true velocities of each individual lightcone mock galaxy of $r\sim0.7$  including about 10\% of satellite galaxies with virial motions. The power spectra of the velocity divergence agree well with theoretical predictions up to $k\sim0.2\,h\,{\rm Mpc}^{-1}$. This is far superior to the results obtained from simple linear reconstructions of the peculiar velocities directly applied on the smoothed galaxy field for which statistical correlation coefficients of the order of 0.5 are obtained \citep[][though this work used the Sloan main sample at lower redshifts being further in the nonlinear regime, making a direct comparison difficult]{2016A&A...586A.140P}. Improved results can be obtained with Wiener-filter based  techniques, which need to correct for the bias in a post-processing way \citep[][]{Schaan2015}. It would be interesting to compare the different methods, in particular considering that the ensemble average is not equal to the maximum of the posterior   for non-Gaussian PDFs, as we consider here.
Although it  may seem surprising to get such accurate results from simply assuming linear theory to derive the peculiar motions, we expect that linearised density fields as the ones obtained from lognormal-Poisson reconstructions (even if one takes the nonlinear transformed one), yield improved velocity fields \citep[see][]{Falck12,kitlin}. Also, while linear theory tends to overestimate the peculiar velocity field, the chosen grid resolution with the additional smoothing compensates for this yielding unbiased reconstructed peculiar motions. We have seen that for a given resolution the additonal Gaussian smoothing radius (and the cell resolution) can be derived from the velocity divergence power spectrum to match the linear power spectrum in the quasi-linear regime ($0.1\lsim k\lsim0.5\,h\,{\rm Mpc}^{-1}$).
We demonstrated that the reconstructed linear component reduces the skewness by two orders of magnitude with respect to the density directly derived from smoothing the galaxy field on the same scale.

 We have furthermore demonstrated how to compute the Zeldovich density field from the lognormal reconstructed density fields based on the tidal field tensor in a parameter free way. The recovered filamentary network remarkably connects the discrete distribution of galaxies. The real space density fields obtained in this work could be used to recover the initial conditions with techniques which rely on knowing the dark matter field at the final conditions \citep[see e.g.][]{Wang2013,Wang2014}. 

We aim to improve the Bayesian galaxy distance estimates going to smaller scales, by using non-Poisson likelihoods and including a correction of the virialised motions \citep[][]{Ata15,Kitaura:2015bca}. One could also explore other priors based on perturbation theory \citep[e.g.][]{KitauraHess2013,HKG13}.

Despite of the potential improvements to this work, the reconstructed density and peculiar velocity fields obtained here can already be used  for a number of studies, such as BAO reconstructions, kinematic Sunyaev-Zeldovich (kSZ), integrated Sachs-Wolfe (ISW) measurements, or environmental studies.

\section*{Acknowledgments}

MA thanks the Friedrich-Ebert-Foundation for its  support. 
MA and FSK thank Uros Seljak for the hospitality at LBNL and UC Berkeley and for enriching feedback, and  the Instituto de F{\'i}sica Te{\'o}rica (IFT UAM-CSIC) in Madrid for its support via the Centro de Excelencia Severo Ochoa Program under Grant SEV-2012-0249.
FSK also thanks Masaaki Yamato for support at LBNL and for encouraging discussions.
CC acknowledges support from the Spanish MICINN’s Consolider-Ingenio 2010 Programme under grant MultiDark CSD2009-00064 and AYA2010-21231-C02-01 grant.
CC were also supported by the Comunidad de Madrid under grant HEPHACOS S2009/ESP-1473.
GY acknowledges  financial support from MINECO (Spain) under research grants AYA2012-31101 and AYA2015-63810-P.
The MultiDark Database used in this paper and the web application \url{www.cosmosim.org/} providing online access to it were constructed as part of the activities of the German Astrophysical Virtual Observatory as result of a collaboration between the Leibniz-Institute for Astrophysics Potsdam (AIP) and the Spanish MultiDark Consolider Project CSD2009-00064.
\textsc{CosmoSim.org} is hosted and maintained by the Leibniz-Institute for Astrophysics Potsdam (AIP).

 The \textsc{BigMultiDark} simulations have been performed on the SuperMUC supercomputer at the Leibniz-Rechenzentrum (LRZ) in Munich, using the computing resources awarded to the PRACE project number 2012060963. 

Funding for SDSS-III has been provided by the Alfred P. Sloan Foundation, the Participating Institutions, the National Science Foundation, and the U.S. Department of Energy Office of Science. The SDSS-III web site is http://www.sdss3.org/.

SDSS-III is managed by the Astrophysical Research Consortium for the
Participating Institutions of the SDSS-III Collaboration including the
University of Arizona,
the Brazilian Participation Group,
Brookhaven National Laboratory,
University of Cambridge,
Carnegie Mellon University,
University of Florida,
the French Participation Group,
the German Participation Group,
Harvard University,
the Instituto de Astrofisica de Canarias,
the Michigan State/Notre Dame/JINA Participation Group,
Johns Hopkins University,
Lawrence Berkeley National Laboratory,
Max Planck Institute for Astrophysics,
Max Planck Institute for Extraterrestrial Physics,
New Mexico State University,
New York University,
Ohio State University,
Pennsylvania State University,
University of Portsmouth,
Princeton University,
the Spanish Participation Group,
University of Tokyo,
University of Utah,
Vanderbilt University,
University of Virginia,
University of Washington,
and Yale University.

\bibliographystyle{mnras}
\bibliography{lit}

\vspace{0.5cm}

{\hspace{-0.65cm}$^1$ Leibniz Institut f{\"u}r Astrophysik (AIP), An der Sternwarte 16, D-14482 Potsdam, Germany\\
$^2$ Lawrence Berkeley National Lab, 1 Cyclotron Rd, Berkeley CA 94720, USA\\
$^{3}$ Departments of Physics and Astronomy, University of California, Berkeley, CA 94720, USA\\
$^{4}$ Instituto de Astrofisica de Canarias (IAC), Calle Via Lactea s/n, 38200\\
$^{5}$ Departamento de Astrof{\'i}sica, Universidad de La Laguna (ULL), E-38206 La Laguna, Tenerife, Spain\\
$^{6}$ Instituto de F{\'i}sica Te{\'o}rica, (UAM/CSIC), Universidad Aut{\'o}noma de Madrid, Cantoblanco, E-28049 Madrid, Spain\\
$^{7}$Campus of International Excellence UAM+CSIC, Cantoblanco, E-28049 Madrid, Spain \\
$^{8}$Departamento de F{\'i}sica Te{\'o}rica,  Universidad Aut{\'o}noma de Madrid, Cantoblanco, 28049, Madrid, Spain\\
$^{9}$Centro de Estudios de F{\'i}sica del Cosmos de Arag\'on (CEFCA), Plaza San Juan, 1, planta 2, E-44001 Teruel, Spain,\\
$^{10}$ Sorbonne Universit{\'e}s, Institut Lagrange de Paris (ILP), 98 bis Boulevard Arago, 75014 Paris, France \\
$^{11}$ Laboratoire de Physique Nucl{\'e}aire et de Hautes Energies, Universit{\'e} Pierre et Marie Curie, 4 Place Jussieu, 75005 Paris, France \\
$^{12}$ Institute of Cosmology \& Gravitation, University of Portsmouth, Dennis Sciama Building, Portsmouth PO1 3FX, UK\\
$^{13}$ Department of Physics and Astronomy, University of Utah, 115 S 1400 E, Salt Lake City, UT 84112, USA\\
$^{14}$ Yale Center for Astronomy and Astrophysics, Yale University, New Haven, CT, 06520, USA\\
$^{15}$ Harvard-Smithsonian Center for Astrophysics, 60 Garden St., Cambridge, MA 02138, USA\\
$^{16}$ Key Laboratory for Research in Galaxies and Cosmology, Shanghai Astronomical Observatory, Shanghai 200030, China\\
$^{17}$ Department of Physics, Carnegie Mellon University, 5000 Forbes Avenue, Pittsburgh, PA 15213, USA\\
$^{18}$ Department of Physics and Astronomy, The Johns Hopkins University, Baltimore, MD 21218, USA\\
$^{19}$ Department of Chemistry and Physics, King's College, 133 North River St, Wilkes Barre, PA 18711, USA\\
$^{20}$Instituto de Astrof{\'i}sica de Andaluc{\'i}a (CSIC), Glorieta de la Astronom{\'i}a, E-18080 Granada, Spain\\
$^{21}$ Department of Astronomy and Space Science, Sejong University, Seoul, 143-747, Korea\\
$^{22}$ Max-Planck-Institut f{\"u}r extraterrestrische Physik, Postfach 1312, Giessenbachstr., D-85741 Garching, Germany\\
$^{23}$ Department of Astronomy and Astrophysics, The Pennsylvania State University, University Park, PA 16802\\
$^{24}$  Institute for Gravitation and the Cosmos, The Pennsylvania State University, University Park, PA 16802\\
$^{25}$Department of Physics and Astronomy, Ohio University, 251B Clippinger Labs, Athens, OH 45701, USA\\
- La Laguna, Tenerife, Spain\\
$^{26}$Center for Cosmology and Particle Physics, New York University, 4 Washington Place, New York, NY 10003, USA\\
$^{27}$ University of St Andrews, North Haugh, St Andrews Fife, KY16 9SS, UK}\\
$^{28}$ Instituto de F{\'i}sica, Universidad Nacional Aut{\'o}noma de M{\'e}xico, Apdo. Postal 20-364, 01000 M{\'e}xico,  D.F., M{\'e}xico

\appendix

\section{Cosmic density and velocity reconstruction based on the full CMASS BOSS DR12 sample}

\label{sec:app}

\begin{figure*}
\vspace{-0.5cm}
\begin{tabular}{cc}
\hspace{-.12cm}
\includegraphics[width=8.cm]{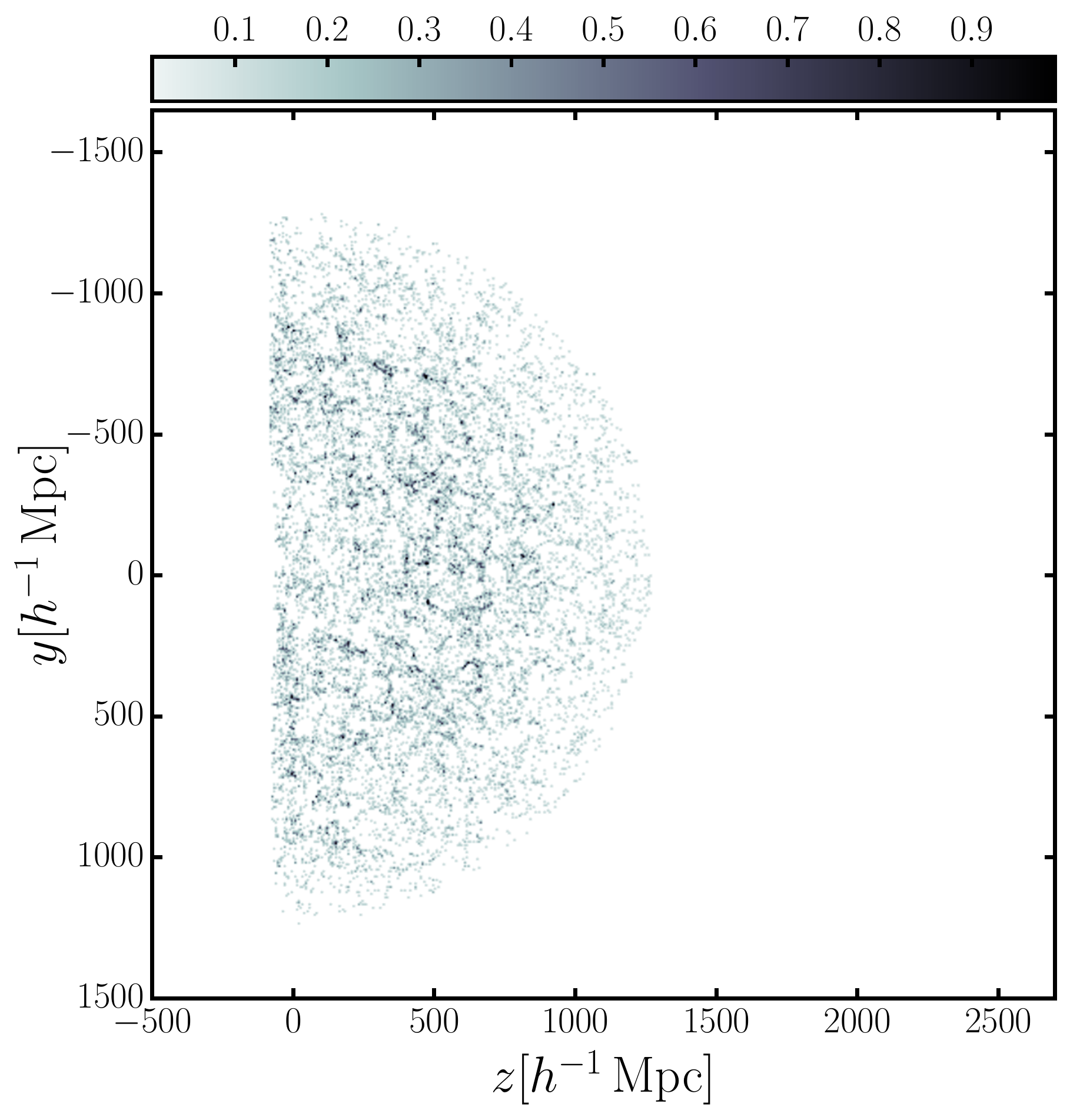}  
\hspace{-15.12cm}
\includegraphics[width=8.cm]{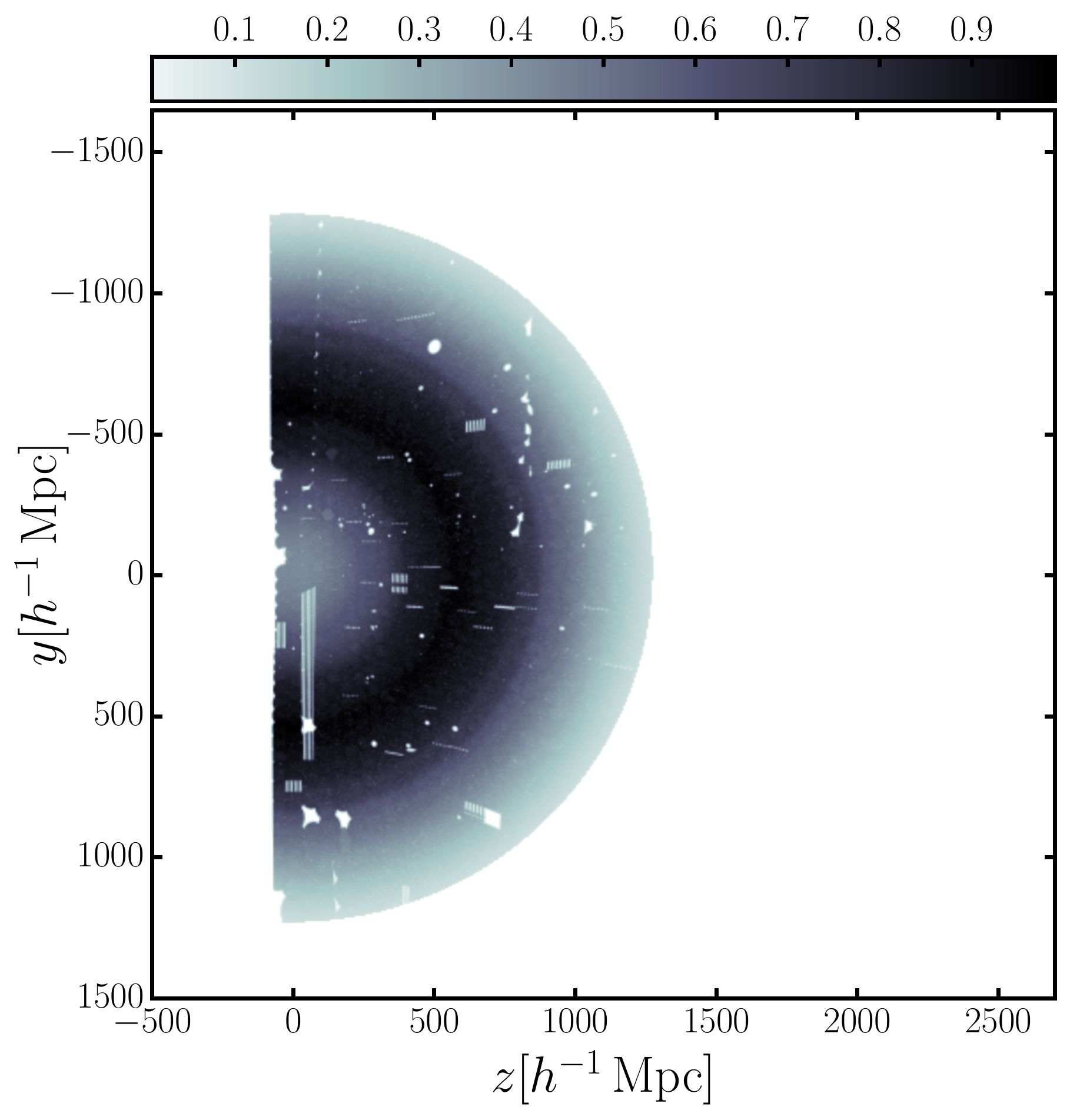} 
\put(-5,18.5){--}
\vspace{-.94cm}
\\
\includegraphics[width=8.cm]{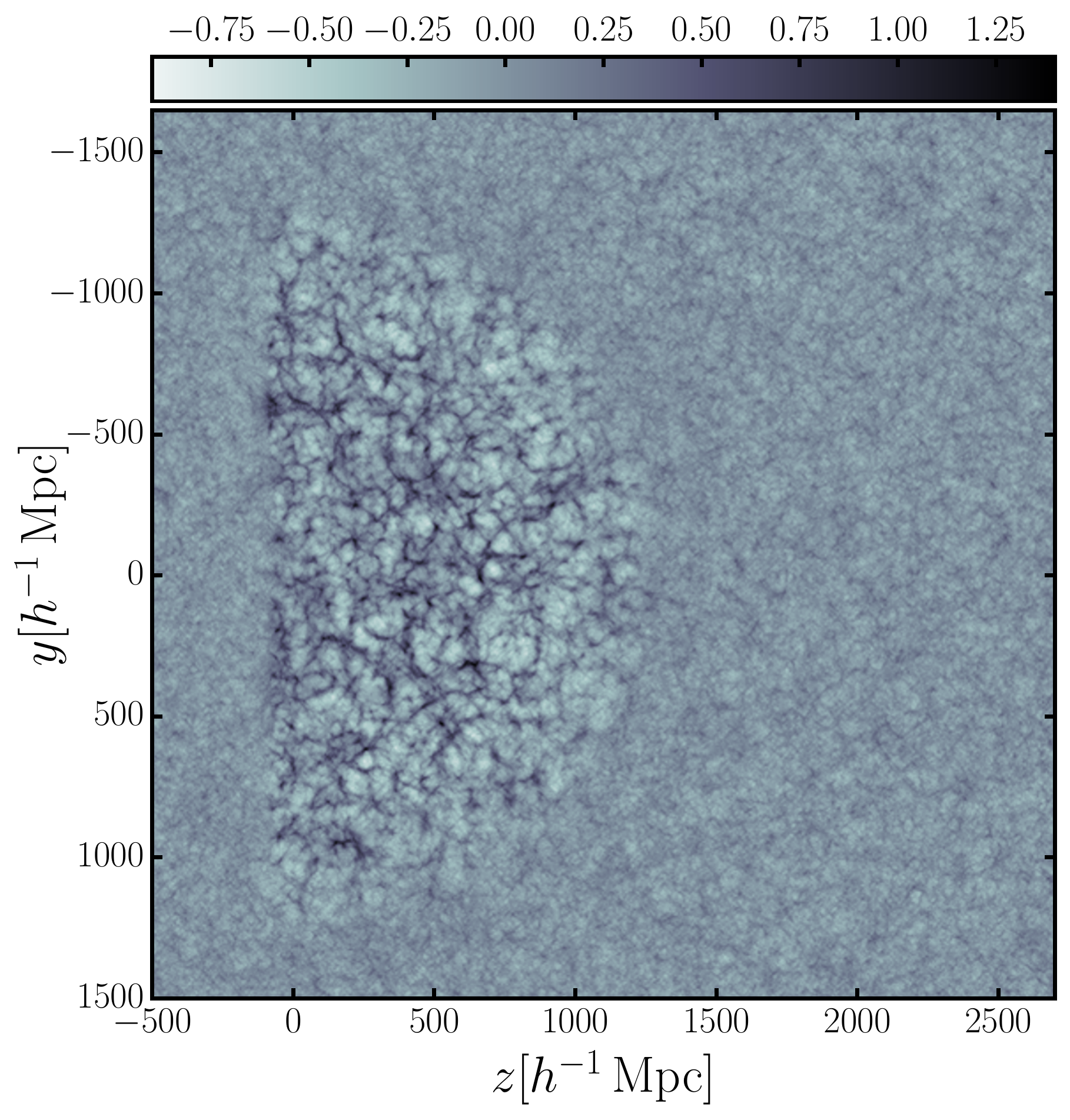}
\hspace{-15.125cm}
\includegraphics[width=8.cm]{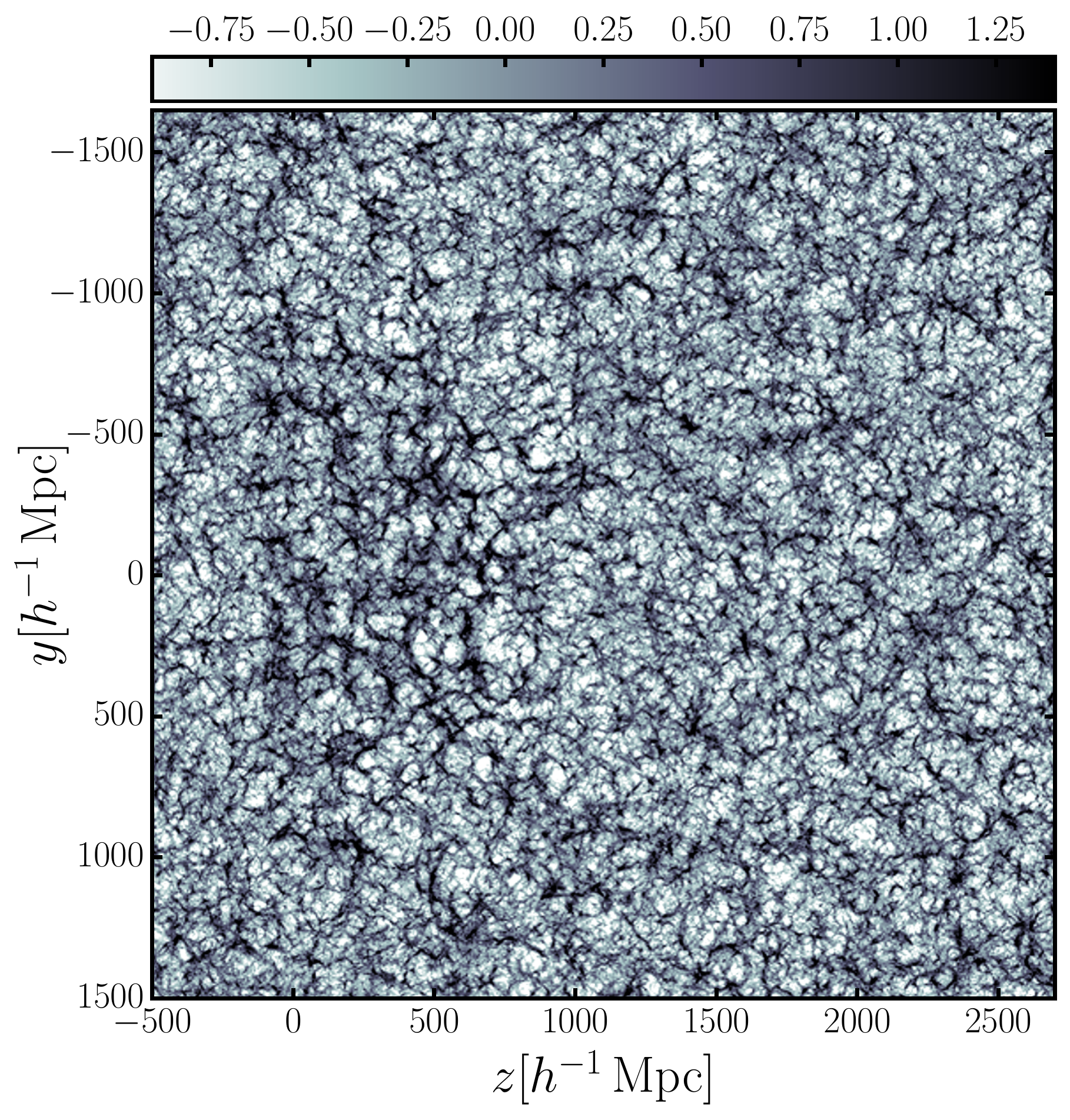}
\put(-5,18.5){--}
\vspace{-.94cm}
\\
\includegraphics[width=8.cm]{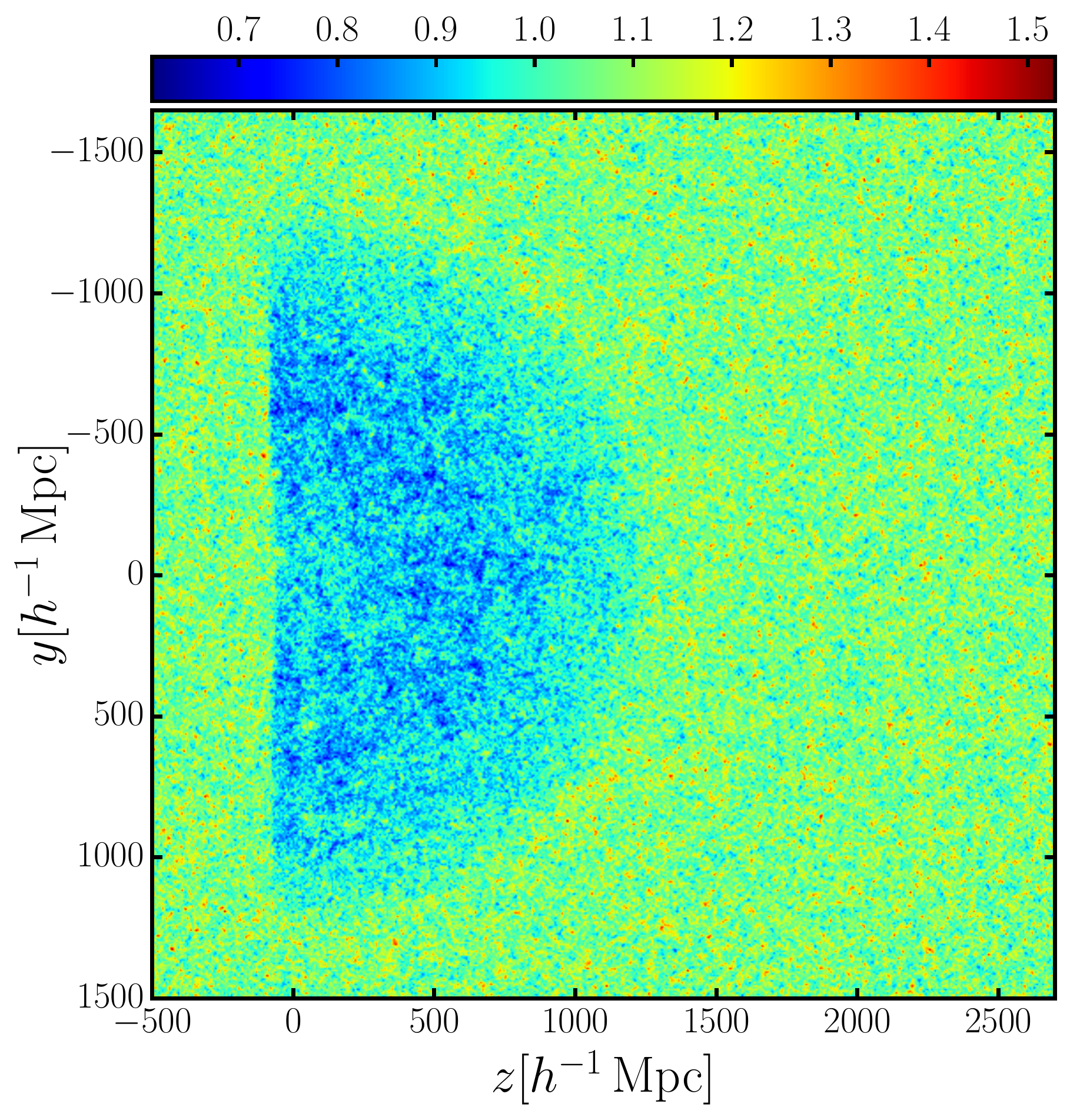}
\hspace{-15.125cm}
\includegraphics[width=8.cm]{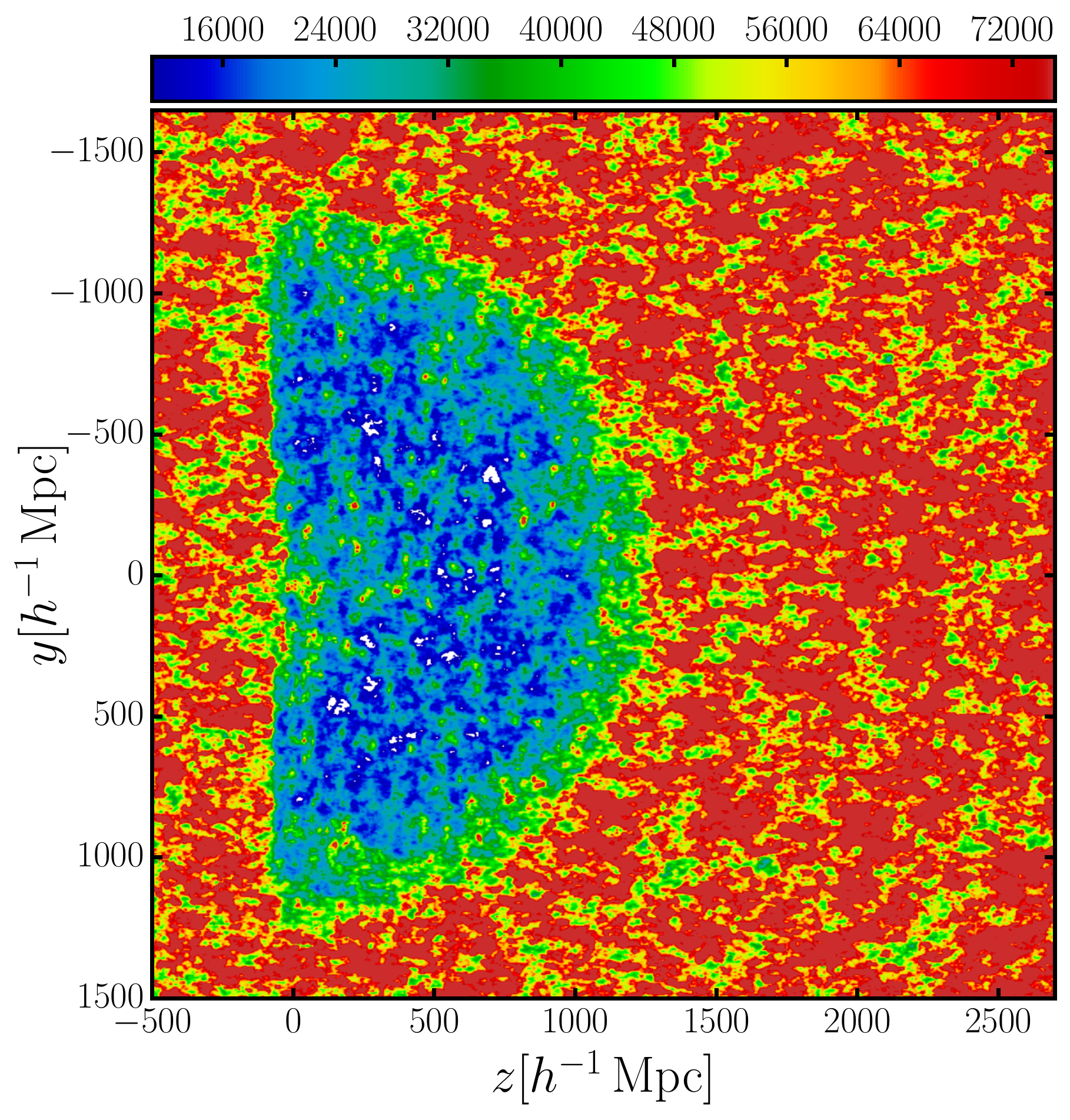}
\put(-5,18.5){--}
\end{tabular}
\vspace{-0.3cm}
 \caption{\label{fig:app1} Based on the CMASS BOSS DR12 catalogue with 6.25$\Mpc$ resolution: slices of the 3D cubical mesh of side 3200 $\Mpc$ and 512$^3$ cells. {\bf Upper left panel:} the 3D completeness or window function  function, {\bf upper right panel:} galaxy number count per cell. {\bf Middle left panel:} one reconstructed Zel'dovich transformed  density field, {\bf middle right panel:} mean over 6000 reconstructed Zel'dovich transformed  density fields. {\bf Lower left panel:} corresponding variance of the peculiar velocity field, {\bf lower right panel:} corresponding variance of the Zel'dovich transformed density field.}
\label{fig:}
\end{figure*}

\begin{figure*}
\begin{tabular}{cc}
\includegraphics[width=8.5cm]{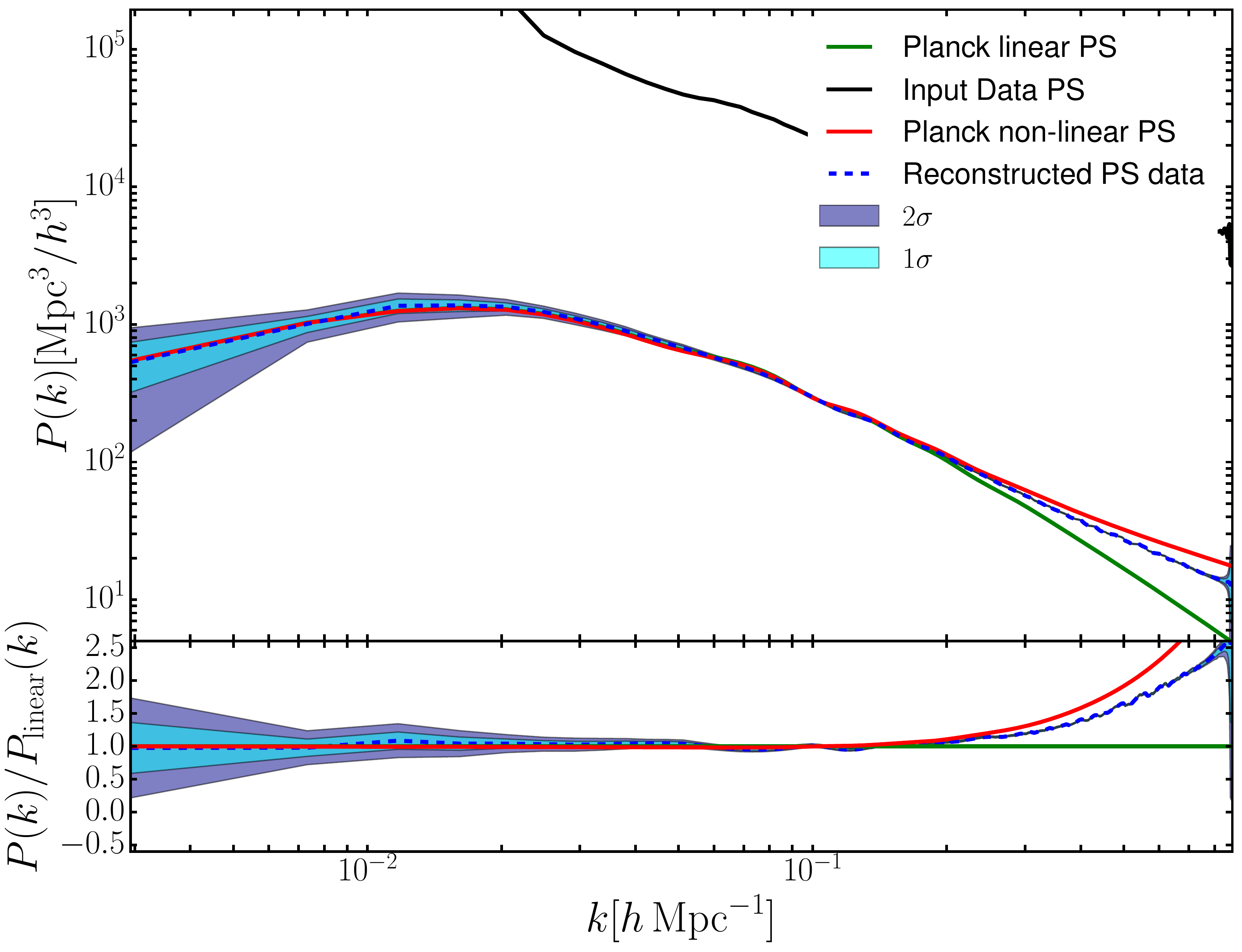}
\put(-210,170){${\rm L}=3200\,h^{-1}\,{\rm Mpc}$}
\put(-210,160){$d_{\rm L}=6.25\,h^{-1}\,{\rm Mpc}$}
\put(-210,150){\rotatebox[]{0}{\text{BOSS DR12}}}
\put(-210,140){with velocity dispersion}
\hspace{-16.25cm}
\includegraphics[width=8.5cm]{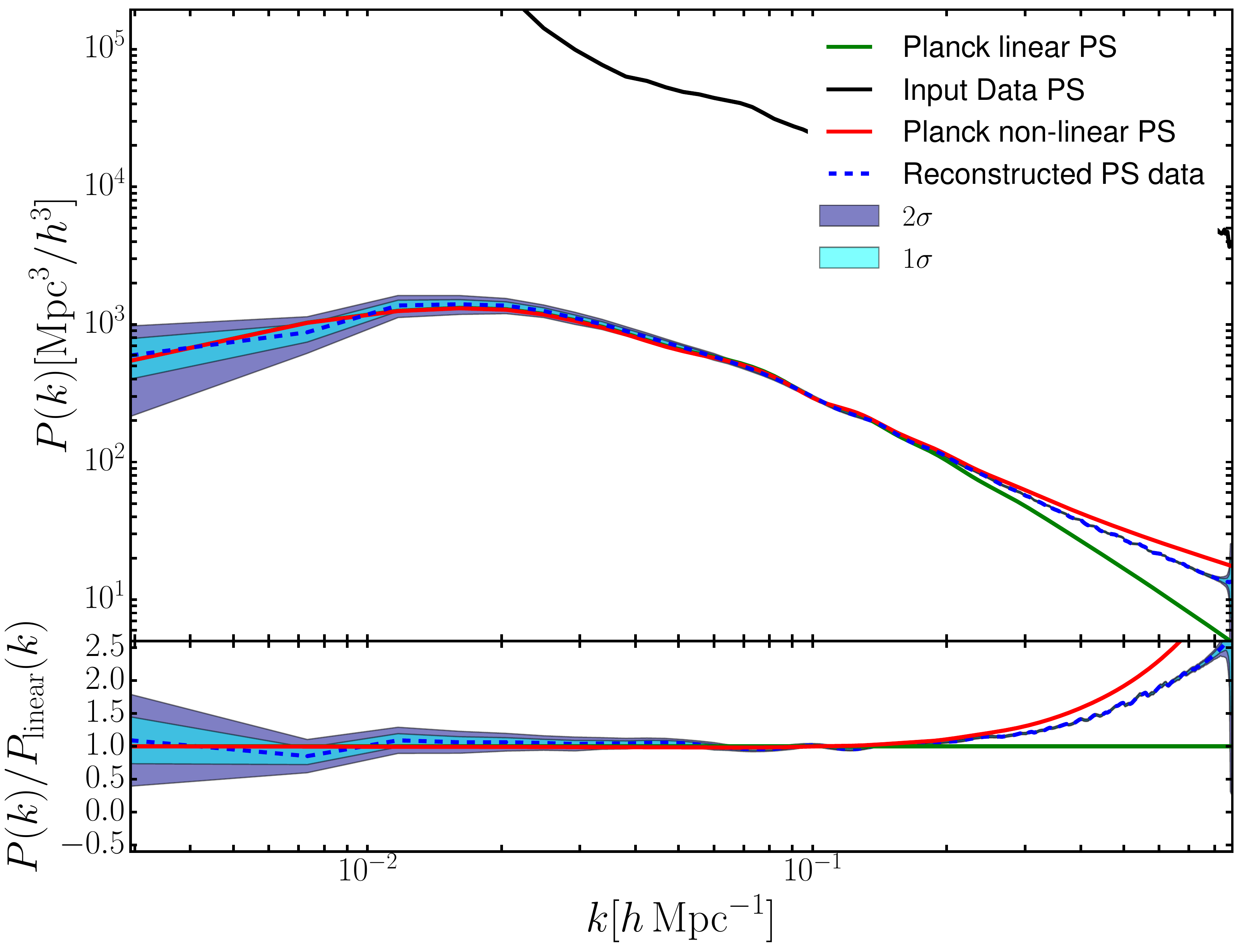}
\put(-210,170){${\rm L}=3200\,h^{-1}\,{\rm Mpc}$}
\put(-210,160){$d_{\rm L}=6.25\,h^{-1}\,{\rm Mpc}$}
\put(-210,150){\rotatebox[]{0}{\text{MOCKS BOSS DR12}}}
\put(-210,140){with velocity dispersion}
\\
\includegraphics[width=8.5cm]{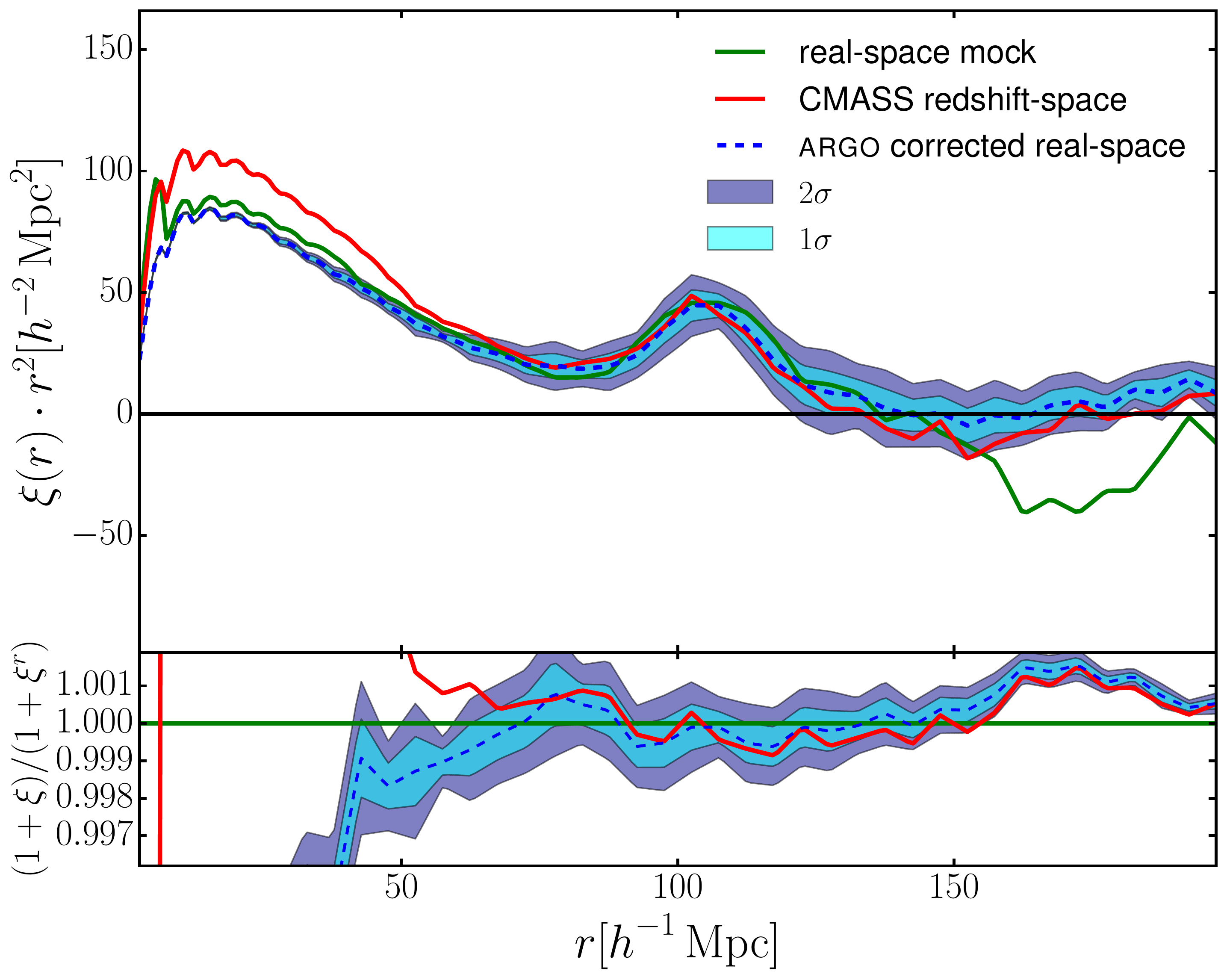}
\put(-210,100){${\rm L}=3200\,h^{-1}\,{\rm Mpc}$}
\put(-210,90){$d_{\rm L}=6.25\,h^{-1}\,{\rm Mpc}$}
\put(-210,80){\rotatebox[]{0}{\text{BOSS DR12}}}
\put(-210,70){with velocity dispersion}
\hspace{-16.17cm}
\includegraphics[width=8.5cm]{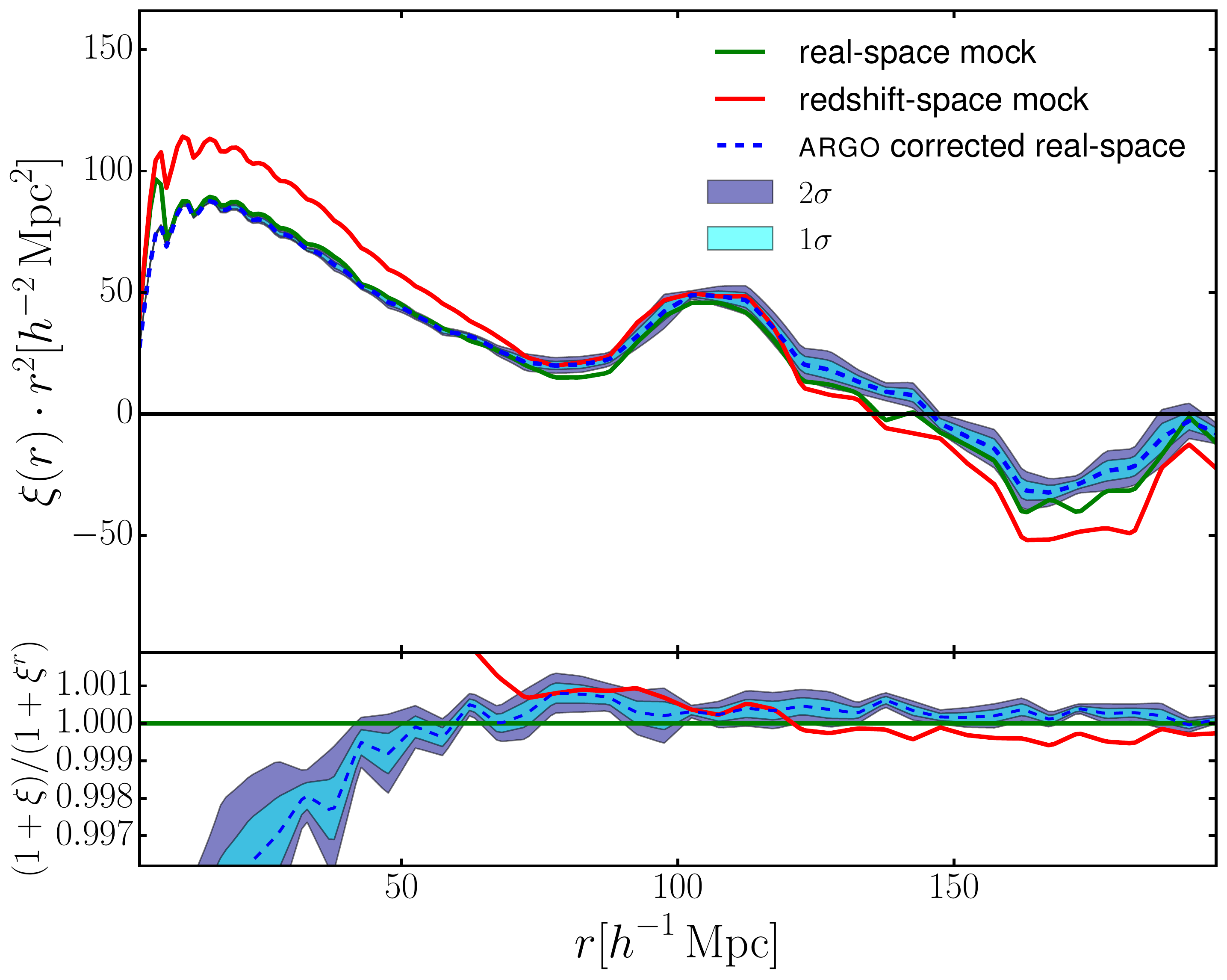}
\put(-210,100){${\rm L}=3200\,h^{-1}\,{\rm Mpc}$}
\put(-210,90){$d_{\rm L}=6.25\,h^{-1}\,{\rm Mpc}$}
\put(-210,80){\rotatebox[]{0}{\text{MOCKS BOSS DR12}}}
\put(-210,70){with velocity dispersion}
\vspace{-.845cm}
\\
\includegraphics[width=8.5cm]{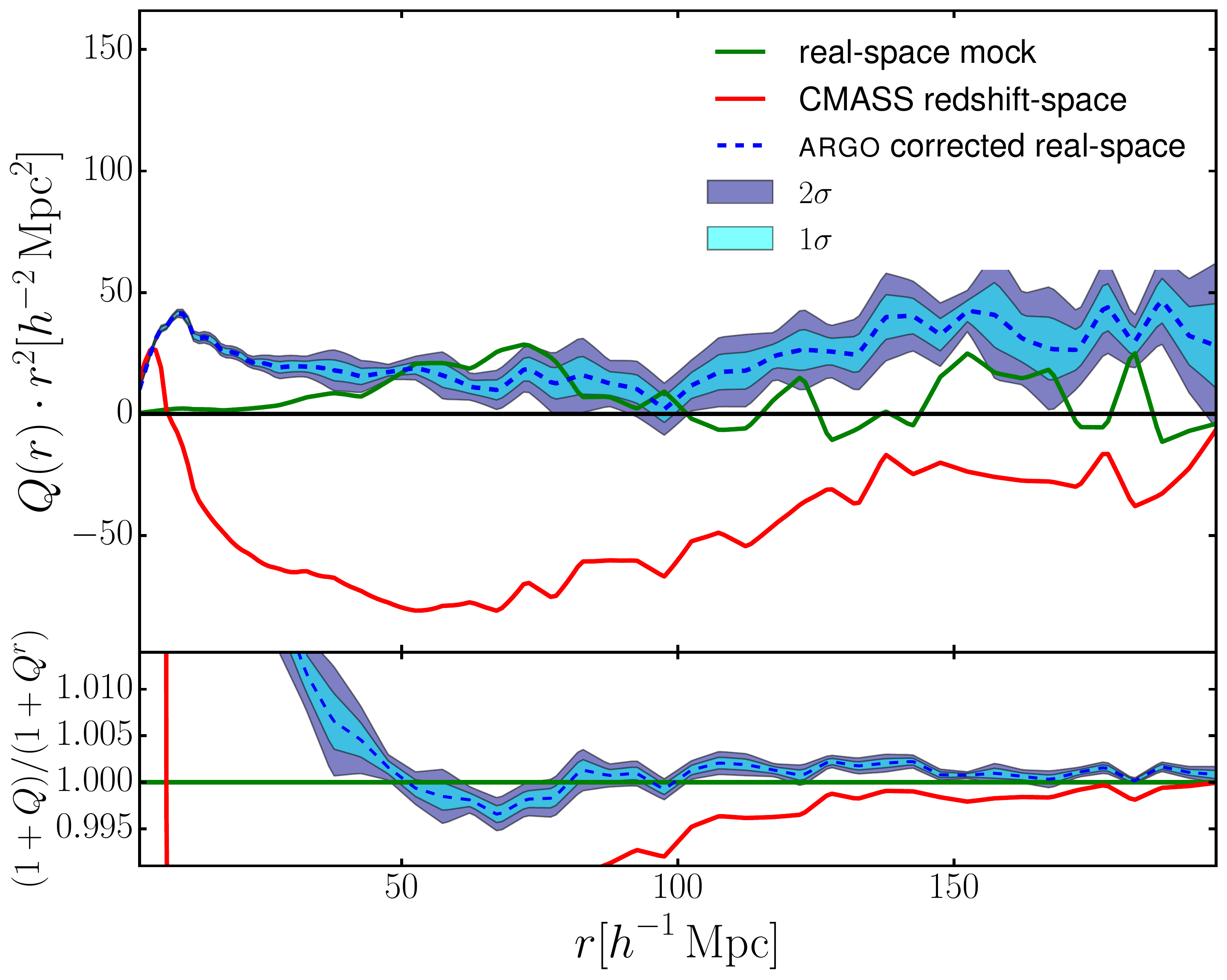}
\put(-210,170){${\rm L}=3200\,h^{-1}\,{\rm Mpc}$}
\put(-210,160){$d_{\rm L}=6.25\,h^{-1}\,{\rm Mpc}$}
\put(-210,150){\rotatebox[]{0}{\text{BOSS DR12}}}
\put(-210,140){with velocity dispersion}
\hspace{-16.17cm}
\includegraphics[width=8.5cm]{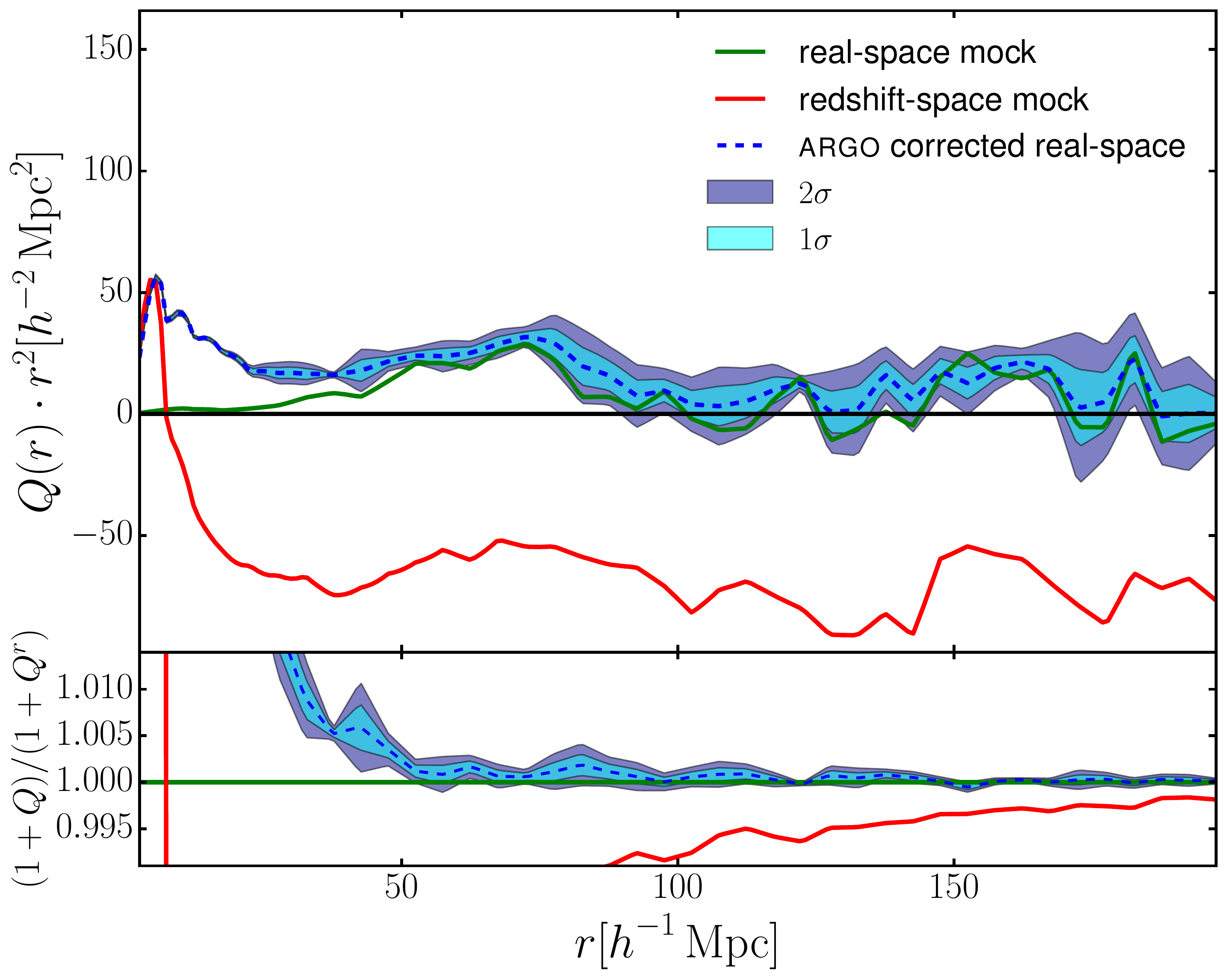}
\put(-210,170){${\rm L}=3200\,h^{-1}\,{\rm Mpc}$}
\put(-210,160){$d_{\rm L}=6.25\,h^{-1}\,{\rm Mpc}$}
\put(-210,150){\rotatebox[]{0}{\text{MOCKS BOSS DR12}}}
\put(-210,140){with velocity dispersion}
\end{tabular}
 \caption{\label{fig:app2} Monopoles and quadrupoles based on the ({\bf left panels:}) light-cone mock catalogue, ({\bf right panels:}) CMASS BOSS DR12 data. Same setting as in previous figure.
{\bf Upper panels:} power spectra  showing  the mean  (dashed blue line) over 6000 reconstructed dark matter fields on a mesh with 1 and 2 $\sigma$ contours (light and dark blue shaded areas, respectively), as compared to the raw galaxy power spectrum (black solid line), the nonlinear  (red solid line) and the linear power spectrum (green solid line) assuming the fiducial cosmology, and below the corresponding ratio with respect to the linear power spectrum. {\bf Middle panels:} two-point correlation functions of the galaxy distribution showing the  mean  (dashed blue line) over 6000 reconstructed real space catalogues with 1 and 2 $\sigma$ contours (light and dark blue shaded areas, respectively), in addition,  the real (green line for mocks only) and redshift space  (red line) catalogues, and below the corresponding ratio with respect to the real space correlation function ($\xi^{\rm r}$). {\bf Lower panels:} corresponding quadrupole correlation functions to the middle panels. }
\label{fig:}
\end{figure*}

{\color{black}
Here we present results from the Bayesian reconstruction of cosmic density and peculiar velocity fields applied to the full volume covered by CMASS BOSS DR12 data. The chosen resolution is 6.25$\Mpc$ on a 3D cubical mesh of side 3200 $\Mpc$ and 512$^3$ cells. A random velocity dispersion term was included in the iterative procedure, as explained in \S \ref{sec:vel}. The completeness, galaxy number counts, and  reconstructions are shown in Fig.~\ref{fig:app1}. The variance of the density and peculiar velocity field reconstructions in the bottom panels show that the uncertainty in the reconstructed fields increase towards lower completeness, being largest in the unobserved regions. 
The analysis of the reconstructed dark matter density field power spectrum shown in the upper panels of Fig.~\ref{fig:app2} demonstrates for both mock and observation based catalogues that the reconstructed dark matter fields are unbiased towards large scales ($k\lsim0.2\hperm$). The correlation function study based on the galaxy catalogues depicted in the middle panels qualitatively demonstrates that we recover the real space correlation function including the real space baryon acoustic peak (see left panel based on mock data). 
The recovered quadrupole shown in the lower panel has the same features as the original real space mock galaxy catalogue on large scales. These results show that the method presented in this work is handling correctly the selection effects, biasing, and peculiar motions on large scales.}

\end{document}